\documentclass[iop]{emulateapj}
\usepackage{natbib}
\usepackage{amsmath}

\usepackage{graphicx}               
\usepackage{amsmath,amsthm,amsfonts,amsopn,amssymb} 

\newcommand{\cms}{\mbox{cm s$^{-1}$}}
\newcommand{\ms}{\mbox{m s$^{-1}$}}
\newcommand{\kms}{\mbox{km s$^{-1}$}}

\newcommand{\kepler}{\mbox{\it{Kepler}~}}

\def\code#1{\texttt{#1}}
\received{}
\accepted{}

\slugcomment{to appear in PASP}

\shortauthors{Fischer {\it et~al.}}
\shorttitle{State of the Field: EPRV}
\begin{document}

\title{State of the Field: Extreme Precision Radial Velocities}\altaffilmark{1}
\author{Debra A. Fischer\altaffilmark{2},
Guillem Anglada-Escude\altaffilmark{3, 4},
Pamela Arriagada\altaffilmark{5}, 
Roman V. Baluev\altaffilmark{6,7},
Jacob L. Bean\altaffilmark{8},
Francois Bouchy\altaffilmark{9},
Lars A. Buchhave\altaffilmark{10}, 
Thorsten Carroll\altaffilmark{11},
Abhijit Chakraborty\altaffilmark{12},
Justin R. Crepp\altaffilmark{13}, 
Rebekah I. Dawson\altaffilmark{14,15},
Scott A. Diddams\altaffilmark{16, 17},
Xavier Dumusque\altaffilmark{18},
Jason D. Eastman\altaffilmark{19},
Michael Endl\altaffilmark{20},
Pedro Figueira\altaffilmark{21},
Eric B. Ford\altaffilmark{14, 15},
Daniel Foreman-Mackey\altaffilmark{22, 23},
Paul Fournier\altaffilmark{24},
Gabor F\H{u}r\'{e}sz\altaffilmark{25},
B. Scott Gaudi\altaffilmark{26},
Philip C. Gregory\altaffilmark{27},
Frank Grundahl\altaffilmark{28},
Artie P. Hatzes\altaffilmark{29},
Guillaume H\'{e}brard\altaffilmark{30, 31},
Enrique Herrero\altaffilmark{32},
David W. Hogg\altaffilmark{22, 33},
Andrew W. Howard\altaffilmark{34},
John A. Johnson\altaffilmark{19},
Paul Jorden\altaffilmark{35},
Colby A. Jurgenson\altaffilmark{2},
David W. Latham\altaffilmark{19},
Greg Laughlin\altaffilmark{36},
Thomas J. Loredo\altaffilmark{37},
Christophe Lovis\altaffilmark{18},
Suvrath Mahadevan\altaffilmark{14, 15, 38},
Tyler M. McCracken\altaffilmark{2},
Francesco Pepe\altaffilmark{18},
Mario Perez\altaffilmark{39},
David F. Phillips\altaffilmark{19},
Peter P. Plavchan\altaffilmark{40},
Lisa Prato\altaffilmark{41},
Andreas Quirrenbach\altaffilmark{42},
Ansgar Reiners\altaffilmark{43},
Paul Robertson\altaffilmark{14, 15},
Nuno C. Santos\altaffilmark{21, 44},
David Sawyer\altaffilmark{2},
Damien Segransan\altaffilmark{18},
Alessandro Sozzetti\altaffilmark{45},
Tilo Steinmetz\altaffilmark{46},
Andrew Szentgyorgyi\altaffilmark{19},
St{\'e}phane Udry\altaffilmark{18},
Jeff A. Valenti\altaffilmark{47},
Sharon X. Wang\altaffilmark{14, 15},
Robert A. Wittenmyer\altaffilmark{48, 49},
Jason T. Wright\altaffilmark{14, 15, 38} } 

\email{debra.fischer@yale.edu}

\altaffiltext{1}{The content of this publication emerged from presentations and discussion by the 150 participants in the Second Workshop on Extreme Precision Radial Velocities, held at Yale University. The talks and posters from this meeting are archived at http://eprv.astro.yale.edu}
\altaffiltext{2}{Department of Astronomy, Yale University, New Haven, CT 06511 USA}
\altaffiltext{3}{Centre for Astrophysics Research, Science and Technology Research Institute, University of Hertfordshire, College Lane, Hatfield AL10 9AB, UK}
\altaffiltext{4}{School of Physics and Astronomy, Queen Mary University of London, 327 Mile End Road, London E1 4NS, UK}
\altaffiltext{5}{Department of Terrestrial Magnetism, Carnegie Institute of Washington, Washington, DC 20015, USA}
\altaffiltext{6}{Central Astronomical Observatory at Pulkovo of Russian Academy of Sciences, Pulkovskoje shosse 65, St Petersburg 196140, Russia}
\altaffiltext{7}{Sobolev Astronomical Institute, St Petersburg State University, Universitetskij prospekt 28, Petrodvorets, St Petersburg 198504, Russia}
\altaffiltext{8}{Department of Astronomy and Astrophysics, University of Chicago, 5640 S. Ellis Avenue, Chicago, IL 60637, USA}
\altaffiltext{9}{Aix Marseille Universit{\'{e}}, CNRS, LAM (Laboratoire d'Astrophysique de Marseille) UMR 7326, 13388 Marseille, France}
\altaffiltext{10}{Centre for Star and Planet Formation, Natural History Museum of Denmark \& Niels Bohr Institute, University of Copenhagen, \O ster Voldgade 5-7, DK-1350 Copenhagen K, Denmark}
\altaffiltext{11}{Leibniz-Institut f{\"u}r Astrophysik Potsdam, An der Sternwarte 16, 14482 Potsdam, Germany}
\altaffiltext{12}{Astronomy and Astrophysics Division, Physical Research Laboratory (PRL), Ahmedabad 380009, India}
\altaffiltext{13}{Department of Physics, University of Notre Dame, South Bend, IN, USA}
\altaffiltext{14}{Department of Astronomy and Astrophysics, The Pennsylvania State University, 525 Davey Laboratory,
University Park, PA 16802, USA}
\altaffiltext{15}{Center for Exoplanets and Habitable Worlds, The Pennsylvania State University, University Park, PA 16802, USA}
\altaffiltext{16}{National Institute of Standards and Technology, 325 Broadway, Boulder, CO 80305, USA}
\altaffiltext{17}{Department of Physics, University of Colorado, 2000 Colorado Ave, Boulder, CO, 80309, USA}
\altaffiltext{18}{Observatoire Astronomique de l'Universit\'{e} de Gen\'{e}ve, 51 chemin des Maillettes, CH-1290 Versoix, Switzerland}
\altaffiltext{19}{Harvard-Smithsonian Center for Astrophysics, Cambridge, MA 02138, USA}
\altaffiltext{20}{The University of Texas at Austin and Department of Astronomy and McDonald Observatory, 2515 Speedway, C1400, Austin, TX 78712, USA}
\altaffiltext{21}{Instituto de Astrof\'isica e Ci\^encias do Espa\c{c}o, Universidade do Porto, CAUP, Rua das Estrelas, 4150-762 Porto, Portugal}
\altaffiltext{22}{Center for Cosmology and Particle Physics, Department of Physics, New York University, USA}
\altaffiltext{23}{Astronomy Department, University of Washington, Box 951580, Seattle, WA 98195, USA}
\altaffiltext{24}{Fibertech Optica Inc., Canada}
\altaffiltext{25}{MIT Kavli Institute for Astrophysics and Space Research 77 Mass Ave 37-515, Cambridge, MA, 02139, USA}
\altaffiltext{26}{Department of Astronomy, Ohio State University, 140 W. 18th Ave., Columbus, OH 43210, USA}
\altaffiltext{27}{Physics and Astronomy, University of British Columbia, 6244 Agricultural Rd, British Columbia, Canada V6T 1Z1}
\altaffiltext{28}{Stellar Astrophysics Centre, Department of Physics and Astronomy, Aarhus University, Ny Munkegade 120, DK-8000 Aarhus C, Denmark}
\altaffiltext{29}{Th\"{u}ringer Landessternwarte, Sternwarte 5, Tautenburg 5, 07778 Tautenburg, Germany}
\altaffiltext{30}{Institut d'Astrophysique de Paris, UMR7095 CNRS, Universit\'{e} Pierre \& Marie Curie, 98bis boulevard Arago, F-75014 Paris, France}
\altaffiltext{31}{Observatoire de Haute-Provence, Universit\'{e} d'Aix-Marseille \& CNRS, F-04870 Saint Michel lÕObservatoire, France}
\altaffiltext{32}{Institut de Ci\`{e}ncies de lÕEspai (CSICÐIEEC), Campus UAB, Facultat de Ci\`{e}ncies, Torre C5 parell, 2a pl, 08193 Bellaterra, Spain}
\altaffiltext{33}{Center for Data Science, New York University, USA}
\altaffiltext{34}{Institute for Astronomy, University of Hawaii, Honolulu, HI 96822, USA}
\altaffiltext{35}{e2v Technologies, 106 Waterhouse Lane, Chelmsford, Essex, UK}
\altaffiltext{36}{UCO Lick Observatory, Department of Astronomy and Astrophysics, University of California at Santa Cruz, Santa Cruz, CA 95064, USA}
\altaffiltext{37}{Center for Radiophysics and Space Research, Space Sciences Building, Cornell University Ithaca, NY 14853-6801, USA}
\altaffiltext{38}{Penn State Astrobiology Research Center, The Pennsylvania State University, University Park, PA 16802, USA}
\altaffiltext{39}{NASA Headquarters, Washington D.C., USA}
\altaffiltext{40}{Department of Physics, Astronomy and Material Science, 901 S National Avenue, Missouri State University, Springfield, MO 65897, USA}
\altaffiltext{41}{Lowell Observatory, 1400 West Mars Hill, Road, Flagstaff, AZ 86001, USA}
\altaffiltext{42}{Landessternwarte, Zentrum f\"{u}r Astronomie der Universit\"{a}t Heidelberg, K\"{o}nigstuhl 12, 69117 Heidelberg, Germany}
\altaffiltext{43}{Institut f\"{u}r Astrophysik, Georg-August-Universit\"{a}, Friedrich-Hund-Platz 1, 37077 G\"{o}ttingen, Germany}
\altaffiltext{44}{Departamento de F\'isica e Astronomia, Faculdade de Ci\^encias, Universidade do Porto, Rua do Campo Alegre, 4169-007 Porto, Portugal}
\altaffiltext{45}{INAF-Osservatorio Astrofisico di Torino, Via Osservatorio 20, I-10025 Pino Torinese, Italy}
\altaffiltext{46}{Menlo Systems GmbH, Am Klopferspitz 19a, 82152 Martinsried, Germany}
\altaffiltext{47}{Space Telescope Science Institute, 3700 San Martin Dr., Baltimore, MD 21218, USA}
\altaffiltext{48}{School of Physics and Australian Centre for Astrobiology, University of New South Wales, Sydney, NSW 2052, Australia}
\altaffiltext{49}{Computational Engineering and Science Research Centre, University of Southern Queensland, Toowoomba, Queensland 4350, Australia}

\begin{abstract}
The Second Workshop on Extreme Precision Radial Velocities defined circa 2015 the state of the art Doppler 
precision and identified the critical path challenges for reaching 10 \cms\ measurement precision. The presentations 
and discussion of key issues for instrumentation and data analysis and the workshop recommendations for 
achieving this bold precision are summarized here. 

Beginning with the HARPS spectrograph, technological advances for precision radial velocity measurements
have focused on building extremely stable instruments. To reach still higher precision, future 
spectrometers will need to improve upon the state of the art, producing even higher fidelity 
spectra. This should be possible with improved environmental control, 
greater stability in the illumination of the spectrometer optics, 
better detectors, more precise wavelength calibration, and broader bandwidth spectra. 
Key data analysis challenges for the precision radial velocity community include distinguishing 
center of mass Keplerian motion from photospheric velocities (time correlated noise) and the proper 
treatment of telluric contamination. Success here is coupled to the instrument design, but also requires 
the implementation of robust statistical and modeling techniques. Center of mass velocities produce 
Doppler shifts that affect every line identically, while photospheric velocities produce line profile 
asymmetries with wavelength and temporal dependencies that are different from Keplerian signals. 

Exoplanets are an important subfield of astronomy and there has been an impressive rate of discovery 
over the past two decades. However, higher precision radial velocity measurements are required 
to serve as a discovery technique for potentially habitable worlds, to confirm and characterize 
detections from transit missions, and to provide mass measurements for other space-based missions. 
The future of exoplanet science has very different trajectories depending on the precision that can 
ultimately be achieved with Doppler measurements.

\end{abstract}


\section{Introduction}
The past two decades have been a golden era for the discovery of exoplanets --- thousands of exoplanets 
have been detected using Doppler measurements, transit photometry, microlensing, and direct imaging. 
Pioneering technology has driven remarkable acceleration in the rate of detections --- of special note: the clever 
use of an iodine reference cell for wavelength calibration and modeling of the instrumental profile 
to achieve radial velocity (RV) precisions of a few meters per second using general purpose spectrometers 
\citep{Butler1996, Marcy1992}; the stabilized HARPS spectrometer, 
which set a new standard in RV precision \citep{Mayor2003, Pepe2002}; ground-based transit surveys that evolved rapidly from 
demonstrating the existence of transiting extrasolar planets \citep{Henry2000, Charbonneau2000} 
to achieving 1-mmag photometric precision \citep{Johnson2009}; and the \kepler space mission with 
a dramatic improvement over the precision of ground-based photometry.  The \kepler mission 
confirmed earlier suggestions \citep{MayorUdry2008} that most of the stars in our galaxy have planetary systems and that small planets are 
ubiquitous \citep{Howard2012, Fressin2013, Buchhave2014}. However, while ever smaller and lower mass 
planets have been detected in the past few years, all current techniques fall just short of detecting 
so-called exo-Earths (planets that are nearly the mass of the Earth, orbiting at 
habitable zone distances) around nearby stars. 

A generation of students has grown up thinking of exoplanet science as a booming field where the rate of 
discovery and characterization has a steep and positive trajectory. However, the key to growth 
in this field has been the sequential improvements in technology and measurement precision and 
this is also a fundamental requirement for the future of the field.  If we keep making the same 
observations with the same instruments, we will improve the SNR and population size, but we will 
basically have the same results and the field will stagnate; advances 
in strategy, analysis and instrumentation are required to reach higher RV precision and to push into 
new parameter space.

New space missions including the Transiting Exoplanet Survey Satellite \citep[TESS;][]{Ricker2014}, the CHaracterizing ExOPlanet Satellite \citep[CHEOPS;][]{Fortier2014} and PLAnetary Transits and Oscillation of stars \citep[PLATO;][]{Rauer2014} will have the precision to detect transiting planets with small radii and will observe bright nearby stars. 
Doppler observations will provide critical support for these missions and will remain a key asset in the 
search for exo-Earths \citep{Mayor2014}. If we can improve our current instrumental precision and if we can distinguish 
stellar photospheric velocities from Keplerian motion, then we have a 
chance of detecting, confirming and characterizing a bounty of exoplanets in new mass and period regimes 
around nearby stars.   

In July 2015, we held the Second Workshop for Extreme Precision Radial Velocities (EPRVs) at 
Yale University with 
three goals: to examine the current state of the Doppler precision, to discuss recent advances, and to 
chart a course for eliminating the remaining obstacles on the road towards 10 \cms.  We discussed 
the future game-changers and the current bottlenecks. 
This paper summarizes the presentations and discussions from this workshop. 
Section \ref{section:soa} presents the state of the art for fourteen different RV surveys highlighted at the workshop 
and compares some of the key attributes correlated with high Doppler precision. In Section \ref{ic}, we  
discuss instrumentation challenges and solutions. In Section \ref{jit} we discuss 
the challenges to high precision that are not instrumental in nature, such as stellar atmospheric 
velocities and telluric lines in the Earth's atmosphere and capture the discussion about statistical 
analysis techniques that show promise in distinguishing stellar photospheric signals from 
Keplerian motion.  

\section{State of the Art} 
\label{section:soa}
It is important to first clarify what we mean by Doppler precision. Radial velocity measurements are currently 
carried out by one of two methods: the iodine technique or the cross 
correlation technique \citep{LF2010}. The estimated single measurement 
precision (or averaged nightly velocity precision) is a good indication of the photon-limited ``on sky" precision. 
The long-term velocity rms (several days to years) is yet another indicator, exposing both systematic instrumental errors and the limitations in analysis techniques for treating variability from the stellar 
photosphere (so-called stellar ÒjitterÓ). 

Participants at the workshop discussed the Doppler precision for current radial velocity 
planet search programs.  In standardized formats, presenters were asked to show the dependence of single 
measurement precision (SMP) on the signal to noise ratio (SNR) for the equivalent of a 3 \kms\ pixel bin at 550 nm and to provide 
the radial velocity rms of their program stars without removing trends or Keplerian signals.  This is the 
first side-by-side comparison that has ever been made for Doppler planet search programs and we owe a 
great debt of gratitude to the teams that contributed data for this comparison\footnote{Additional current or 
planned PRV spectrographs can be found in Tables 2 and 3 of \citet{Plavchan2015}}.
Their generosity in sharing accumulated data allows the community to 
consider the possible relative importance of environmental stability, 
spectral resolution, wavelength coverage, calibration techniques and observing cadence. 
No comparison metric is perfect and there will still be discrepancies 
that arise because of differences in the stellar samples and the scientific goals of the programs.

\begin{deluxetable*}{lrrrrccrl}
\tablecolumns{9}
\tabletypesize{\scriptsize}
\tablecaption{Current Doppler Planet Search Programs\label{RVsurveys} }
\tablehead{
\colhead{Spectrograph}  & \colhead{slit or}  & \colhead{Temp}  & \colhead{Spectral}  & \colhead{Wavelength}      & \colhead{Wavelength} &
\colhead{SMP [\ms]}   & \colhead{Number}   & \colhead{Duration}    \\
\colhead{ } & \colhead{fiber}  & \colhead{Control}  & \colhead{Resolution}   & \colhead{range [nm]}     & \colhead{calibrator} &
\colhead{SNR = 200}    &  \colhead{of stars}   & \colhead{of program}    \\
}
\startdata
HARPS      & f & Y & 115,000 & 380 -- 690   & ThAr  &  0.8           &  2000  & 2003 --            \\
HARPS-N  & f & Y & 115,000 &  380 -- 690  & ThAr & 0.8              & ~500 & 2012 --  \\
PARAS      & f & Y & 67,000 & 380 -- 690  & ThAr & 1.0                & 27   & 2012 --  \\
CHIRON   & f & Y &  90,000   & 440 -- 650    & Iodine &  1.0                    &  35    & 2011 --           \\
SOPHIE    & f & Y  & 75,000 & 387 -- 694  &   ThAr   &  1.1                     & 190    & 2011 --            \\
PFS           & s &  Y & 76,000 & 390 -- 670  &   Iodine &  1.2                    & 530   & 2010 --            \\
HIRES      & s &  Y & 55,000 & 364 -- 800   &   Iodine &  1.5                   & 4000  & 1996 --           \\
Levy (LCPS)  & s &  Y & 110,000 & 376 -- 970 &   Iodine & 1.5 &  100    & 2013 --          \\
Levy (CPS)  & s & Y & 100,000 & 376 -- 940 &    Iodine & 2.0                  &  300    & 2013 --           \\
SONG      & s & N & 90,000 & 440 -- 690 &   Iodine &  2.0                &  12      & 2014 --           \\
HRS         & s & Y & 60,000 & 408 -- 784  &   Iodine &  3.0                   & 100    &  2001 -- 2013 \\
Hamilton    & s & N & 50,000 & 390 -- 800 &    Iodine &  3.0          &  350    & 1987 - 2011  \\
UCLES      & s  & N & 45,000 & 478 -- 871  & Iodine  & 3.0     & 240  &  1998 -- \\
Tull            & s & N & 60,000 & 345 -- 980 &   Iodine &  5.0          &  200    & 1998 --           \\
\enddata
\end{deluxetable*}

Table \ref{RVsurveys} lists the key parameters for key Doppler surveys. The programs are sorted by the demonstrated single 
measurement precision.  We summarize some of the fundamental parameters for the programs, including 
whether the coupling of light is accomplished with a slit or fiber, whether there is any environmental control (however, the extent of 
environmental control varies widely), the spectral resolution, wavelength range, method of wavelength calibration, 
single measurement precision for SNR of 200 (at 550 nm), the number of stars on the program and the time baseline of the program.   
Fiber fed, stable instruments with high spectral resolution represent 
the state of the art circa 2015.
Future instruments will require even more stringent specifications in order to gain another order of 
magnitude in RV precision.  High resolution and broad bandwidth may end up being critical parameters 
for fitting out perturbing signals from stellar photospheres and telluric contamination. 

The Doppler programs listed in Table \ref{RVsurveys} are discussed in more detail below in order of the date that 
each planet survey began.

\subsection{3-m Lick Hamilton, 1987 - 2011}\label{Hamilton}

 The Hamilton spectrograph \citep{Vogt1987} at Lick Observatory\footnote{Presentation by Debra Fischer} 
 was used to search for exoplanets 
 between 1987 and 2011 \citep{Fischer2014, Butler1997}. The program effectively ended in 2011 when insulating 
 material on the cell burned and changed the iodine spectrum of the cell. The Hamilton echelle 
 spectrograph is a general purpose instrument that can be fed with either the 
 0.6-m Coud{\'e} Auxiliary Telescope  
 or the 3-m Shane telescope. The wavelength range can be selected to include 
 wavelengths from 390 - 900 nm 
 in a single observation. The instrumental resolution depends on the selected slit width, varying from 
 30,000 - 115,000, but the typical resolution used for Doppler planet hunting was R $\sim 50,000$. The 
 instrument is located in a coud{\'e} room and experiences diurnal and seasonal temperature swings of 
 several degrees. One pioneering aspect of this program was the use of an iodine reference 
 cell for the wavelength calibration and modeling the line spread function \citep{Butler1996, Marcy1992} to 
 reach a measurement precision of 3 \ms. The use of a reference cell is ideal for general-purpose instruments 
 because the forward modeling of iodine accounts for variations in the SLSF 
 and instrumental drifts. 
 The iodine absorption lines only span a wavelength range from 510 to 620 nm; this limits the Doppler analysis 
 to about 110 nm of the spectrum and high SNR is required in order to model the multi-parameter 
 SLSF. 
 
The search for exoplanets at Lick Observatory began with a sample of 109 stars and the program 
was allocated roughly 25 nights per year; that allocation was augmented with as many as 50 additional 
nights per year on the smaller Coud{\'e} Auxiliary Telescope.  
The first assessment of the occurrence rate of exoplanets  \citep{Cumming1999} 
and the first assessment of the impact of the impact of photospheric magnetic activity on radial velocity 
precision \citep{Saar1998} was made using data from the Lick planet search program. The sample 
size was increased \citep{Fischer1999} to include an additional $\sim 300$ stars, with a focus on a subset 
of metal-rich stars. \citet{Johnson2007} carried out a search for planets orbiting subgiants. The 
Hamilton spectrograph was also used to study the impact of optical scrambling on radial velocity 
measurements \citep{Spronck2010, Spronck2012, SKFSS2012, Spronck2013}. 

\begin{figure*}[ht]     
\includegraphics[width=9cm, height=9cm, angle=-90, clip]{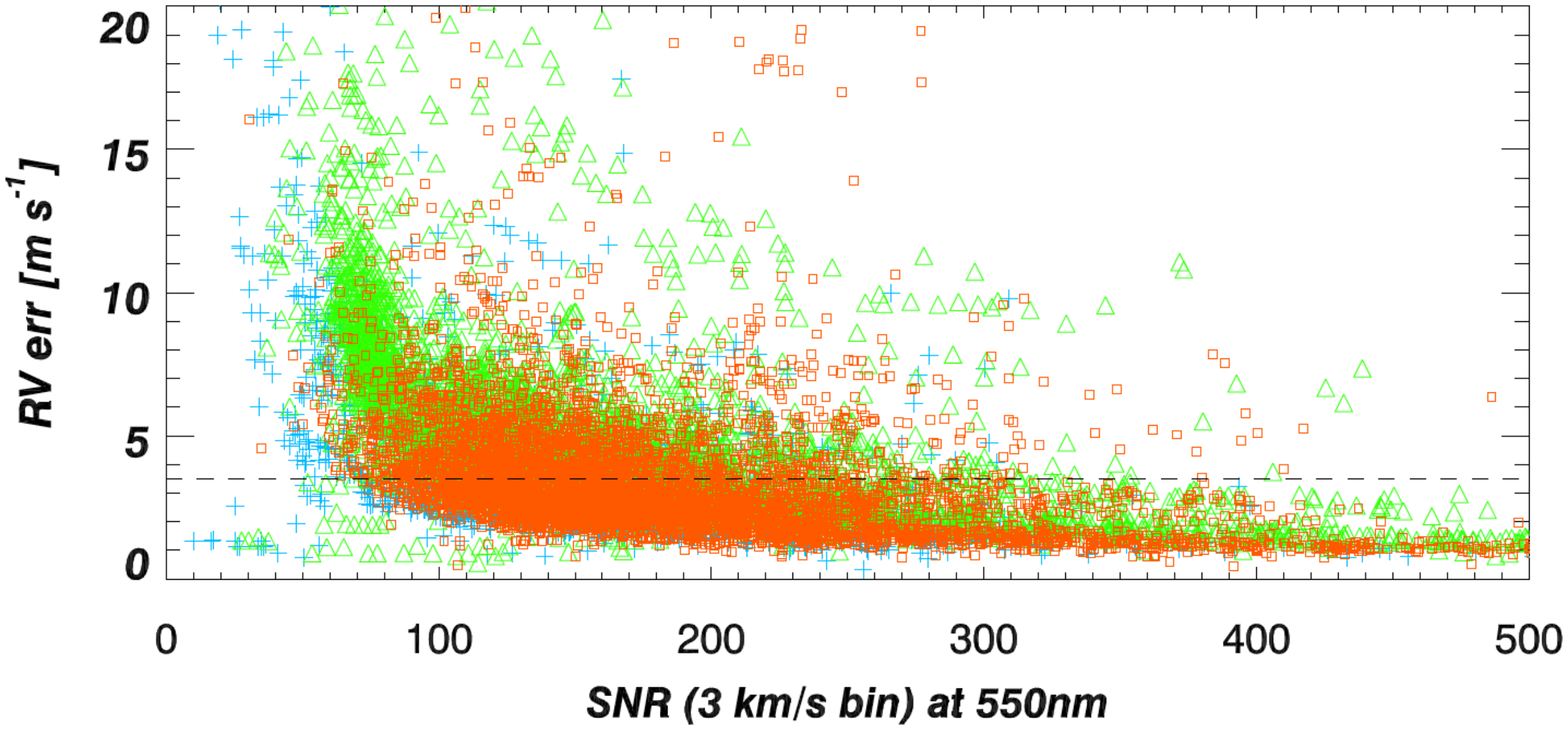}
\includegraphics[width=9cm, height=7.5cm, angle=-90, clip]{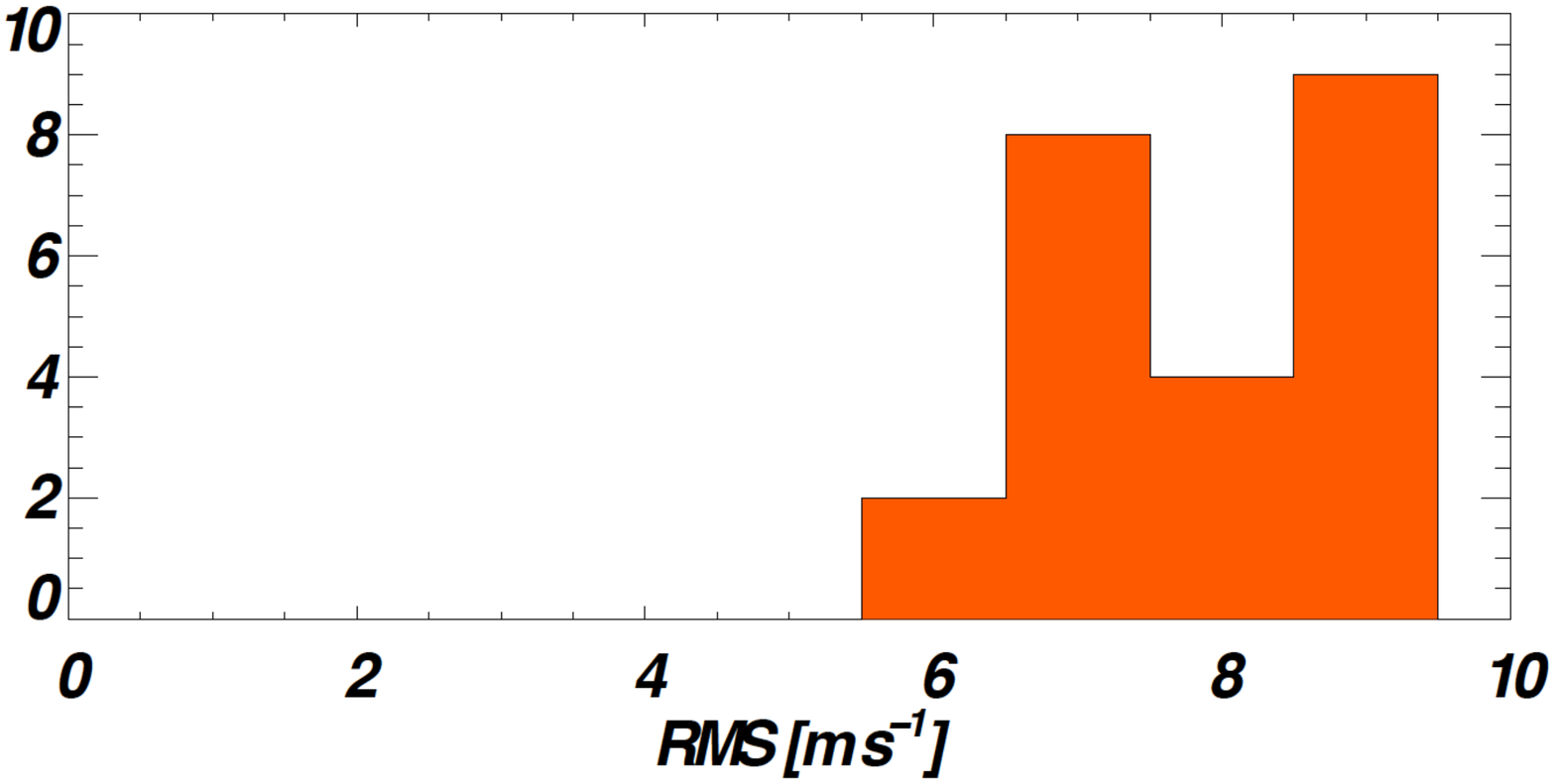}
\caption{Single measurement precision as a function of SNR (left) for dewar 13 (blue triangles) 
used from 1994 - 1997, dewar 6 (green diamonds) from 1998 - 2000, and dewar 8 
(red circles) from 2001 - 2011. The radial velocity rms scatter (right) is plotted
for data from the Hamilton spectrograph at Lick Observatory; most stars have 
rms scatter greater than the 10 \ms\ cutoff (Figures provided by Debra Fischer).
\label{lick} }                                                                                                          
\end{figure*}  

With the iodine modeling technique for the Lick Hamilton and CTIO CHIRON programs, the spectrum is 
divided into many small chunks, typically 2 or 3 \AA\ wide, and the wavelength solution, the spectral 
line spread function, and the Doppler shift are modeled for every chunk.  Each spectral chunk 
provides an independent estimate of the relative Doppler shift that is good to about 30 -- 50 \ms\ and 
$3\sigma$ outliers are rejected. Then, a weighted uncertainty is calculated for each chunk as follows.  
First, the difference, $d_i$, between the velocity for each $i^{th}$ chunk and the median velocity of all 
chunks in a given observation is calculated: 
\begin{equation}\label{dchnk}
d_i =  vel_i - median(vel)  
\end{equation}

\noindent
The uncertainty in the measurement for each chunk, $\sigma_i$, reflects the ability of that chunk to consistently report 
a Doppler shift and it is determined over several observations as the standard deviation of $d_i$ for $n$ observations:

\begin{equation}\label{sigobs}
\sigma_i =  stddev( d_i[1], ..., d_i[n_{obs}]  )
\end{equation}

\noindent
Thus, $d_i$ for a particular chunk may be large (or small) with many observations, but as long as it is consistently 
large (or small), $\sigma_i$ for the chunk will be small.  This allows for offsets between chunks (e.g., that might occur 
because of stitching errors in the CCD or other instrumental issues). Chunks with values of $d_i$ that are erratic from one 
observation to the next will have a large uncertainty, $\sigma_i$. A subtle corollary is that the sigma for a given chunk 
is better determined as more observations are accumulated, i.e., the uncertainty for a specific observation can change 
slightly as future observations are analyzed. 

The single measurement precision (SMP) at Lick was limited by instrumental errors, rather than photon statistics. 
The Hamilton spectrograph was not a stabilized instrument and this set the floor for the SMP 
of Doppler measurements to about 3 \ms\ \citep{Butler1996} with rms velocities 
greater than 5 \ms.  Figure \ref{lick} (left) shows the dependence of estimated internal 
errors on the SNR of the observations. The different families of errors (depicted with different 
plot symbols) correspond to eras where the CCD detectors were changed on the Hamilton 
spectrograph. In particular, the worst radial velocity precision for the Lick program occurred 
when a thinned CCD (dewar 6) was used; this device suffered from charge diffusion that broadened 
the SLSF and yielded a significantly degraded velocity precision. Figure \ref{lick} (right) is 
a histogram of the rms of the velocities. No trends or Keplerian signals were fitted out and 
most stars have rms velocities that were greater than the 10 \ms\ boundary of the plot. 

The Lick planet search program demonstrates the importance of using a stable, special-purpose 
instrument. Several different detectors were used over the course of the 25 year program and 
each of these changes to the instrument introduced systematic offsets in the time series radial 
velocity data. Improving the instrument stability was difficult in the large coud{\'e} room where 
large seasonal temperature were observed. 
The Lick planet search program was retired as the new Automated Planet Finder was being 
brought online.

\subsection{10-m Keck-1 with HIRES spectrograph, 1996 --}

In the early days of radial velocity planet searches, HIRES \citep{Vogt1994} on the 
Keck 10-m telescope\footnote{Presentation by Andrew Howard} was 
one of the instruments that made high impact contributions to exoplanet detections.  The 
program began in 1996 and has continued operations to this day. The spectrometer was 
designed as a slit-fed instrument and originally configured for a Doppler planet search with a 
wavelength range of 389 -- 618 nm and a 0 \farcs 8 slit that provides a moderate resolution of 
about 55,000 (higher resolution can be achieve by narrowing the slit with an accompanying 
loss of light). In 2004, the detector was upgraded to a mosaic of three CCDs, providing 
wavelength coverage of 364 -- 479 nm, 498 -- 642 nm, and 655 -- 797 nm for each detector 
respectively. The wavelength calibration is achieved with an iodine reference cell and the 
Doppler analysis is therefore restricted to wavelengths between 510 -- 620 nm. Although 
HIRES is not in a vacuum enclosure, the temperature is a fairly stable ($1^\circ  \pm 1^\circ$ C) in the instrument room.

The primary science goals are planet detection and characterization for a broad range of 
stellar populations \citep{Vogt2010a, Vogt2010b}. A hallmark of the Keck program is the collaborative 
integration of projects from several principal investigators, using a time-sharing scheme that pools 40 -- 60 nights per year. 
This strategy improves the observing cadence for several programs, including 
the California Planet Search \citep[CPS;][]{Howard2010}, 
the N2K (``next 2000") search for gas giant planets around metal rich stars \citep{Fischer2005}, a 
search for planets orbiting subgiants \citep{Johnson2007} and the ``Eta-Earth" program to 
systematically search for low mass planets ($3 -- 30 M_\oplus$) orbiting in the habitable 
zone of the nearest 230 GKM stars \citep{Howard2009}.  Keck HIRES programs also follow 
up transiting planets (e.g., \kepler, K2, HAT-NET) to determine masses \citep{Marcy2014, Weiss2014, MWP2014} 
and to measure the Rossiter-McLaughlin effect \citep{Sanchis-Ojeda2015, Winn2011, WHJ2011, Winn2010}.  
Keck data have been used to derive planet occurrence rates \citep{Cumming2008, Howard2012} 
and spectra have been used for characterization of host star metallicity and 
activity \citep{FV2005, VF2005, IF2010}. Collectively, more than 4000 stars have 
been observed and analyzed from Keck-HIRES. 

\begin{figure*}[ht]
\includegraphics[height=5.5cm, angle=0, clip]{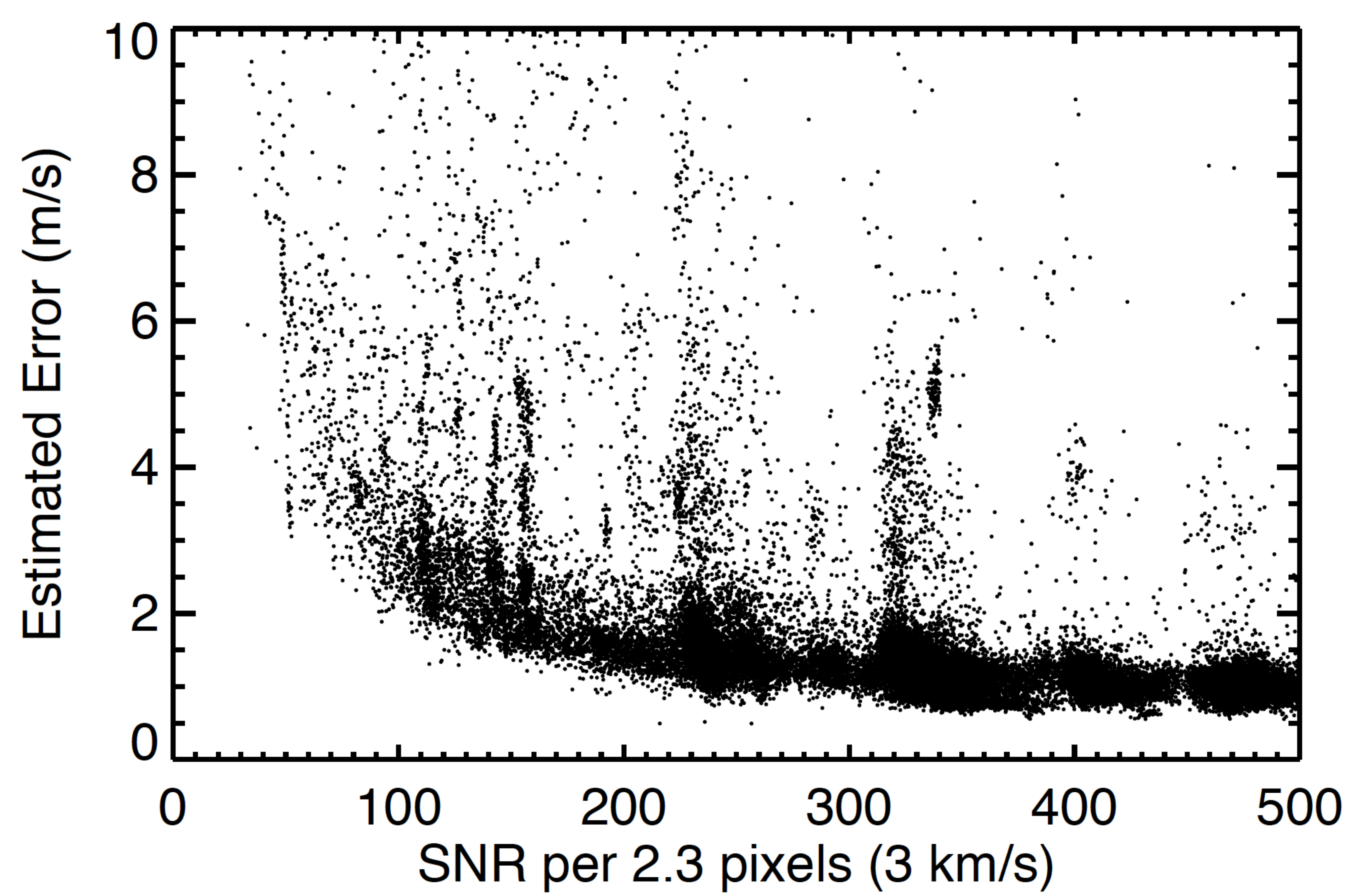}
\includegraphics[width=6cm, angle=90, clip]{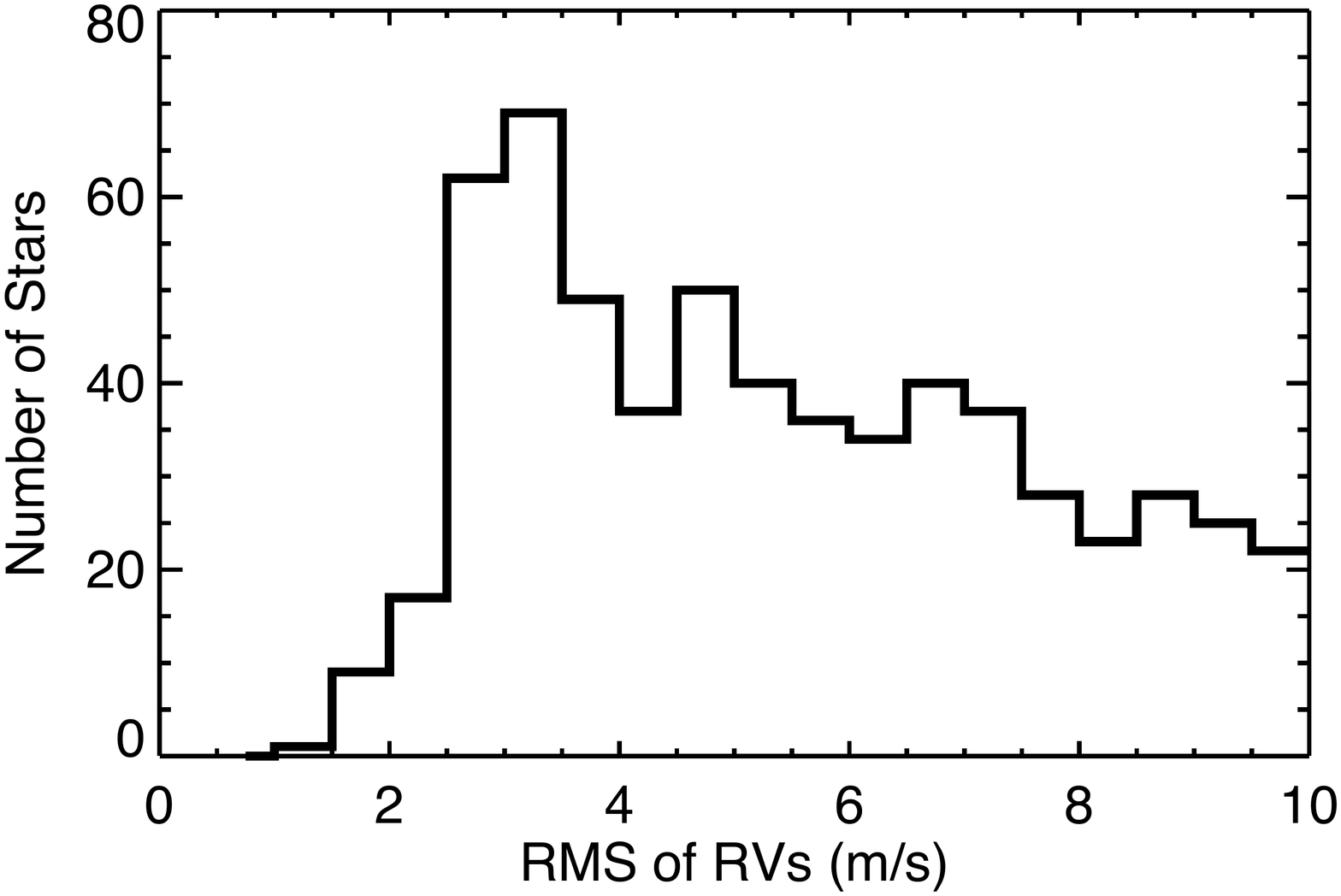}
\caption{Radial velocity precision (left) and rms velocity scatter for Doppler measurements 
by the California Planet Search team using the HIRES at Keck. (Courtesy of Andrew Howard)
\label{keck} }
\end{figure*}

Before the HIRES detector upgrade, a number of detector-related noise sources 
produced flux dependent errors. After correcting for this effect, the SMP was 3 \ms. When the detector was upgraded in August 2004, 
the SMP improved to about 1.5 \ms\ for a SNR of 200 and reached a floor of about 1 \ms\ for SNRs 
approaching 500. The radial velocity rms is typically greater than 2 \ms\ because of instrumental 
errors, limitations of the iodine technique \citep{IF2010, Spronck2015}, and the impact of stellar 
photospheric velocities, or stellar ``jitter." Figure \ref{keck} (left) shows the dependence 
of estimated internal errors for CPS observations as a function of the SNR of the observations. 
Figure \ref{keck} (right) is a histogram of the rms of the velocities (without attempting to remove any 
trends or Keplerian signals).

Although the iodine technique enables the use of general purpose instruments for precise 
Doppler measurements, this method requires high SNR (about 200 per pixel) for the forward 
modeling process. The variable SLSF contributes to increased 
scatter in the radial velocity measurements and limits the ability to take advantage of the 
exceptional signal to noise that is possible with a 10-m telescope. The CPS team is 
now designing SHREK, a new stabilized instrument for the Keck II telescope.

\subsection{2.7 McDonald and Tull Coud{\'e} spectrograph, 1998 --}

The Tull spectrograph \citep{Tull1995} has been used at the 2.7-m Harlan J. Smith telescope at 
McDonald Observatory\footnote{Presentation by Michael Endl} 
for exoplanet searches since 1998 and more than 10,000 RV measurements have been made over the 
past 15 years. The observing time is scheduled roughly once per month. The spectral resolution 
is 60,000 and the wavelength range is 345 -- 1000 nm.  The spectrometer is not environmentally 
stabilized and it is slit-fed; wavelength 
calibration is carried out with an iodine reference cell, restricting Doppler analysis to the wavelength range of 510 -- 620 nm. 

\begin{figure*}[ht]
\includegraphics[width=6cm, height=7.5cm, angle=-90, clip]{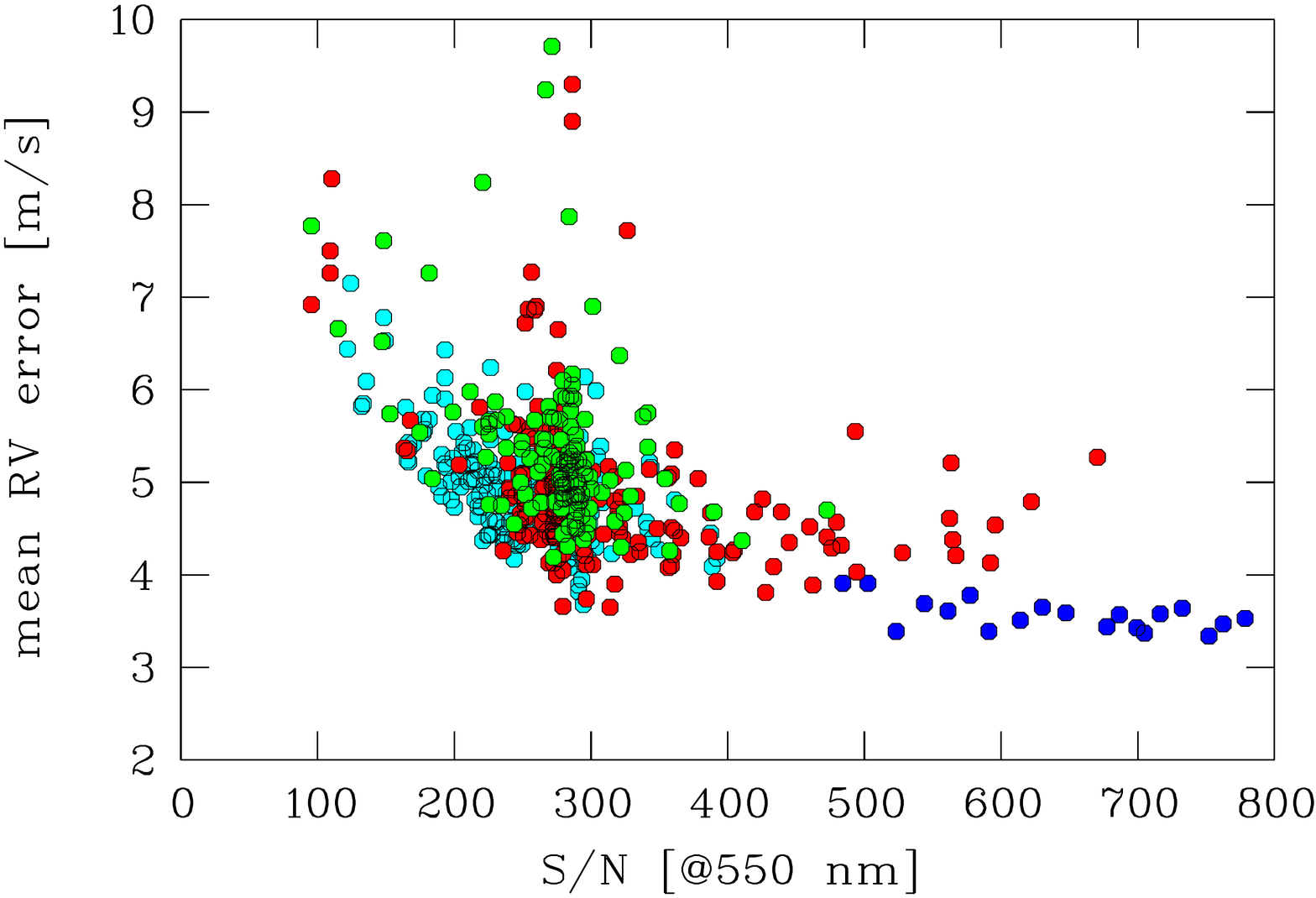}\includegraphics[width=6cm,height=7.5cm,angle=-90, clip]{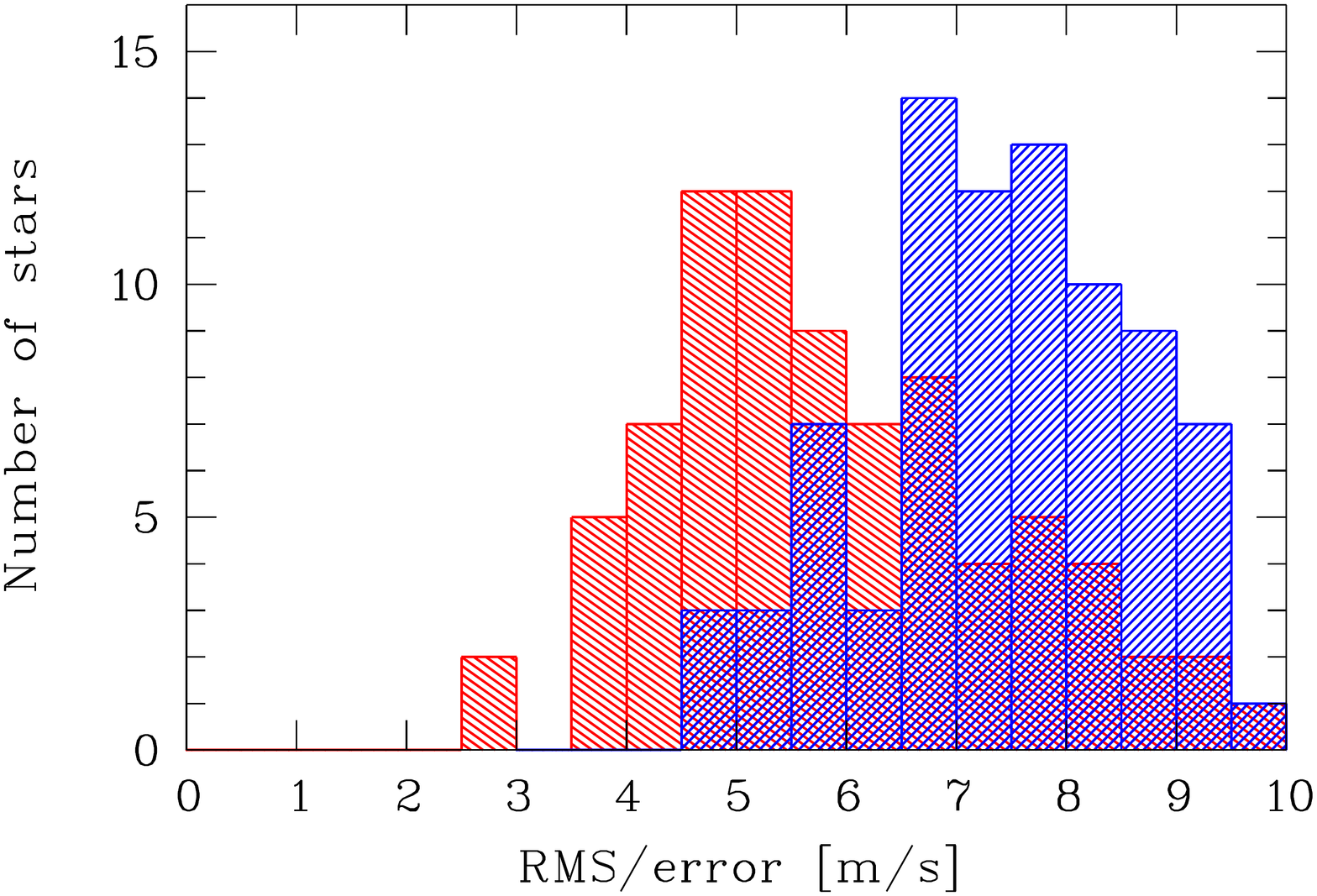}
\caption{(left) Estimated single measurement precision as a function of SNR using the Tull spectrograph on the 2.7-m Harlan Smith Telescope at McDonald Observatory. Different color plot symbols are for different stars.
(right) Radial velocity scatter (right) for the stars observed with this same spectrograph at McDonald Observatory. (Courtesy of Michael Endl)
\label{Tull} }
\end{figure*}

As shown in Figure \ref{Tull} there are about 200 stars on the exoplanet survey  
with the Tull spectrograph; 82 of these stars have an extensive baseline of observations. 
The primary science goal of this program is the detection of Jupiter analogs at 4 -- 5 AU. 

The internal errors represent the formal uncertainty of the weighted mean of the individual RVs that 
are modeled for small spectral chunks. These chunks are typically 2 -- 3 \AA\ wide and the internal 
error depends on the overall width of the distribution of the accepted chunks (extreme deviating chunk 
RVs are rejected) and their total number. A detailed description can be found in \citep{Endl2000}.
The mean SMP errors peak at about 5 \ms\ and the velocity rms scatter peaks at about 6 -- 7 \ms. When 
observing the sun through a solar port where there is a more homogeneous illumination of the slit, the RV data 
routinely improves to a precision of 2 -- 3 \ms\, demonstrating the importance of uniform illumination of the spectrometer optics. 

The most important weaknesses of this program are that this is a large multi-setup, multi-user instrument with 
complex, large optics, relatively poor telescope image stability and guiding and a 2-pixel sampling of the 
instrumental profile. The detector is a Tektronix CCD (circa 1990). The project could likely be improved 
with a new 15-micron CCD, octagonal fiber link, wavefront sensor for focus, and fast tip/tilt for image stabilization. 

\subsection{3.9-m AAT and UCLES Coud{\'e} spectrograph, 1998 --}

The Anglo-Australian Planet Search (AAPS) has been in continuous operation since 1998, using the University College 
London Echelle Spectrograph (UCLES) at the 3.9-m Anglo-Australian Telescope\footnote{Information provided by Rob Wittenmyer}.  
The 1 arcsecond slit is used for the AAPS and corresponds to a resolution of 45,000. The instrument wavelength range extends from 
4780 -- 8415 \AA\ and the instrument 
employs an iodine reference cell for wavelength calibration. For flexible scheduling, the 
program uses an automated queue-based observing system, which controls instrument configuration, removes 
operator error, and implements parallel detector readout and telescope slewing.

UCLES on the AAT has concentrated on the same sample of 240 stars from 1998 to 2012, detecting 
more than 45 new planets orbiting these stars 
\citep{Butler2001, Carter2003, Bailey2009, OToole2009, Tinney2011, Wittenmyer2014}. The AAPS has entered its final 
phase of operations and is concentrating on the key science question: What is the occurrence rate of Jupiter-analogs, gas 
giant planets still orbiting beyond the frost line with no inner gas giant planets? Owing to the legacy of the 17-year 
continuously-observed sample, the AAPS remains a leader in the detection of long-period planets.
 
\begin{figure}[ht]
\includegraphics[width=6cm, height=8cm, angle=0, clip]{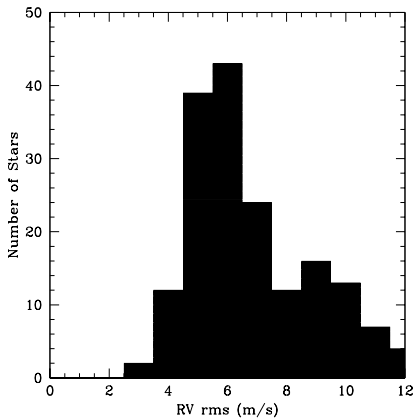}
\caption{Radial velocity  scatter (right) for UCLES. (Courtesy of Robert Wittenmyer)
\label{UCLES} }
\end{figure}

Although UCLES is a non-stabilized spectrograph, the AAPS has achieved a velocity precision of $\sim$3 \ms\ 
per epoch at a signal-to-noise of 200 per pixel and a resolution of 45,000. The quoted velocity uncertainties 
from the AAPS include the effects of photon-counting uncertainties, residual
errors in the spectrograph PSF model, and variation between the underlying
iodine-free template spectrum and spectra observed through the iodine cell.
The spectrum is broken up into several hundred 2 \AA\ chunks and velocities are derived 
from each chunk (Butler et al. 1996).  The final radial 
velocity measurement is determined as the mean of the chunks, and the quoted internal uncertainties are estimated 
from the scatter of the individual chunks \citep{Butler1996}. Figure \ref{UCLES} is a histogram of the rms of the velocities (without attempting to remove any 
trends or Keplerian signals).

Over the lifetime of the AAPS, the team 
has worked to improve the stability and precision of the radial velocity measurements.  In 2005 the program began using 
longer integrations, which averaged over stellar oscillation noise \citep{OToole2008} and 
delivered higher SNR (typically 400). As a result, the systematic and stellar uncertainties were reduced 
below 2 \ms\, as demonstrated for the bright triple-planet-hosting star 61 Vir \citep{Vogt2010b}.  This long-term 
stability is reflected in the number of long-period systems detected by AAPS; the majority of AAPS planets 
(62$\pm$13\%) have periods longer than one year.

To advance the precision at the AAT, a next-generation spectrograph, \textit{Veloce}, is being built as a replacement for 
UCLES, with commissioning planned in late 2016.  \textit{Veloce} is an 
asymmetric white-pupil echelle spectrograph that employs the innovations used by other PRV instruments (e.g. G-CLEF, 
SALT-HRS, and KiwiSpec) to maximize velocity stability: dual-shell thermal enclosure, vacuum-enclosed echelle 
grating, octagonal fibres, an innovative ball-lens double scrambler, and 
wavelength calibration with an ultra-stablized pulsed-laser frequency comb \citep{Murphy2007}.  This
calibration source will be injected into the simultaneous calibration fiber feed to achieve a wavelength 
calibration better than 20 \cms.

\subsection{9.2-m Hobby-Eberly telescope and HRS spectrometer, 2001 --}

The High Resolution Spectrograph \citep[HRS;][]{Tull1998} was commissioned at the 9.2-m Hobby-Eberly Telescope (HET) at McDonald 
Observatory\footnote{Presentation by Michael Endl} in 2001 and carried out a search for exoplanets through 2013.  The instrument and telescope are currently closed for the HET upgrade project, but will be recommissioned in 2016. The wavelength range of HRS is 408 -- 784 nm and the spectral resolution is R=60,000. An iodine reference cell is used for wavelength calibration. The instrument is not stabilized for temperature or pressure. Observations are carried out in queue mode scheduling and stars are observed between 1 and 5 times per month. Over the past 10 years, more than 1600 RV measurements have been obtained. 

The primary targets on the HRS program are 41 faint stars; these are typically M dwarfs with apparent magnitudes of 
$10 < V < 12$. There is also a program to carry out follow-up of \kepler transiting planet candidates for mass determination. 

\begin{figure*}[ht]
\includegraphics[width=6cm, height=8cm, angle=-90, clip]{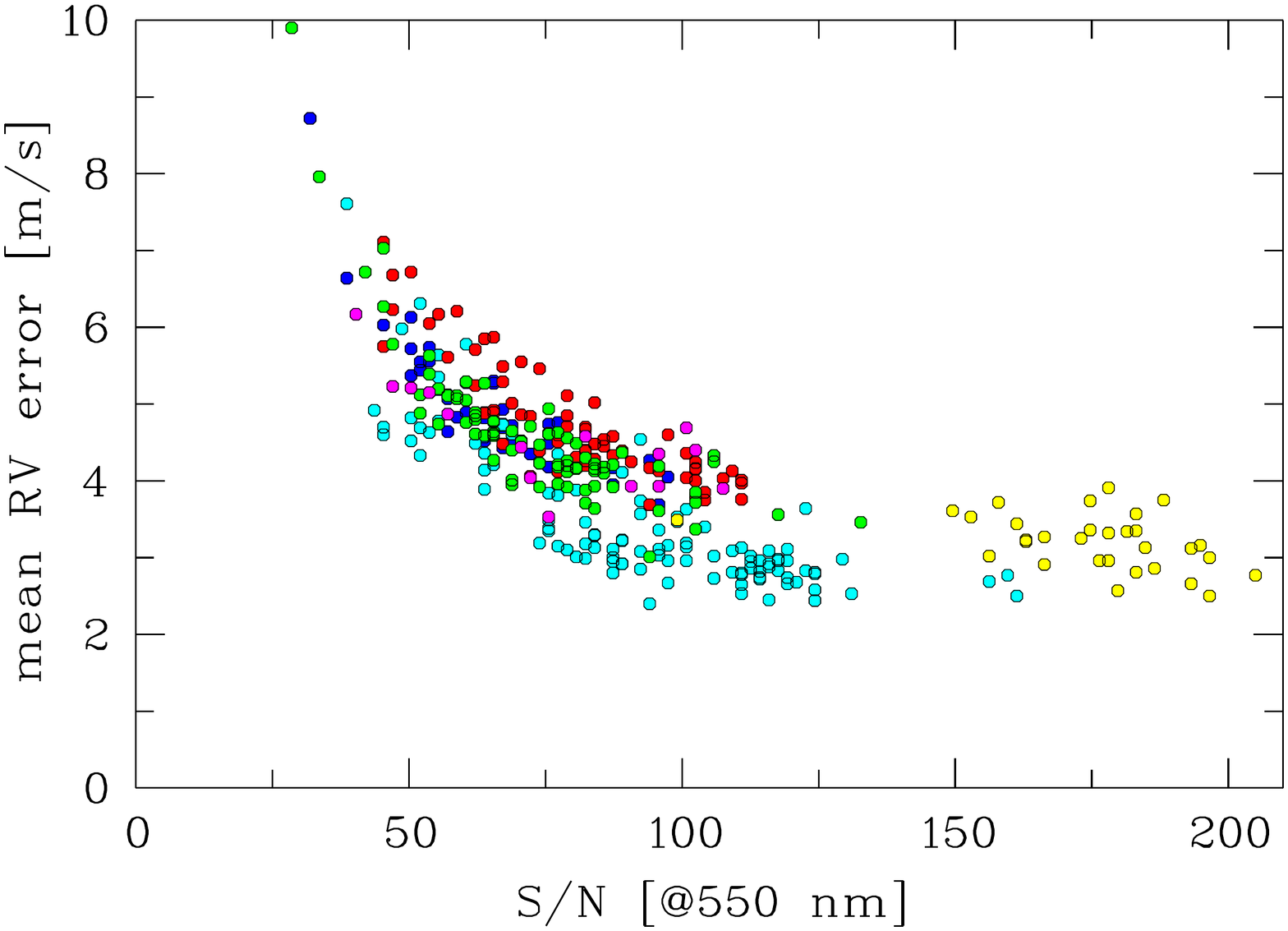}
\includegraphics[width=6cm, height=8cm, angle=-90, clip]{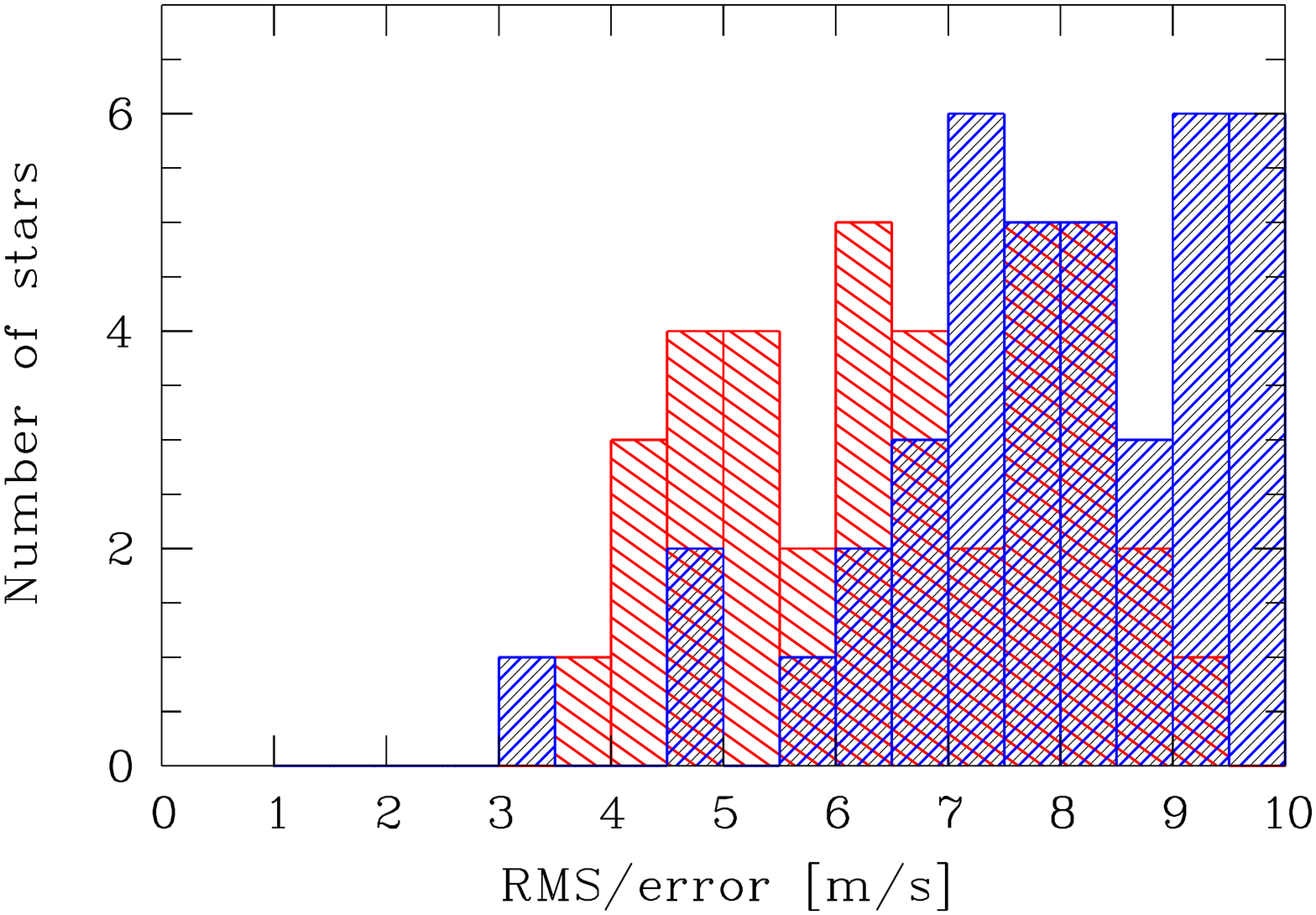}
\caption{Radial velocity precision as a function of SNR (left) and radial velocity scatter (right) for HRS. (Courtesy of Michael Endl)
\label{HRS} }
\end{figure*}

There are about 65\% slit losses with the HRS. This occurs because the slit must be narrowed to reach a resolution of 60,000.  As a result, the HRS observations have a SNR that is typically around 100; this reduces the Doppler precision when modeling with the iodine technique and yields estimated SMP errors that range from 4 to 8 \ms\ with a long term velocity rms that is greater than 6 \ms\ (Figure \ref{HRS}).

The instrument does not have an exposure meter; this is particularly critical for HET because the telescope design produces variable illumination at the telescope pupil.  The HRS-upgrade project is now underway with plans to provide image slicers for an increase in resolution to 70,000, a new efficient cross disperser, octagonal fiber feeds to provide some scrambling, and an exposure meter.  The improved efficiency should allow for observations of stars that are about two magnitudes fainter with the same SNR. The team notes that the last twelve RV measurements of sigma Draconis show a low rms of 1 \ms. These observations were taken when the outer ring of the HET mirror segments were already taken off, pointing to a possible problem with the optical quality of the camera that was improved by stopping down the pupil to stabilize the PSF.

\subsection{3.6-m ESO telescope and HARPS, 2003 --}\label{HARPS_Lovis}

The High Accuracy Radial Velocity Planet Searcher (HARPS) spectrometer was commissioned in 2003 \citep{Mayor2003, Pepe2002} on the 3.6-m telescope at La Silla\footnote{Presentation by Christophe Lovis} in Chile and was the first spectrograph to be specifically designed for a Doppler exoplanet survey. The instrument operates in a vacuum enclosure with a stable temperature of $17^\circ \pm 0.01^\circ$ C and the operating pressure is kept stable to better than 0.01 mbar since pressure changes will introduce drifts in the spectrum at the level of about 100 \ms\ per mbar. The fiber-fed instrument has a resolution of 115,000 and the calibration of the entire wavelength range from 380 nm to 690 nm is achieved with simultaneous reference from a Thorium-Argon lamp with a second fiber to track instrumental drifts during the night.  Doppler measurements are obtained by cross-correlating each spectrum with a template mask. HARPS uses a fiber double scrambler \citep{Avila2008} to invert the near and far fields producing a scrambling gain of about 10,000 for consistent illumination of the optics. HARPS ushered in a new era for radial velocity measurements, and was the first instrument to deliver better than 1 \ms\ RV measurement precision. 

Several exoplanet programs are carried out with HARPS, including a search of metal-rich and metal-poor stars \citep{Santos2014}, an M dwarf exoplanet search \citep{Bonfils2013}, and a search for super-Earth and Neptune mass planets around FGK stars \citep{Pepe2011}.  The particular HARPS program that was highlighted at the EPRV workshop was the high precision program that focuses on detection of the lowest mass exoplanets to reveal population statistics around FGK type stars.

\begin{figure*}[ht]
\includegraphics[width=8cm, height=6cm, clip]{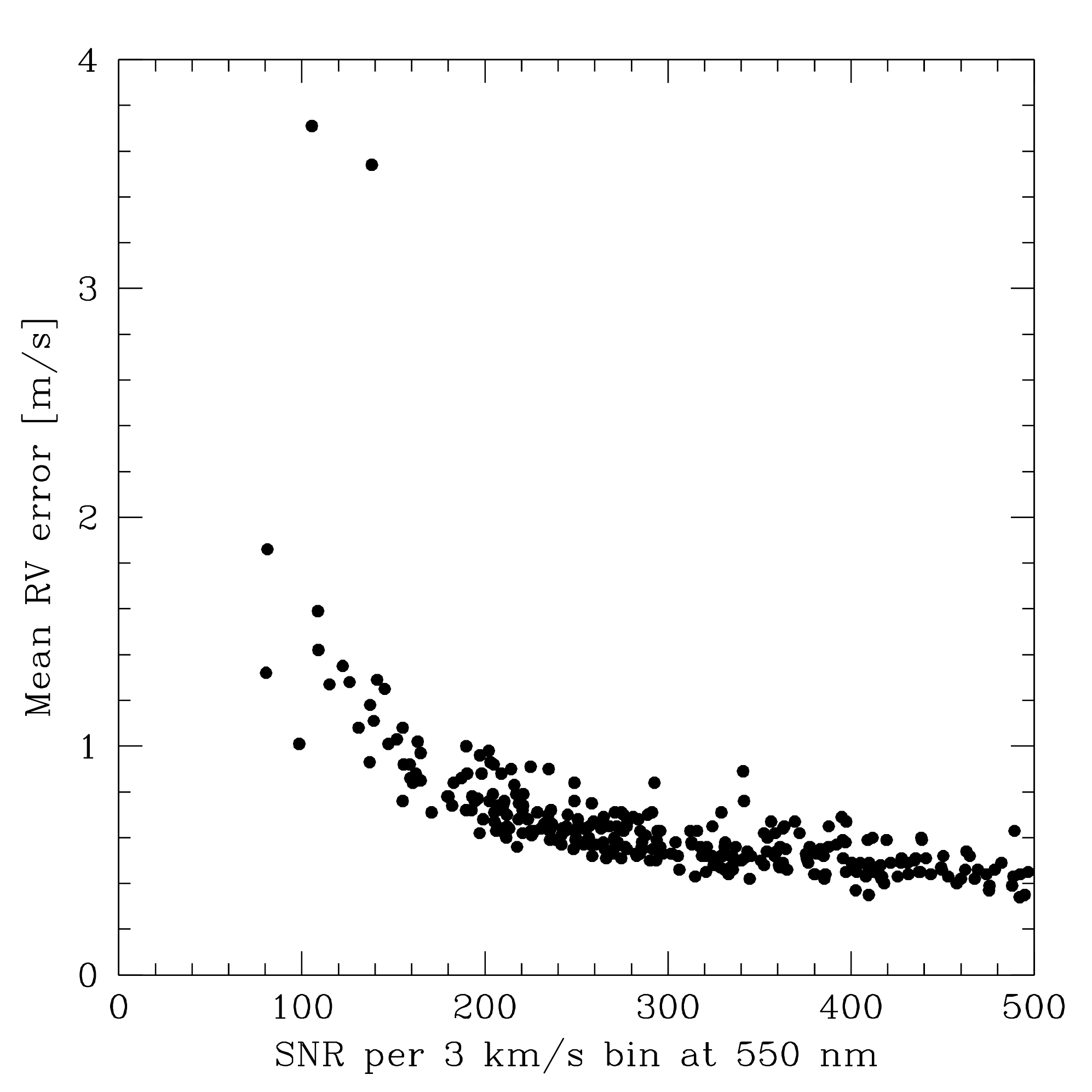}
\includegraphics[width=8cm, height=6cm, clip]{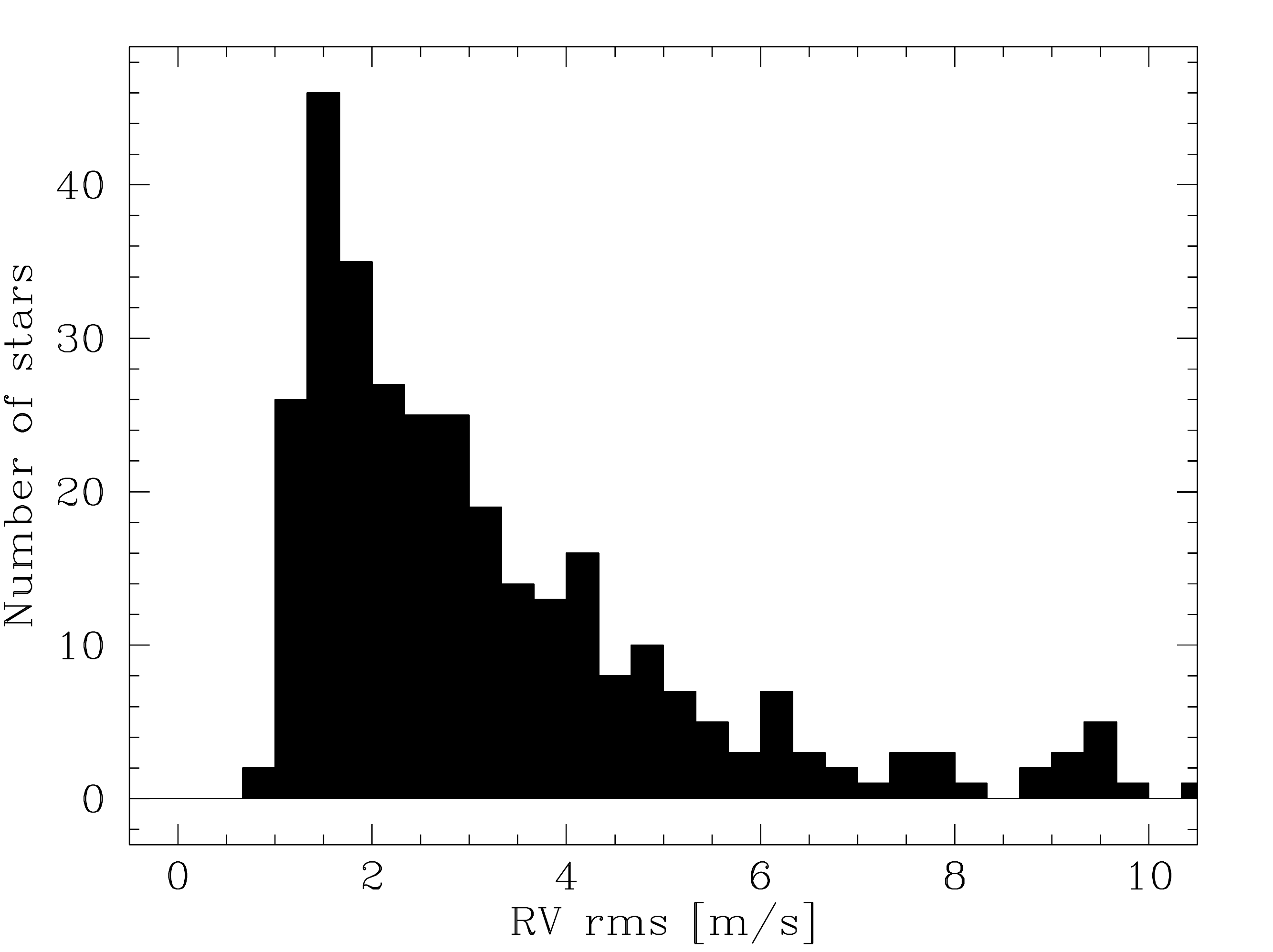}
\caption{Single measurement precision as a function of SNR (left) and radial velocity scatter (right) at HARPS for the target sample of FGK dwarfs to search for super Earth and Neptune mass exoplanets with HARPS. (Courtesy of Christophe Lovis)
\label{harps}}
\end{figure*}

HARPS and HARPS-N both use the cross correlation technique \citep{Baranne1996, Pepe2002}. The internal uncertainties are calculated as 
the quadratic sum of three different contributions: 1) photon and readout noise, 2) wavelength calibration error, 
and 3) instrumental drift error. The radial velocity (RV) uncertainty caused by photon and readout noise is obtained by 
propagating error bars from the spectrum to the weighted CCF \citep{Pepe2002}, and computing the fundamental RV 
uncertainty on the CCF using the formula in \citet{Bouchy2001}. A global uncertainty on the wavelength calibration is 
estimated from the rms dispersion of Thorium Argon (ThAr) lines around the wavelength solutions, as well as the 
total number of fitted ThAr lines. Finally, a drift uncertainty is also computed as the fundamental RV uncertainty on 
the simultaneous reference fiber. Combined together, these error sources provide a lower bound to the true error bars 
on the RVs. Additional contributions from stellar photospheric velocities, background contamination, and other instrumental effects 
should also be added in quadrature to the internal errors \citep{Diaz2016, Santerne2015}.

It is worth noting that HARPS was a significant financial investment for the European Southern Observatories (ESO), reflecting a strong vision and political commitment to the improvement of spectrometers for exoplanet detection. As Figure \ref{harps} (left) shows, a velocity precision of 0.8 \ms\ is achieved with SNR of 200.  Higher SNR improves the single measurement precision to better than 0.5 \ms.  The long term velocity rms scatter shown in Figure \ref{harps} (right) is also significantly lower than the velocity scatter with the iodine technique; many stars show an rms that is less than 2 \ms. HARPS pushed the instrumental errors lower by a factor of two relative to the general-purpose spectrometers with iodine reference cells for wavelength calibration. 

Importantly, the team sees evidence for photospheric activity that is imprinted in the stellar spectrum by using several indicators: line bisector shapes and variations in the full width half max for the CCF, and emission in the cores of Ca II H \& K lines \citep{Queloz2001, Figueira2013, GomesdaSilva2012}
The precision on HARPS is affected by guiding errors and variations in the focus, both of which lead to variable illumination of the spectrometer optics and instabilities in the SLSF  Last summer, the team implemented new octagonal fibers to produce a higher scrambling gain for the light injected into the spectrograph. The instrument is also now equipped with a laser frequency comb (LFC) for a better wavelength calibration. The commissioning tests with the LFC demonstrate a dramatic improvement in precision and also identified stitching errors in the pixel format of the CCD (discussed in Section \ref{ccd}). 

An important lesson learned from HARPS is that higher SNR, higher resolution, and higher fidelity spectra are needed to make significant progress on disentangling stellar activity from the center of mass velocities for chromospherically quiet stars.  In response to this, the team is building ESPRESSO on the Very Large Telescope (VLT) at Paranal in Chile. 

\subsection{Haute Provence 1.93-m telescope and SOPHIE spectrometer, 2006 --}

The ELODIE spectrograph \citep{Baranne1996} that was used to discover 51 Peg b \citep{MQ1995} and other exoplanets with the 1.93-m telescope at Observatoire de Haute-Provence was replaced by the SOPHIE\footnote{Presentation by Francois Bouchy} spectrograph in 2006 \citep{Perruchot2008}. That instrument has a resolution of 75,000 with spectral sampling of two pixels FWHM. The wavelength range is 387 -- 694 nm and wavelength calibration is carried out with  thorium argon injected in a second fiber for simultaneous wavelength calibration. In 2011, a significant upgrade was made to the spectrometer, and an octagonal fiber was added to provide better scrambling and to improve the precision \citep{Bouchy2013}.

Among the different SOPHIE planet surveys, the high-precision program focuses on 190 G \& K dwarfs using $\sim 16$\% of the telescope time distributed in fractional nightly allocations. The primary science goal is to search for Super-Earth and Neptune mass planets. There are also follow up programs for transit candidates and plans for future TESS follow-up. 

\begin{figure*}[ht]
\includegraphics[width=7.5cm, height=6cm, clip]{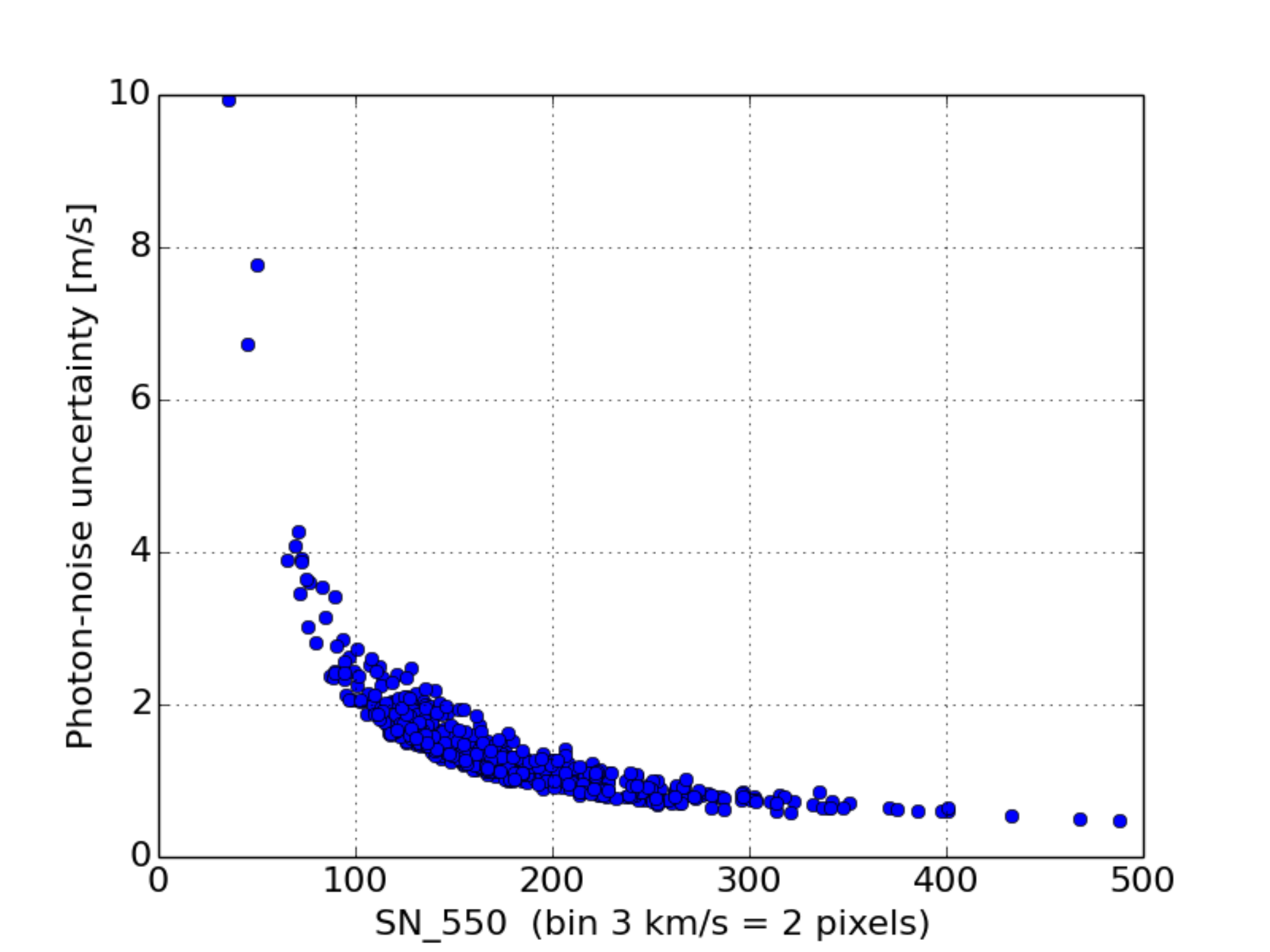}
\includegraphics[width=7.5cm, height=5.5cm, clip]{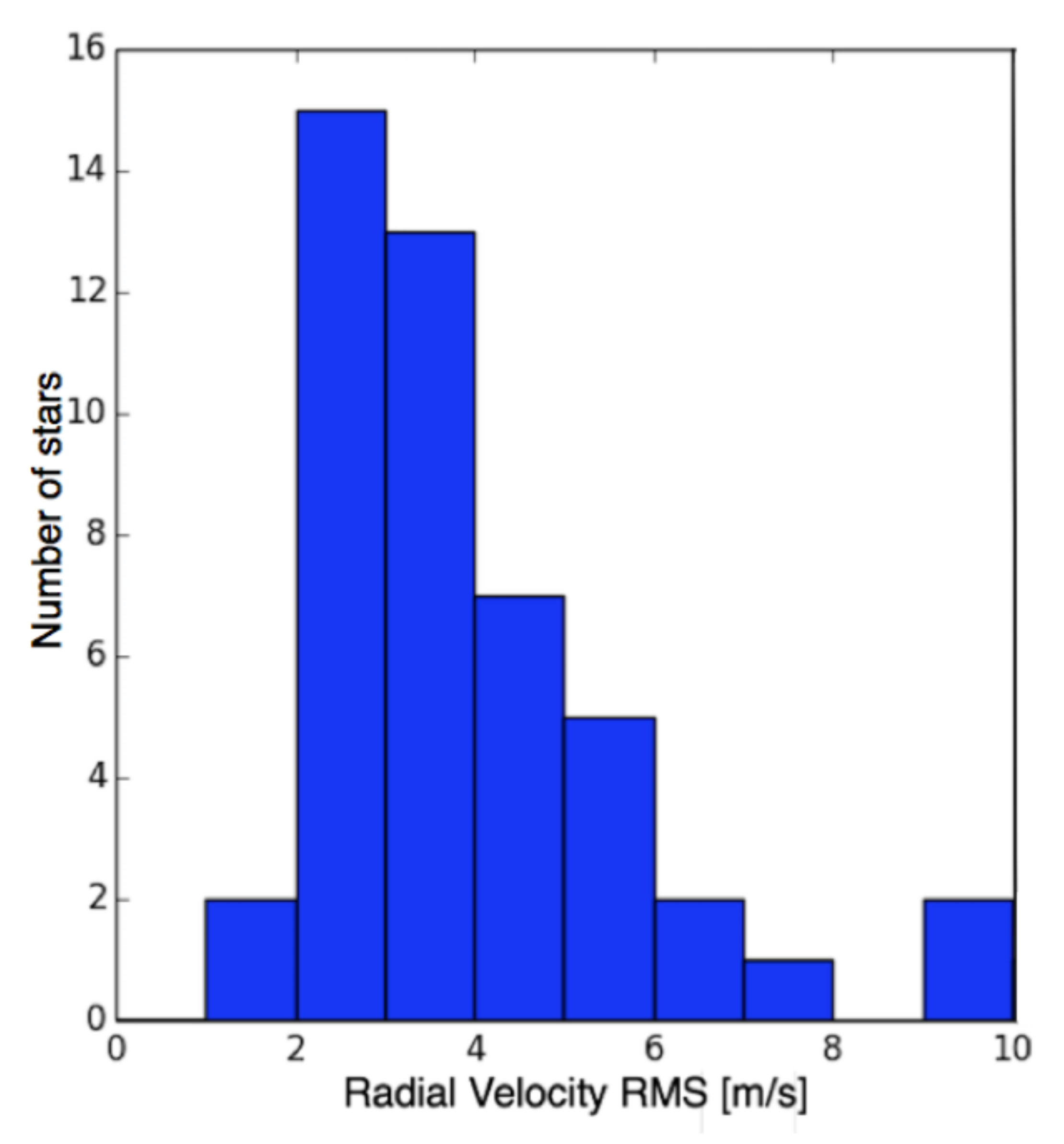}
\caption{Theoretical uncertainty based on SNR and radial velocity scatter for the SOPHIE 
spectrograph. (Courtesy of Francois Bouchy)
\label{sophie}}
\end{figure*}

As illustrated in Figure \ref{sophie} (left) the estimated SMP errors are about 1 \ms\ for SNR of 200. This SNR is reached in about fifteen minutes on V=7.5 stars. This exposure time is adequate to average over p-mode oscillations in solar type stars. There are 49 stars with more than 20 measurements and Figure \ref{sophie} (right) shows that the velocity rms peaks at $\sim 3$ \ms. 

One challenge with SOPHIE is the thermal coupling between the spectrograph and the telescope pillar, which 
introduces annual uncontrolled temperature variations. There are plans to improve the thermal control.  
The ThAr wavelength calibration is one of the current limiting factors in the RV precision and there are plans to implement a Fabry-P{\'e}rot etalon for simultaneous drift and wavelength calibration.  A new data reduction package is also being installed with all of the improvements from the most recent HARPS pipeline. 

One lesson from SOPHIE is that the charge transfer inefficiency of the CCD is a critical parameter.  In addition, scrambling of both the near and far field of the fiber output beam is important; the octagonal fiber was critical for helping to keep the RV precision of SOPHIE competitive. Finally, all available spectroscopic indicators (e.g., line bisectors, FWHM of the cross correlation function, Ca II H \& K line core emission) should be used to help interpret any signals in the radial velocity measurements.

\subsection{6-m Magellan telescope with the PFS spectrometer, 2010 --}

The Planet Finder Spectrograph \citep[PFS;][]{Crane2006} was commissioned at the 6.5-m Magellan telescope\footnote{Presentation by Pamela Arriagada} in 2010 and continues to be used to search for exoplanets. About 30 nights are allocated in 10-day runs each semester. The echelle grating is in a vacuum enclosure and the instrument has active thermal control. This slit-fed spectrograph has a resolution of 76,000 with a 0.5 arc second slit. The wavelength range extends from 390 to 670 nm and an iodine cell is used to calibrate the spectrum between 510 and 620 nm for Doppler measurements. 

There are about 530 stars on the Magellan PFS Doppler survey. The primary goal of this program is the detection of low mass companions around stars that are closer than 50 parsecs. There is also a collaborative program to obtain Rossiter-McLaughlin measurements to determine the coplanarity of transiting exoplanet orbits and follow-up programs for the Hungarian Automated Telescope - South (HAT-S) and the \kepler ``K2'' mission. 

\begin{figure*}[ht]
\includegraphics[width=8cm, height=10cm, clip]{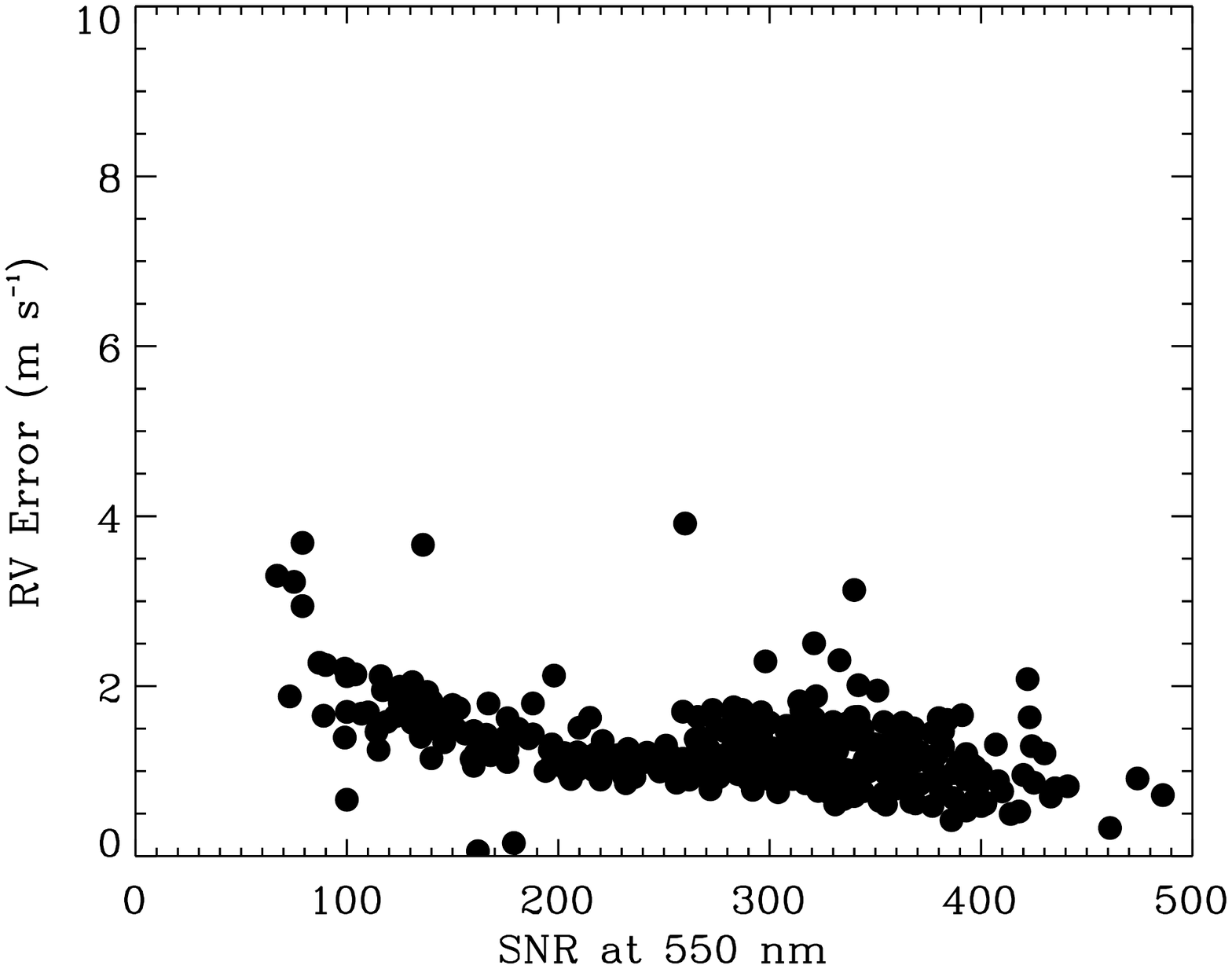}
\includegraphics[width=8cm, height=10cm, clip]{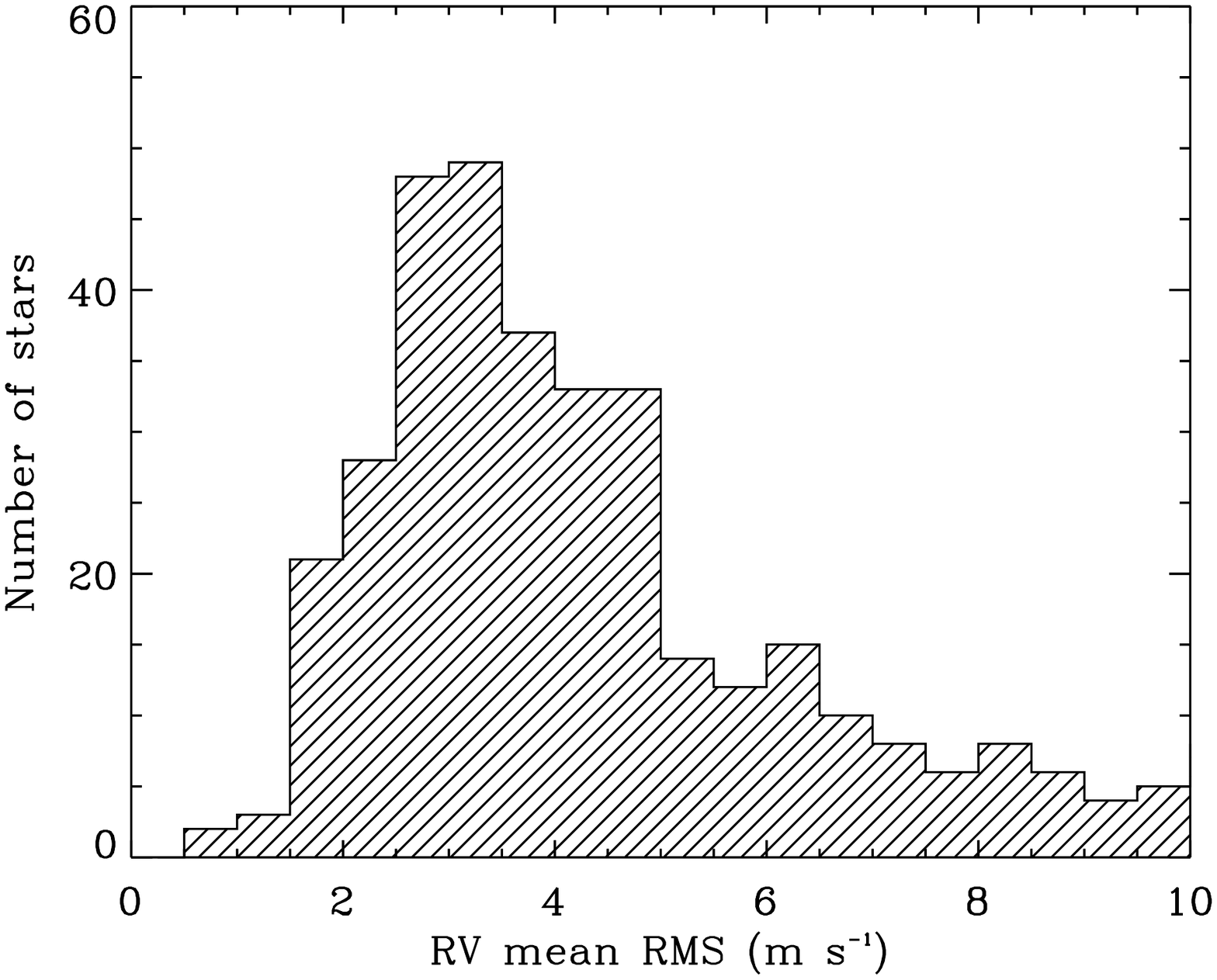}
\caption{Estimated single measurement precision as a function of SNR (left) and radial velocity scatter for PFS.
(Courtesy of Pamela Arriadada)
\label{pfs}}
\end{figure*}

Figure \ref{pfs} (left) shows that the internal SMP for PFS is about 1.2 \ms\ at a SNR level of 200, with an rms floor (Figure \ref{pfs}, right) that typically ranges between 2 -- 4 \ms. The typical SNR is about 300 for most Magellan PFS observations.

There are plans to refine the optical alignment of the spectrograph to achieve better image quality. Additional upgrades are being implemented this year, including replacing the CCD system with a device that has smaller pixels, better electronics and faster readout.  A pupil slicer is planned that will feed six optical fibers to increase the resolving power and to introduce some scrambling.  

The team is also improving the analysis to reduce the number of free parameters in the forward modeling of the iodine spectra.  The higher resolution template spectra provide higher precision especially for late-type stars.  High cadence observations are an advantage for detecting short-period planets during the 10-day observing runs. However, the gaps in these data sets also cause statistical sampling problems for detecting lower amplitude systems where the prospective Keplerian velocity amplitudes are comparable to the photospheric velocity jitter.

\subsection{1.5-m CTIO telescope with the CHIRON spectrometer, 2011 --}

The CHIRON spectrometer \citep{Tokovinin2013} was commissioned at the SMARTS 1.5-m CTIO telescope\footnote{Presentation by Debra Fischer} in 2011 and is still being used for exoplanet searches. The instrument was upgraded and recommissioned in 2012 with a new echelle grating in a vacuum enclosure, two stages of thermal control (the spectrograph enclosure and an outer room maintain temperature stability to $\pm 1$C), an exposure meter for calculating the observation midpoints for barycentric velocity corrections, an octagonal fiber feed for better scrambling, and a new CCD controller for faster readouts. The wavelength range of CHIRON extends from 440 to 650 nm with partial orders from 620 -- 870 nm. Wavelength calibration is accomplished with an iodine cell, limiting the Doppler information to the wavelength subinterval from 510 -- 620 nm. The bare fiber produces a spectral resolution of 28,000. An image slicer was built to increase the resolution to 90,000 and maximize throughput; however, the Doppler precision with the slicer was limited to a few meters per second, probably because of unstable illumination of the slicer and imperfect flat-fielding through the fiber slicer. Instead, a slit mask in front of the fiber is used for the high precision mode. This increases the resolution to either $R=90,000$ or $R=140,000$, albeit at the cost of lost light. 

\begin{figure*}[htb]
\includegraphics[width=6cm, height=8cm, angle=90, clip]{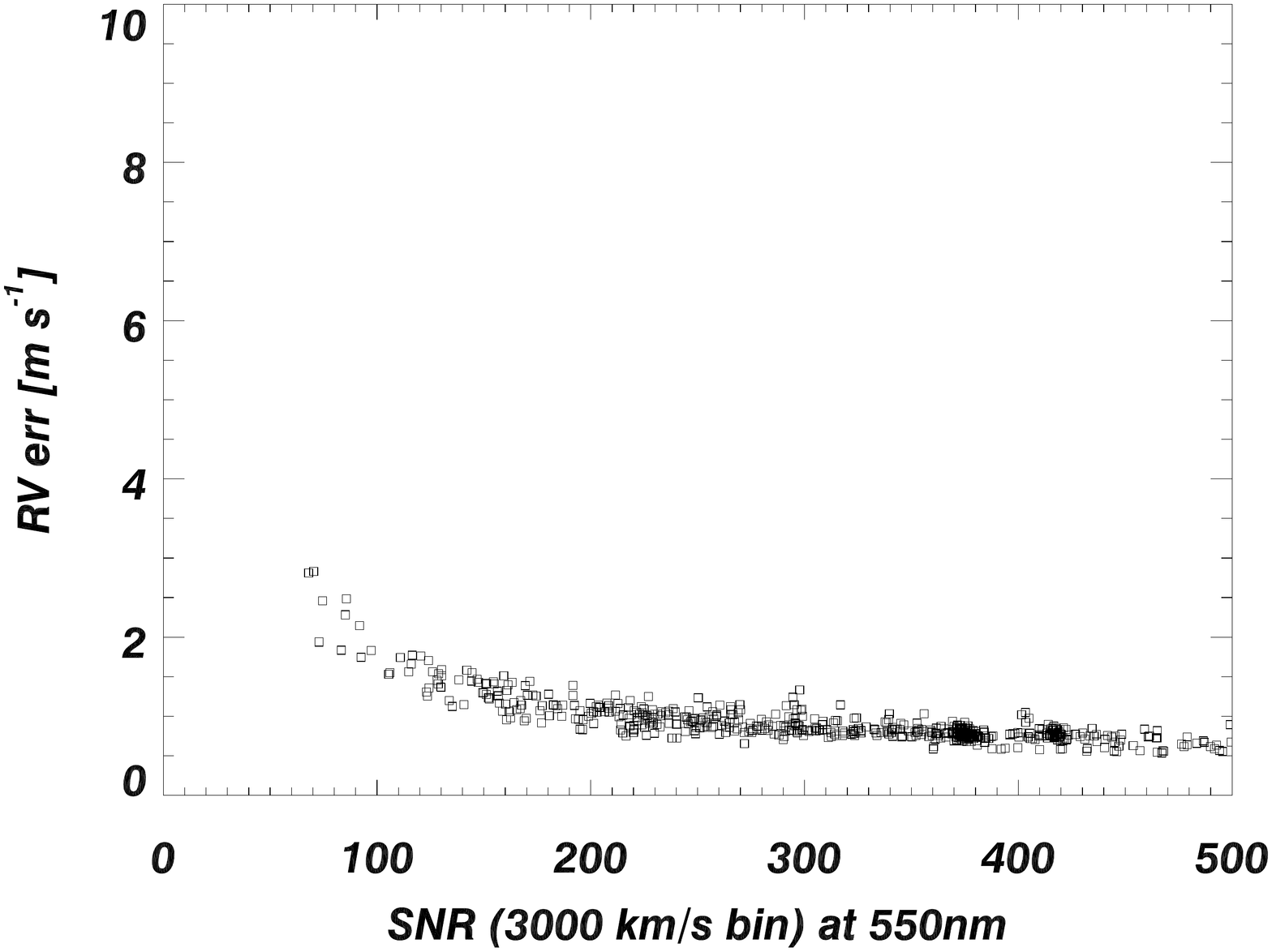}
\includegraphics[width=6cm, height=8cm, angle=90, clip]{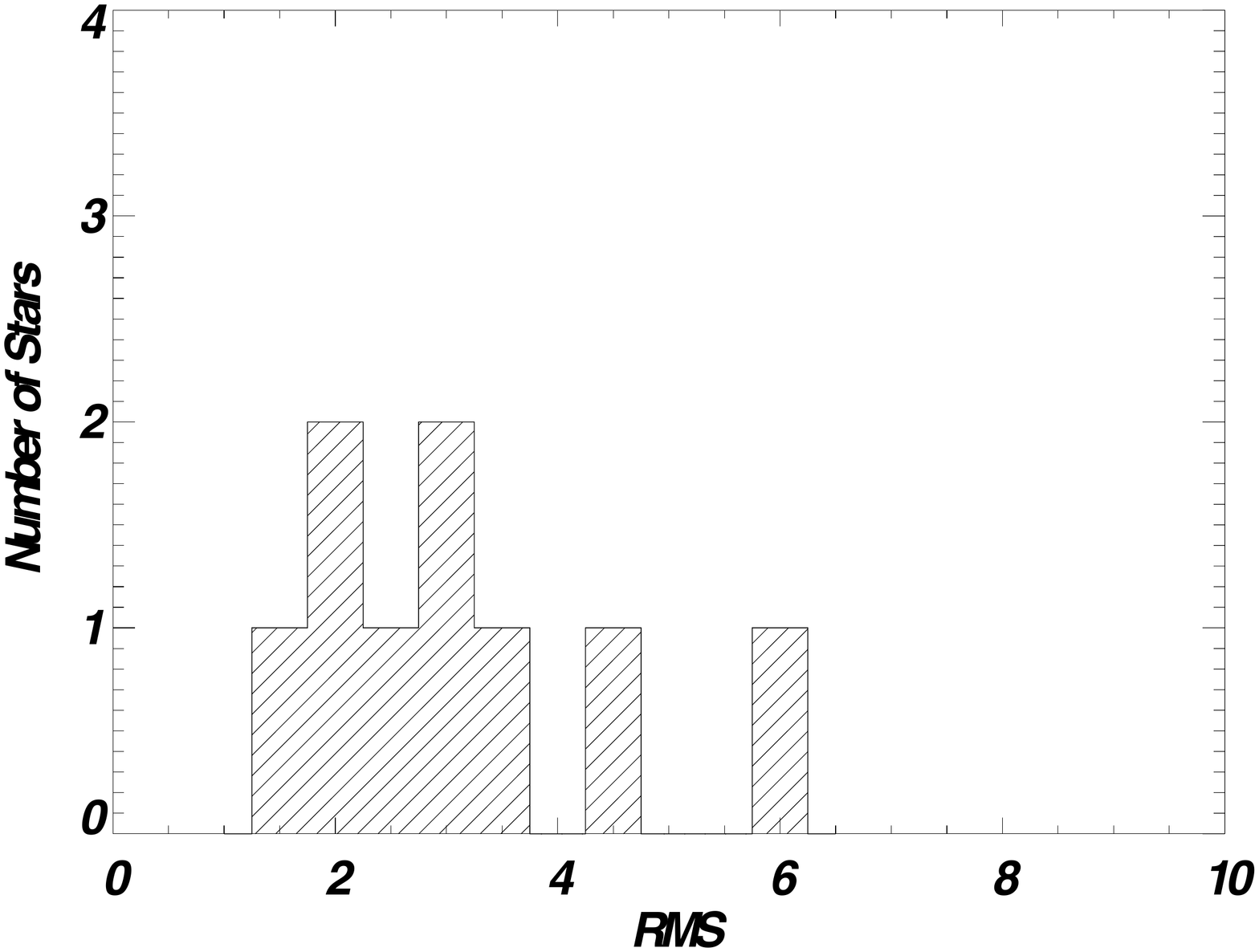}
\caption{Single measurement precision (left) and radial velocity scatter for ten bright stars observed with CHIRON at CTIO.
(Courtesy of Debra Fischer)
\label{chiron}}
\end{figure*}

The primary goal of the CHIRON program was the search for planets around alpha Centauri A and B and the program also included 35 bright stars.  The low throughput associated with a high resolution spectrograph on a small telescope required two hour exposures on stars fainter than $V = 5$. This ultimately lead to a program re-scope to concentrate on the brightest stars with the goal of understanding the limitations of radial velocity precision for detection of lower mass planets. The program uses queue scheduling \citep{Brewer2014} and this enables observations of a few bright stars on virtually every clear night. 

The radial velocities and uncertainties are calculated as described in Section \ref{Hamilton}.
At a SNR of 200, the SMP errors for CHIRON are 1 \ms\ and the SMP improves to 0.5 \ms\ as the SNR approaches 500. Figure \ref{chiron} (left) shows the dependence of estimated internal errors on the SNR of the observations. Figure \ref{chiron} (right) is a histogram of the rms of the velocities (no trends or Keplerian signals were removed) with a minimum rms of 2~\ms. The fairly narrow range in rms velocities occurs because the program concentrated on chromospherically quiet stars. 

The alpha Cen survey suffered from increasing contamination as the separation of the two binary stars decreased from 6 arcseconds in 2011 to 4 arcseconds in  2013 (when the project was put on hold).  The contamination was significant with 2\farcs7 fiber, selected to accommodate the frequent poor seeing at the observatory. A technique was developed to fit for scaled flux contamination when modeling the radial velocity of each star, however this only provided a precision of 3 \ms. 

CHIRON demonstrated the advantage of a fiber-fed instrument and very high cadence Doppler observations. The biggest challenge with CHIRON is the struggle to obtain the high SNR needed for 1 \ms\ precision with the iodine technique; this is difficult on a small aperture telescope and there were additional challenges with poor guiding and telescope control system failures. As a result, the program goals shifted to understanding the limitations of radial velocity precision by focusing on a small subset of the very brightest stars. 

Perhaps the most important lesson from CHIRON is that even with a more stable instrument, parameter degeneracies in forward modeling were identified that ultimately limit the RV precision (in addition to stellar jitter) with the iodine technique \citep{Spronck2015}.  While a more precise wavelength calibrator that spans a broader wavelength range on CHIRON is desirable, the instrument is not in a vacuum enclosure and CHIRON was not designed to be stable enough for simultaneous thorium argon or laser frequency combs. The lesson here is that a change in wavelength calibration is not always an easy retrofit; it is something that needs to be designed into the instrument from the start. 
The team is now building on lessons learned with CHIRON in the design of a new spectrometer, the EXtreme PREcision Spectrograph (EXPRES) for the 4.3-m Discovery Channel Telescope (commissioning is planned for summer 2017). This vacuum-enclosed white pupil design spectrometer will have a resolution of R=150,000, a broad wavelength range, and calibration with a laser frequency comb.

\subsection{PARAS}


PARAS (PRL Advanced Radial-velocity Abu-sky Search) is currently India's only exoplanet search and characterization 
program\footnote{Presentation by Abhijit Chakraborty} and started science observations in 2012 
\citep{Chakraborty2010, Chakraborty2014}. PARAS obtains a single-shot 
spectral coverage of 3800 -- 9500\AA\ at a resolution of 67,000 with 4-pixel sampling. Radial velocity 
measurements are calculated using simultaneous wavelength calibration with a Thorium-Argon (ThAr) hollow cathode lamp. 
The spectrograph is maintained under stable temperature conditions and enclosed in a vacuum vessel. 
The enclosure has active temperature control to correct for thermal conduction through the concrete pier 
under the spectrograph; temperature is monitored both inside the vacuum chamber and in the room, and 
corrections for the heat loss or gain are applied by changing the heater power. This provides an rms 
thermal stability of $25 \pm 0.009$C during the night. The typical vacuum stability is about 0.05 millibar 
during a night of observations. An e2v deep depletion 4kx4k CCD is used for imaging the spectra along 
with the ARC Inc., Leach Controller for CCD electronics and interface with Computer. Simultaneous 
ThAr exposure along with a ThAr exposure immediately after the observation is used to correct for 
instrument drifts with an rms error of 90 \cms\ or better \citep{Chakraborty2014}.     

The blaze peak efficiency of the spectrograph between 5000 \AA\ and 6500\AA, including 
the detector, is about 30\%; and about 25\% including the fiber transmission. The total efficiency, including 
spectrograph, fiber transmission, focal ratio degradation (FRD), and telescope (with 81\% reflectivity 
considering primary and secondary mirrors) is $\sim 7$\% in this same wavelength region.  

The PARAS data analysis pipeline \citep{Chakraborty2014} is custom-designed and 
based on the REDUCE routines of \citet{Piskunov2002}. 
The pipeline performs the routine tasks of cosmic ray correction, dark 
subtraction, order tracing, and order extraction. A thorium line list is used to create 
a weighted mask which calculates the overall instrument drift, based on the 
simultaneous ThAr exposures. 

\begin{figure*}[ht]
\includegraphics[width=8cm, height=6cm, clip]{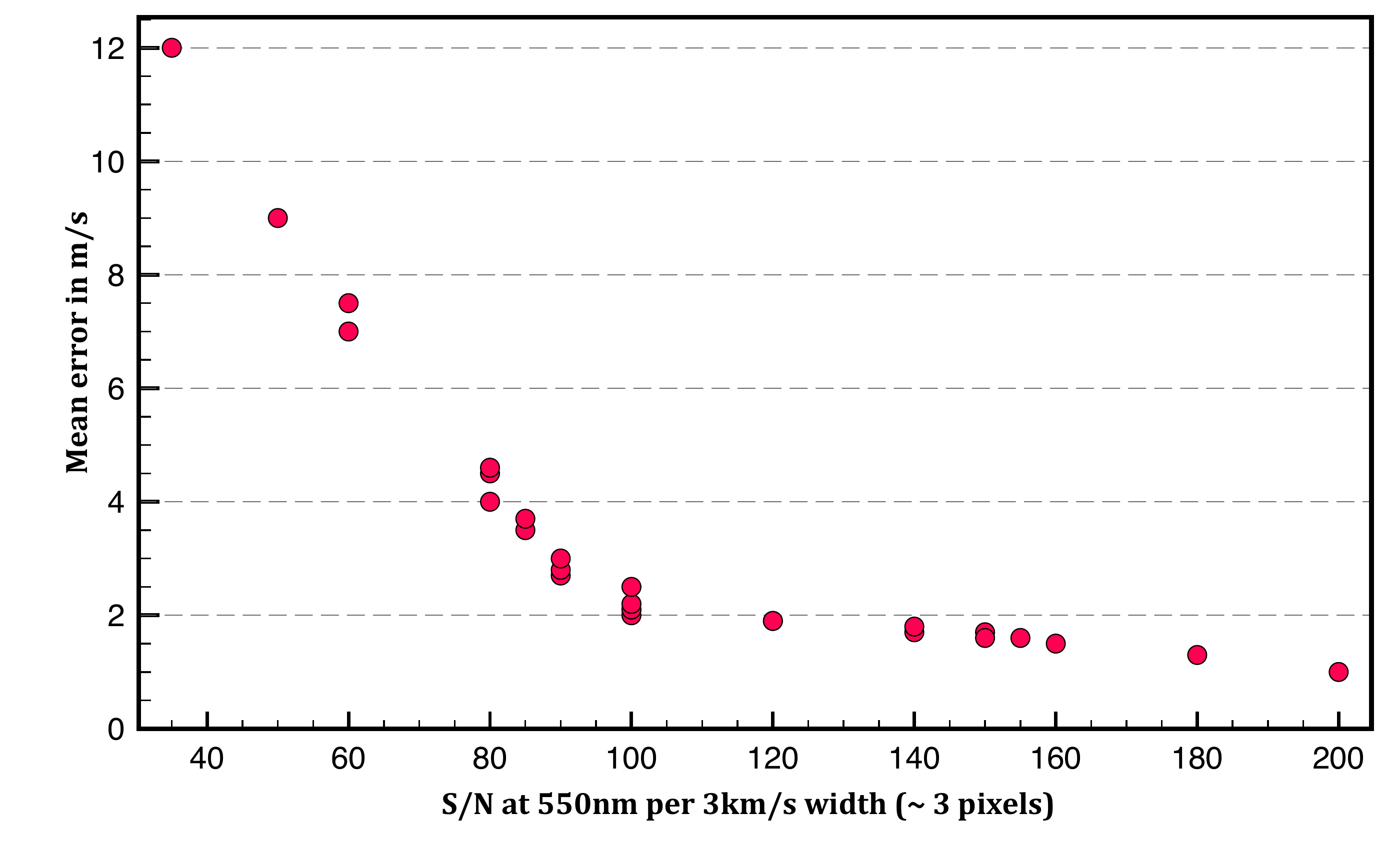}
\includegraphics[width=8cm, height=6cm, clip]{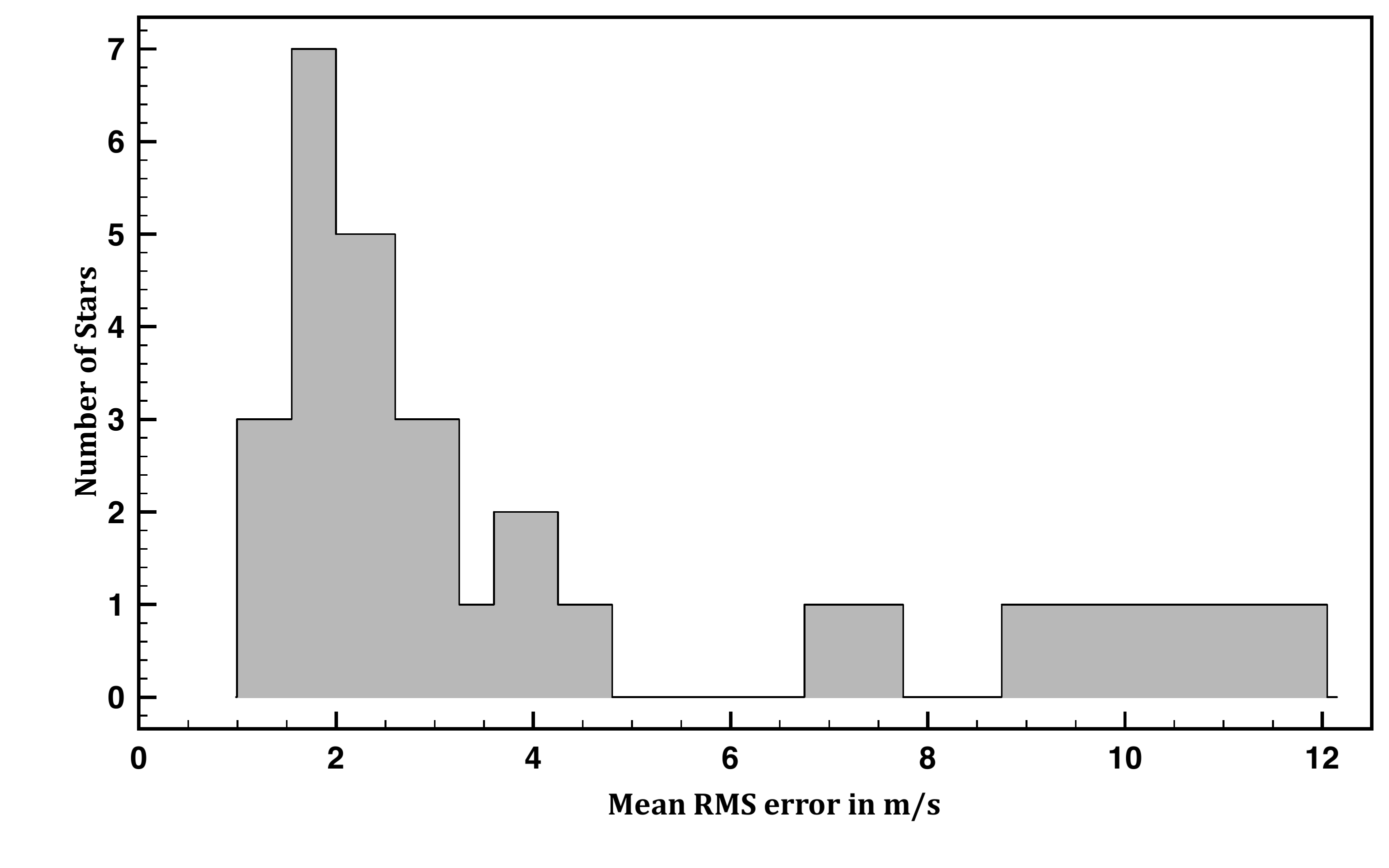}
\caption{Precision and radial velocity scatter for PARAS. (Courtesy of Abhijit Chakraborty)
The plot on the left shows the nightly RV scatter as a function of SNR in 3 \kms\ bins and 
the plot on the right is a histogram of the long term rms scatter for 27 bright stars.
\label{PARAS}}
\end{figure*}

RVs are derived by cross-correlating the target spectra with a suitable numerical 
stellar template mask. The stellar mask is created from a synthetic spectrum of the 
star, and contains the majority of deep photospheric absorption lines \citep[][and references therein]{Pepe2002}. 
RV measurement errors include photon noise errors \citep{Bouchy2001} and other errors like 
those associated with fitting the cross correlation function. Typically, two to three RV points 
are used to calculate the nightly mean velocity. \citet{Chakraborty2014} demonstrate 
a mean nightly rms scatter of 1.2 \ms\ for the star Sigma Draconis over a period of 
seven months and 1 \ms\ for HD 9407 over $\sim30$ days. Additional velocities now show 
1.4 \ms\ over two years for Sigma Draconis and 1 \ms\ for tau Ceti over a period of 40 days. 

Figure \ref{PARAS} shows the estimated measurement precision of PARAS for 27 bright stars 
since the beginning of observations in 2012. These sources have at least 
10 epochs of observations spanning more than 30 days. On a clear photometric night with 90\% reflectivity on 
the telescope primary and secondary mirrors, a SNR of 200 per 3\kms\ RV bin is reached 
in about 15 minutes on a 6.5-magnitude star.

The small telescope aperture of 1.2-m with its large tracking errors (of the order of 1 
to 1.5 arcseconds) is one of the greatest short comings of the project.  A combination of 
octagonal and circular fibers (without a double scrambler) helps to stabilize illumination of 
the spectrograph optics.  Instead of a pinhole at the telescope focal plane, 
a focal reducer is used to focus the F/4.5 beam on the octagonal fiber. The image of the star 
slightly overfills the fiber which is about 2 arcsecs on the sky. This helps to take care 
of telescope tracking and atmospheric scintillation issues at the cost of some light loss. 

The team plans to move PARAS to a new 2.5-m telescope in late 2019 or early 
2020 with a fast tip-tilt image stabilizer and an ADC at the Cassegrain focus 
for proper injection of light into the fiber. The new telescope will have rms tracking 
and jitter errors of less than 0.1 arcsecsonds. The temperature control and spectrograph 
room will be redesigned to alleviate the present deficiencies like the heat conduction 
issues from the pier and to achieve thermal stability of $25 \pm 0.001$C. 
This will eliminate the need for taking large number of ThAr calibrations 
during the night, providing additional time for more stellar observations.

\subsection{3.5-m TNG telescope with the HARPS-N spectrometer, 2012 --}

In 2012, a near twin of HARPS was commissioned at the Telescopio Nazionale Galileo (TNG) observatory\footnote{Presentations by David Phillips and Lars Buchhave}.  The vacuum enclosed instrument, HARPS-N, has temperature and pressure stabilization, a resolution of 115,000 and wavelength range from 383 to 693 nm. HARPS-N was initially commissioned with Thorium Argon simultaneous calibration, with a planned upgrade to an optical frequency comb. The instrument is fed by a combination of octagonal fiber and double scrambler, which inverts the near and far fields. A failure of the CCD delayed the debut of this instrument, but regular operations are now underway. 

HARPS-N is being used for several collaborative exoplanet detection programs \citep{Cosentino2014}, including the Global Architecture of Planetary Systems \citep[GAPS;][]{Covino2013} program. In summer 2015, a novel program was started to carry out solar observations with the goal of reaching sufficient measurement precision to detect Venus \citep{Dumusque2015b}. The solar observations are one example of a project that is not photon starved, and uses high cadence (5-minutes) for about 4 hours each day, using a compact solar telescope to feed the HARPS-N spectrograph. When using the HARPS-N LFC, the error from the wavelength calibration is only 6 \cms\ and the estimated precision is a few meters per second over the course of a week. The project uses a photometric $F-F'$ correction \citep{Aigrain2012, Dumusque2015b} from the Solar Dynamo Observatory (SDO) images to reduce this to 1 \ms. 

\begin{figure*}[ht]
\includegraphics[width=6cm, height=8cm, angle=-90, clip]{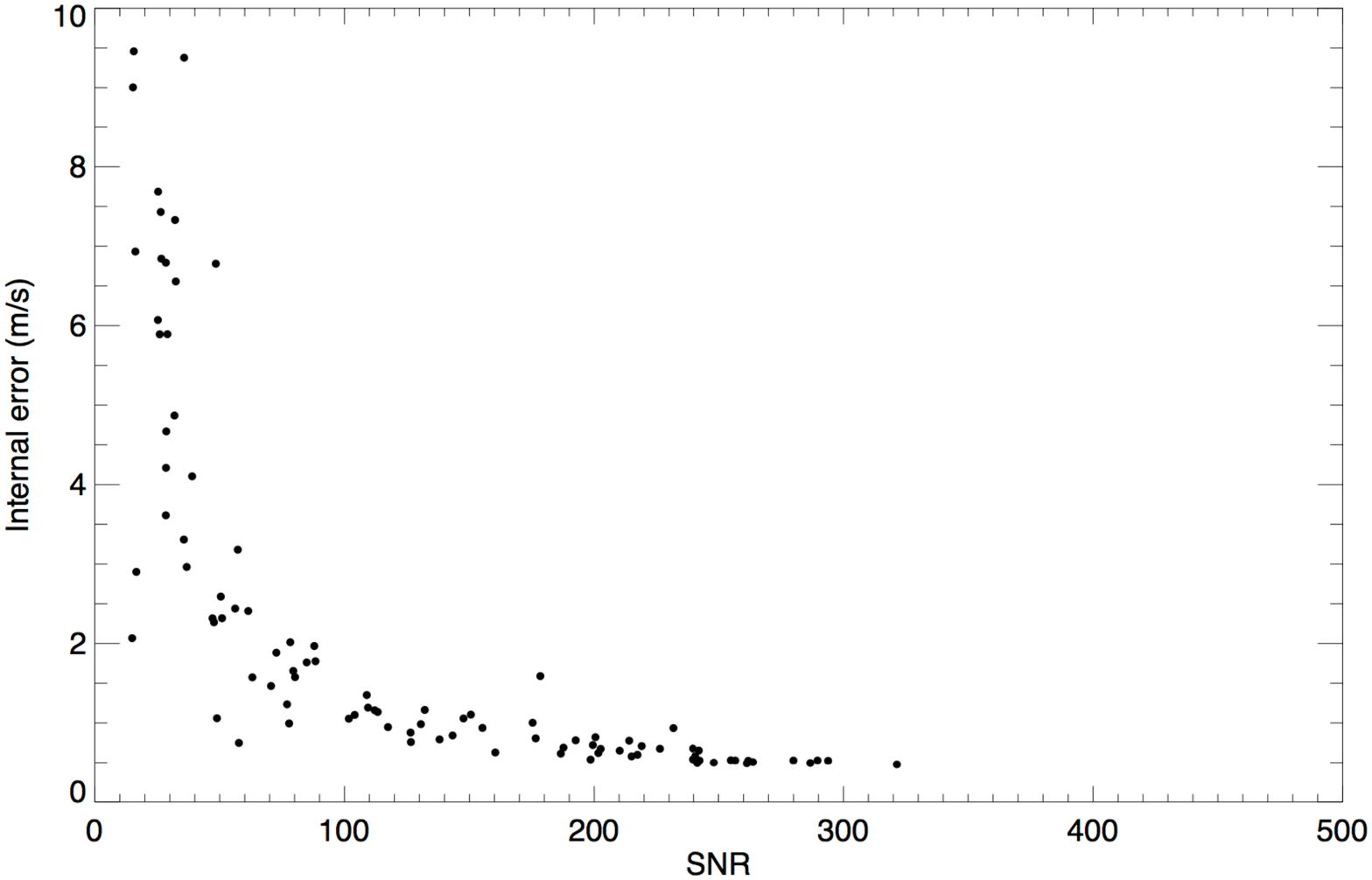}
\includegraphics[width=6cm, height=8cm, angle=-90, clip]{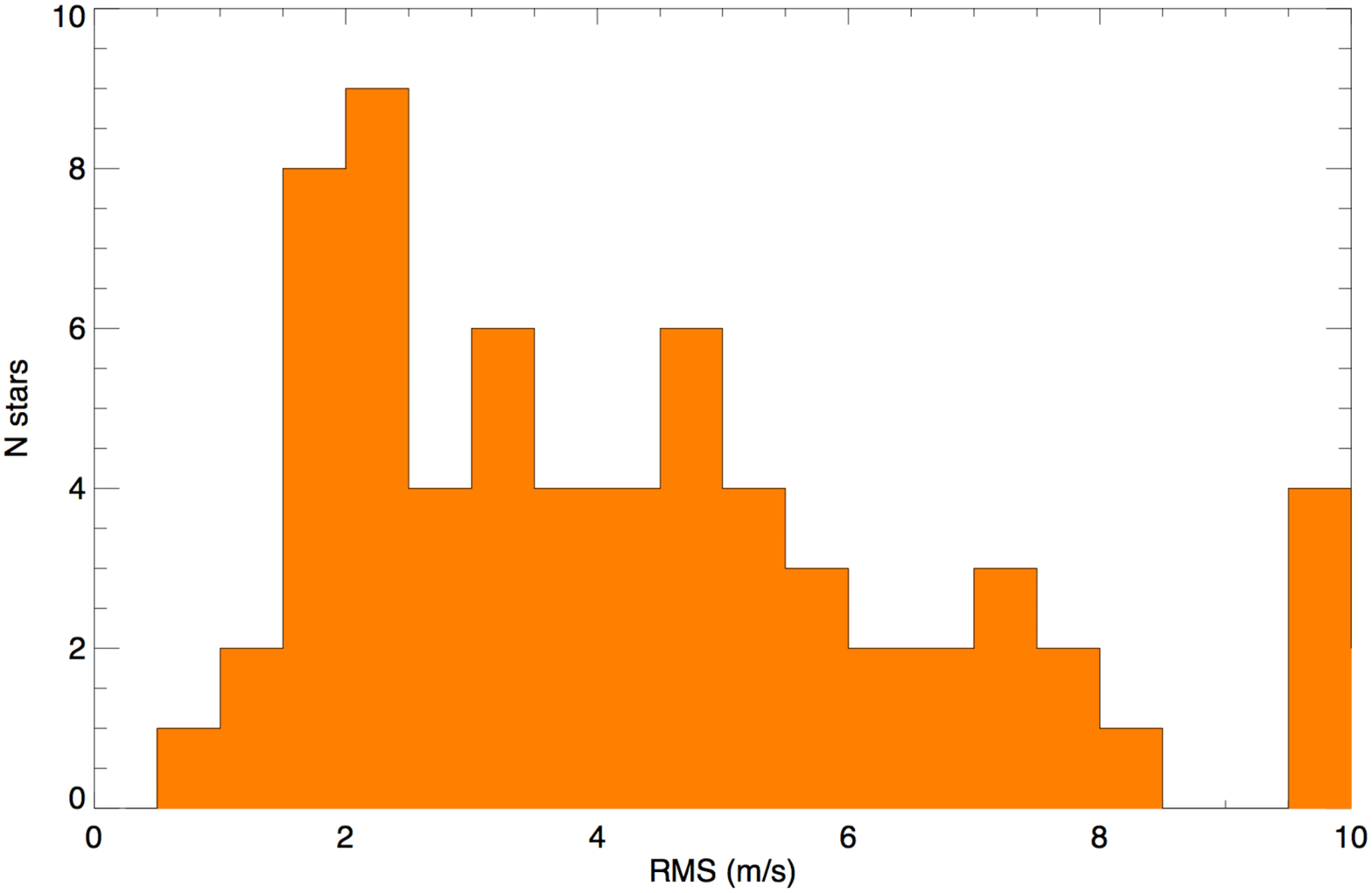}
\caption{Histograms of the estimated SMP and radial velocity scatter for stars with more than 10 observations at HARPS-N. 
(Courtesy of Lars Buchhave)
\label{harps-n}}
\end{figure*}

The velocities and uncertainties are calculated as described in Section \ref{HARPS_Lovis}. Figure \ref{harps-n} shows that the measurement precision at HARPS-N is similar to HARPS. The estimated SMP for SNR of 200 is between 0.5 -- 1 \ms.  The rms velocity scatter peaks at about 2 \ms.  

RePack is the name of a Doppler pipeline for HARPS-N, written by Lars Buchhave. Here, the internal uncertainties are calculated by cross correlating either a high SNR observed spectrum or a co-added combined spectrum (i.e., a template) against the observed spectra. 
Typically, a delta function template is used. This template consists of weighted delta functions based on the measured line centers of known absorption lines, so that each line can contribute to the CCF with some weight. Individual lines have a relatively poor wavelength determination, often with several hundreds of meters per second uncertainties. The final RV of each order can thus be several hundreds of meters per second offset compared to the other orders because of the (combined) uncertainties in the position (wavelengths) of the delta function lines. The offset is the average offset between the weighted delta function lines compared to the actual stellar spectrum. This can create problems when the spectrum is shifted around by the barycentric correction, because lines will move from higher to lower and lower to higher SNR regions on the detector (because of the blaze function). In other words, the combined RV offset of the order can change because e.g. a strong line might move towards lower SNR if it moves towards the edge of the order.

In contrast, an observed template will yield RVs for the individual orders that are more realistically scattered around the RV shift between template and observation. With RePack, a radial velocity measurement is determined for each individual order using a CCF and the internal uncertainty is calculated as the photon weighted rms of the RVs for each order, divided by the square root of the total number of orders used. The final RV is measured by co-adding the CCFs for each order and then determining the position of the co-added CCF peak.

The precision of HARPS-N allows for the detection of additional errors that are normally masked by poorer precision data. A three-minute thermal cycle on the cold plate in the CCD cryostat at HARPS-N was identified and the thermal cycling is believed to produce an expansion-contraction cycle that slightly shifts the position of the CCD. The temperature variations from the cold plate probe correlate directly with wavelength (velocity) shifts in the thorium argon calibration lines for short 20-second exposures.  The stability of the wavelength solution was restored by taking 5-minute thorium-argon observations to average over the cold-plate jitter cycle. Two different data analysis pipelines have been tested at HARPS-N; this was a useful exercise that helped to identify new areas for improving the RV precision.

\subsection{2.4-m APF with the Levy spectrometer, 2013 --}

The 2.4-m Automated Planet Finder \citep[APF;][]{Radovan2010, Vogt2014} is located at Lick Observatory\footnote{Presentations by Greg Laughlin and Andrew Howard} and the Levy spectrograph was commissioned at the APF in 2013. The instrument uses a narrow slit to reach resolutions up to 150,000. Like Keck-HIRES, the wavelength calibration is carried out with an iodine cell and therefore the wavelength range used to derive Doppler velocities is limited to 510 -- 620 nm.  However, the spectra span from 374 -- 950 nm, enabling the use of chromospheric diagnostics like the Ca II H \& K lines. 

The telescope time on the APF is divided between two groups; the Lick-Carnegie Planet Search (LCPS) team at UC Santa Cruz observes about 100 stars and the California Planet Search (CPS) team at UC Berkeley and the University of Hawaii observes a few hundred stars.  The science goals of both teams include detection and characterization of low mass exoplanet systems and follow-up observations to better resolve components of multi-planet systems. In the future, the APF will provide support for the NASA TESS mission. A big advantage of the APF is that it is a robotic instrument and the observing scripts can be launched from remote sites \citep{Burt2015}. The unique combination of large telescope time allocation and small user group allows the facility to obtain very high cadence measurements.

\begin{figure*}[ht]
\includegraphics[width=6cm, height=8cm, angle=90, clip]{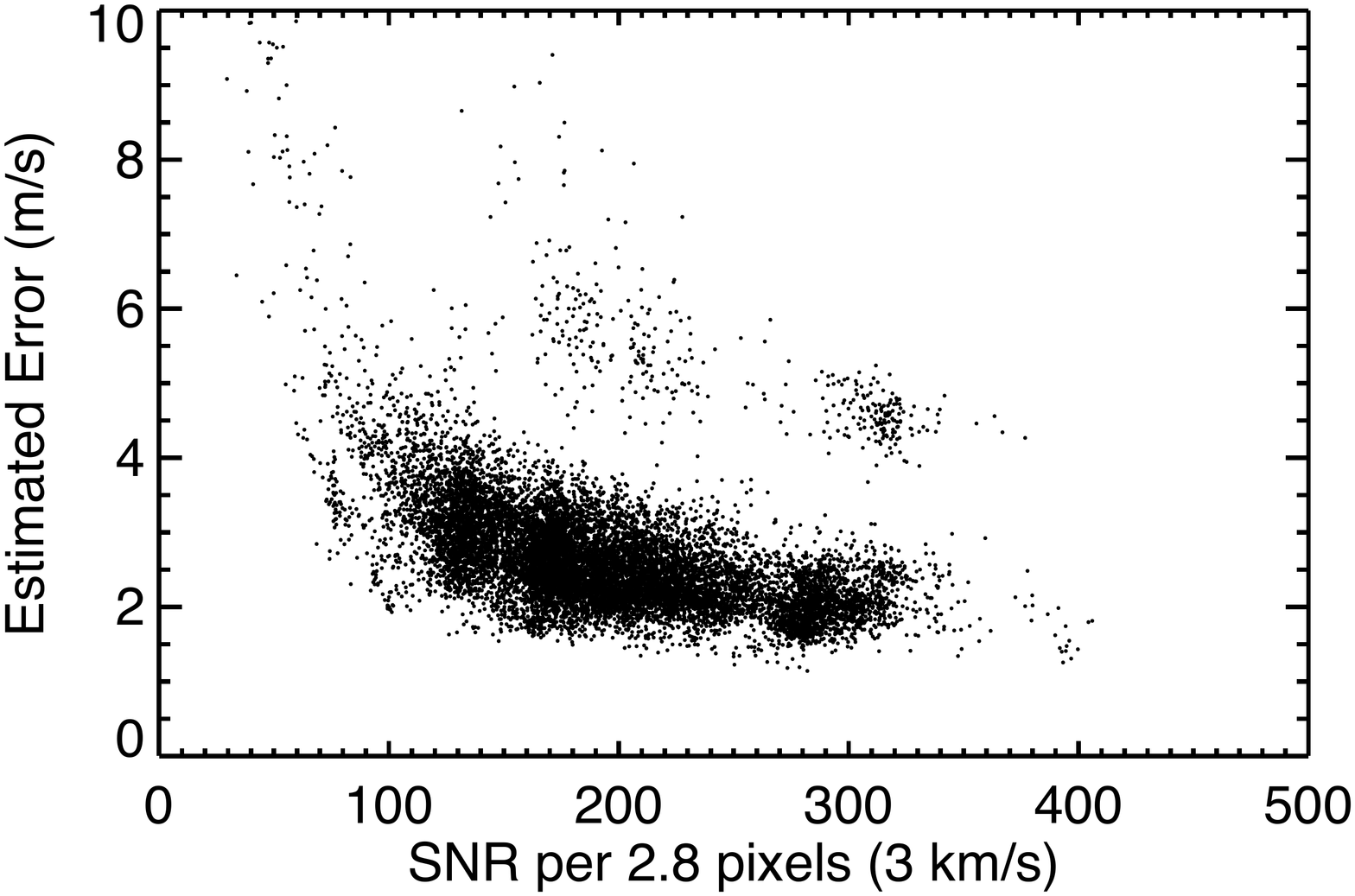}
\includegraphics[width=6cm, height=8cm, angle=90, clip]{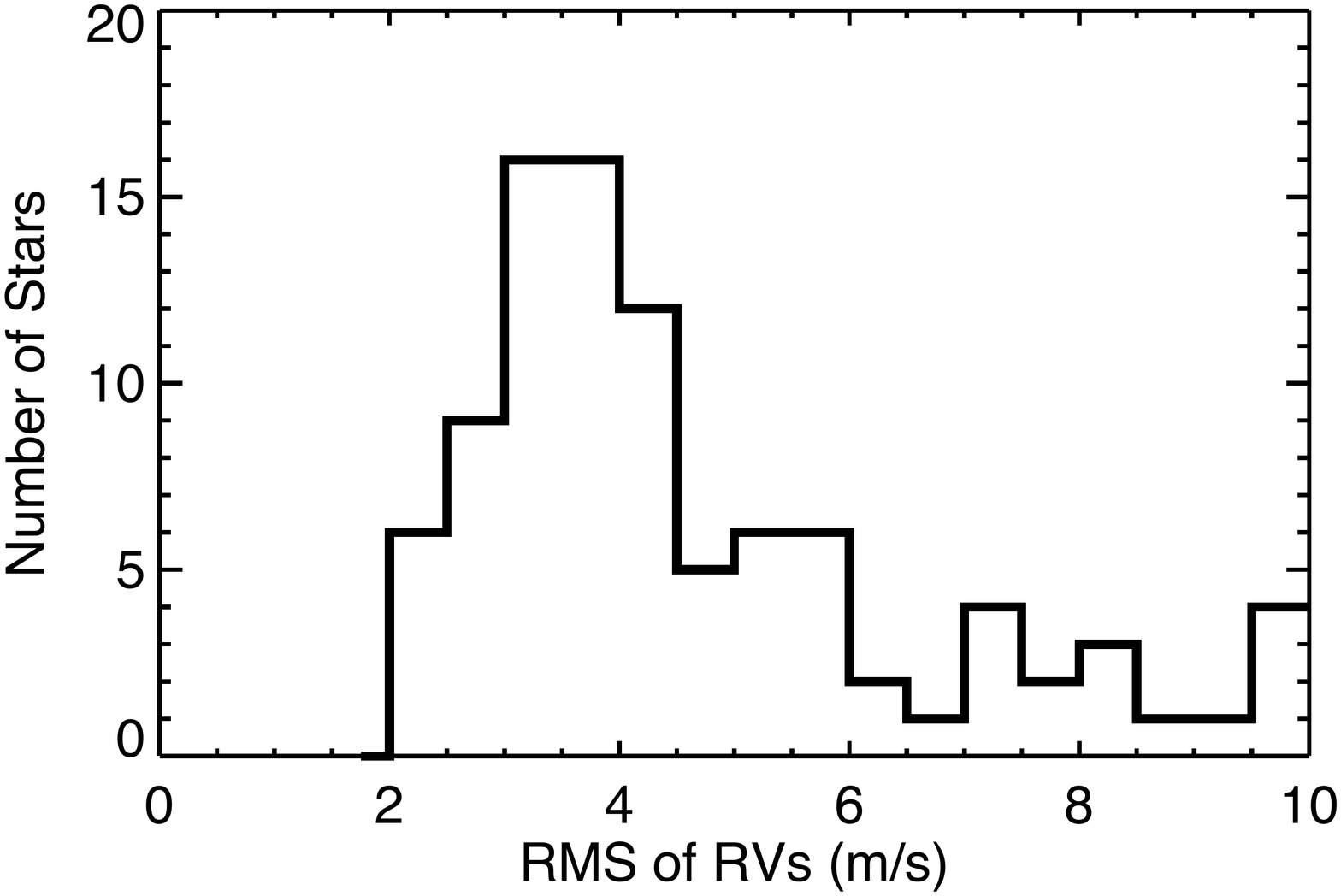}
\caption{Estimated precision (left) and radial velocity scatter (right) for Doppler measurements obtained by the California Planet Search team using the Levy Spectrograph at the Automated Planet Finder. (Courtesy of Andrew Howard)
\label{apf_uh}}
\end{figure*}

\begin{figure*}[ht]
\includegraphics[width=8cm, height=6cm, clip]{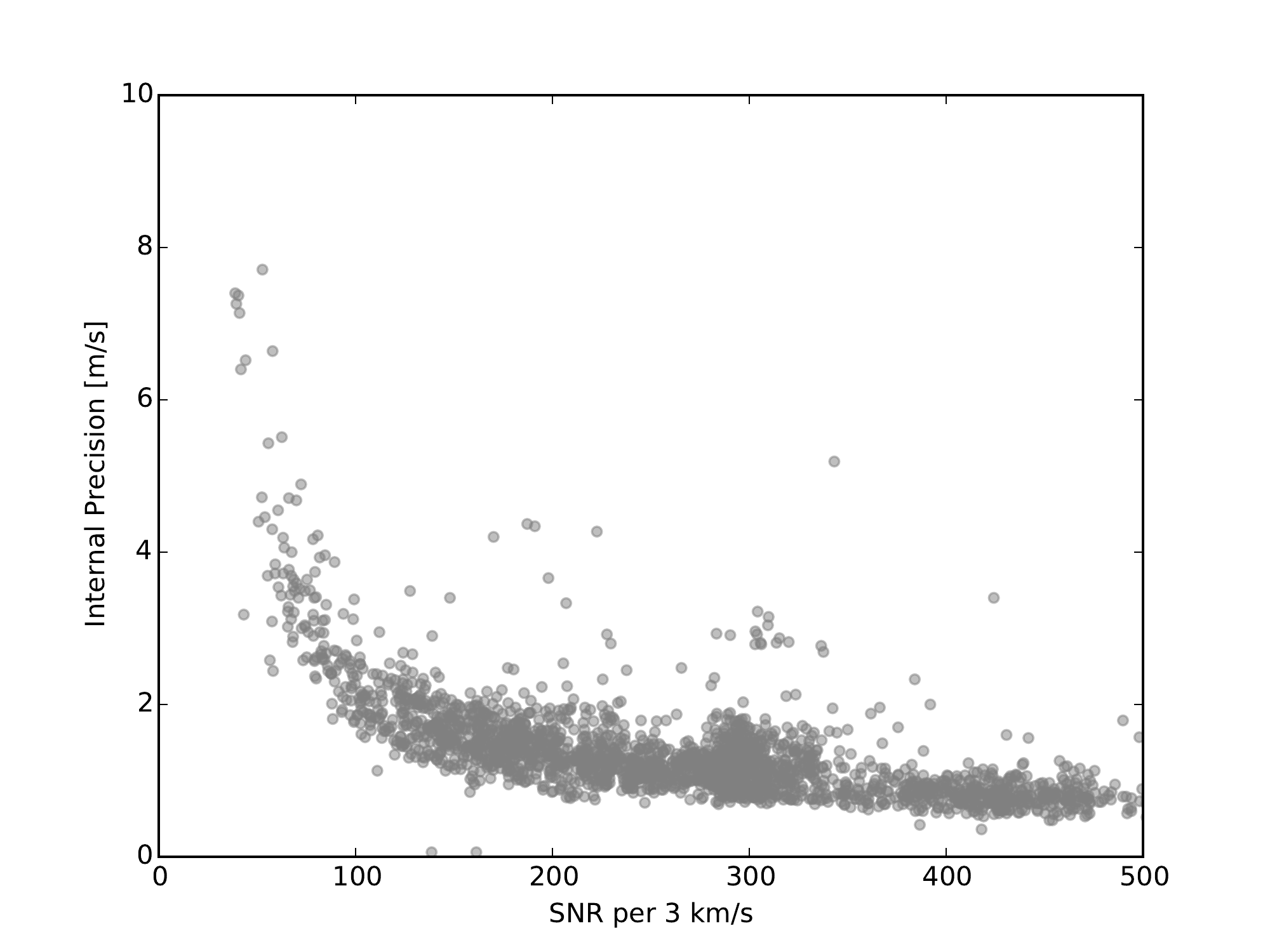}
\includegraphics[width=8cm, height=6cm, clip]{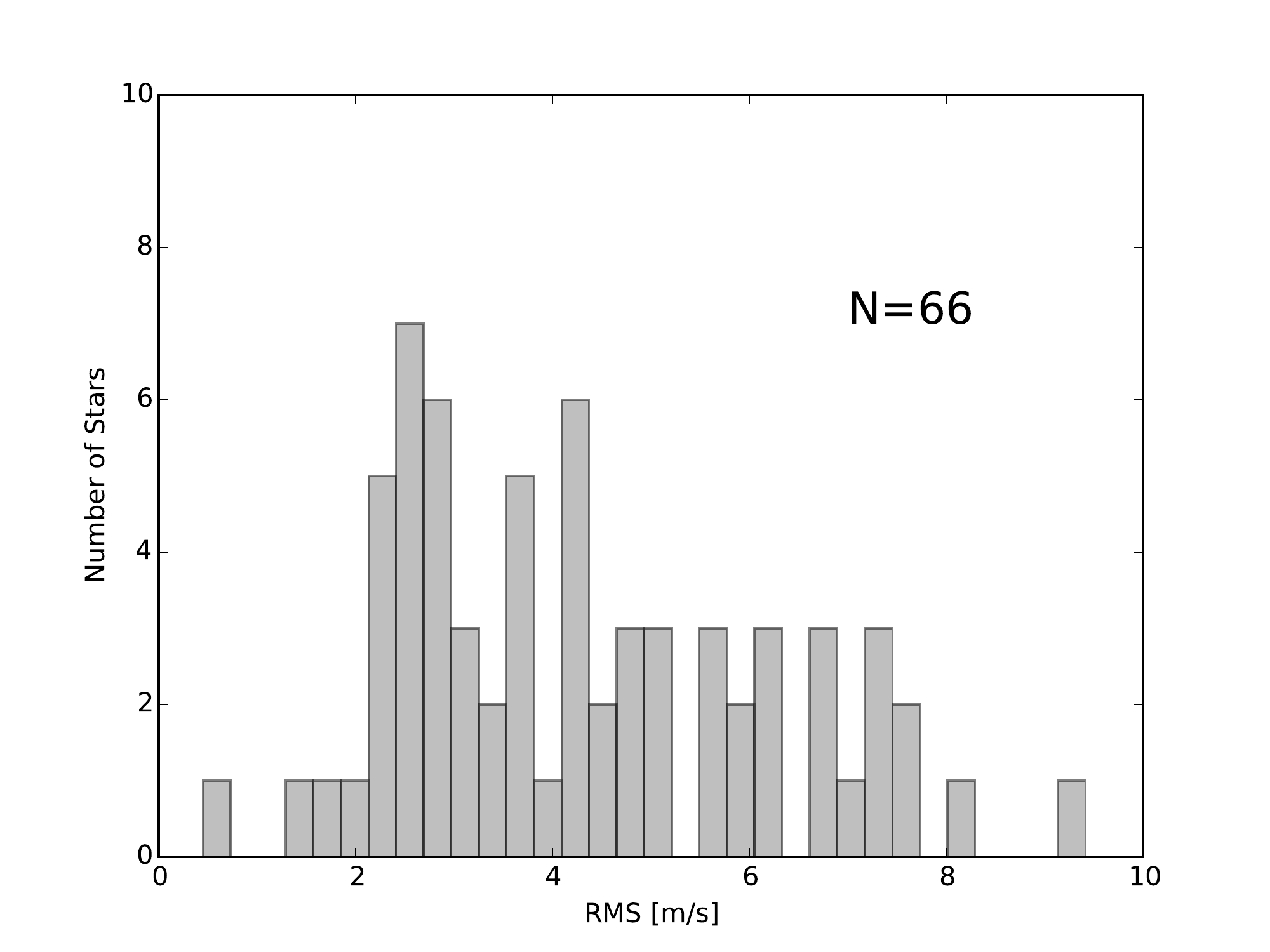}
\caption{Radial velocity precision as a function of SNR (left) and velocity rms (right) for the Lick Planet Search team with the Levy Spectrograph at the Automated Planet Finder. (Courtesy of Greg Laughlin)
\label{apf_ucsc}}
\end{figure*}

The CPS team aims for a fixed target SNR of 200 per observation; this yields estimated internal errors of about 2 \ms. Typically many observations are taken for each star during a given night and the velocity is averaged to give better precision. Because the CPS team is observing stars with low chromospheric activity and a known legacy of Doppler velocities from Keck HIRES, the rms of the velocities is a tighter distribution than Keck HIRES, centered on 3 -- 4 \ms. Figure \ref{apf_uh} (left) shows the dependence of estimated internal errors on the SNR of the observations. Figure \ref{apf_uh} (right) is a histogram of the rms of the velocities (without attempting to remove any trends or Keplerian signals).

The Lick Carnegie team has estimated errors of about 1.5 \ms\ per observation at a SNR of 200 and 1 \ms\ at SNR of 400 and the velocity rms is typically greater than 2.5 \ms. Figure \ref{apf_ucsc} (left) shows the dependence of estimated internal errors on the SNR of the observations. Figure \ref{apf_ucsc} (right) is a histogram of the rms of the velocities (no trends or Keplerian signals were removed).

The precision of the Levy spectrometer is limited to about 1 \ms\ by the iodine reference cell. Although the flexible scheduling has been critical for this program, only a relatively small number of observations can be  
taken each night because of the modest 2.4-m telescope aperture.

\subsection{Hertzprung 1-m telescope with the SONG spectrometer, 2014 --}

The SONG spectrometer \citep{Grundahl2011} was commissioned in mid-2014 at the Mount Teide Observatory in Tenerife\footnote{Presentation by Frank Grundahl}. The instrument has a series of slits and is operated with a spectral resolution of 90,000 and an optical wavelength range of 440 -- 690 nm.  
The instrument is not in a stabilized vacuum enclosure so an iodine reference cell is used to measure Doppler shifts 
between 510 -- 620 nm. The detector is a 2k by 2k Andor device with 2-second readout, suitable for measuring the oscillation periods 
of stars.  There is a fast tip/tilt system to help stabilize the slit illumination. 

\begin{figure*}[ht]
\includegraphics[width=8.5cm, clip=]{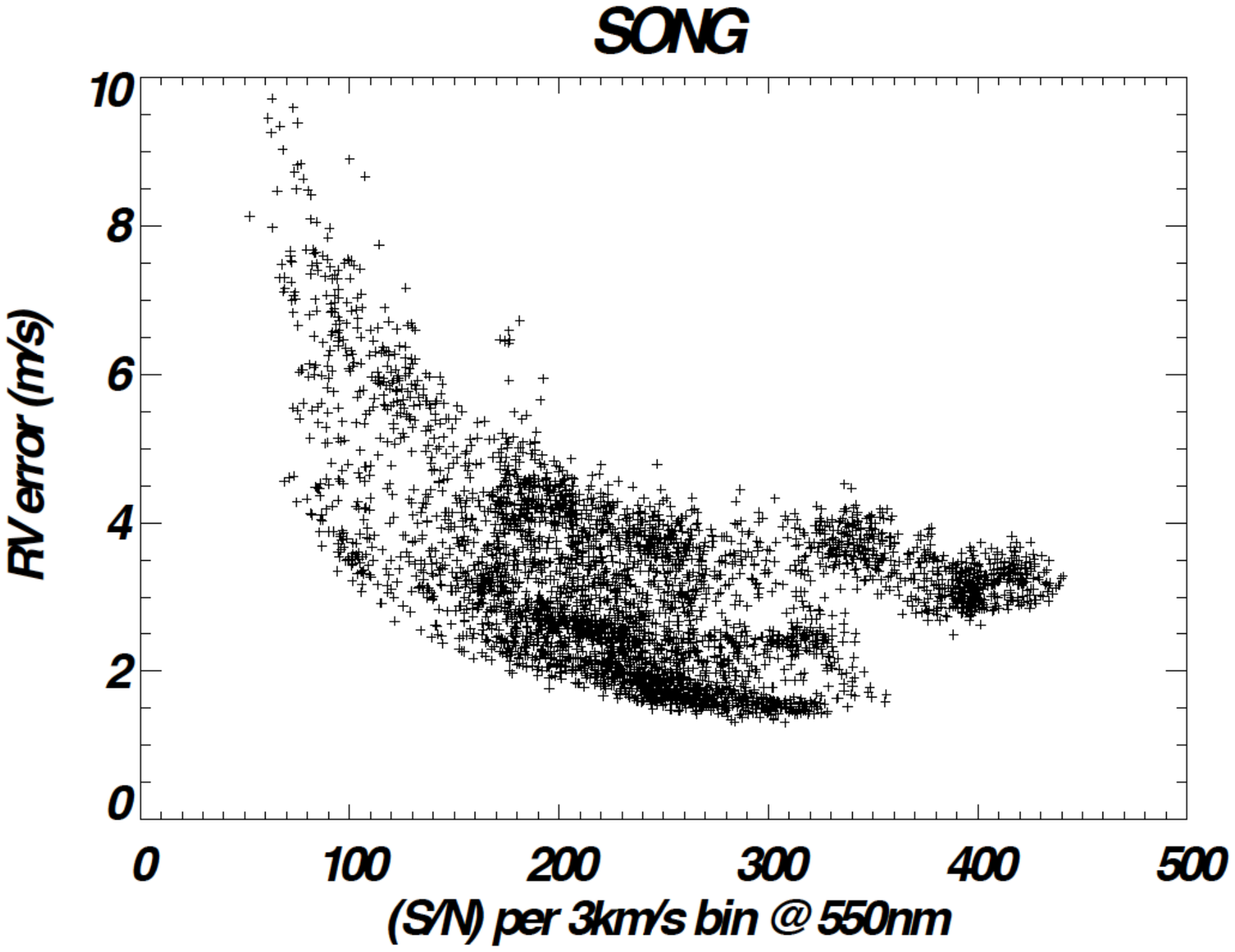}
\hspace*{-1.cm}\includegraphics[width=8.5cm, clip]{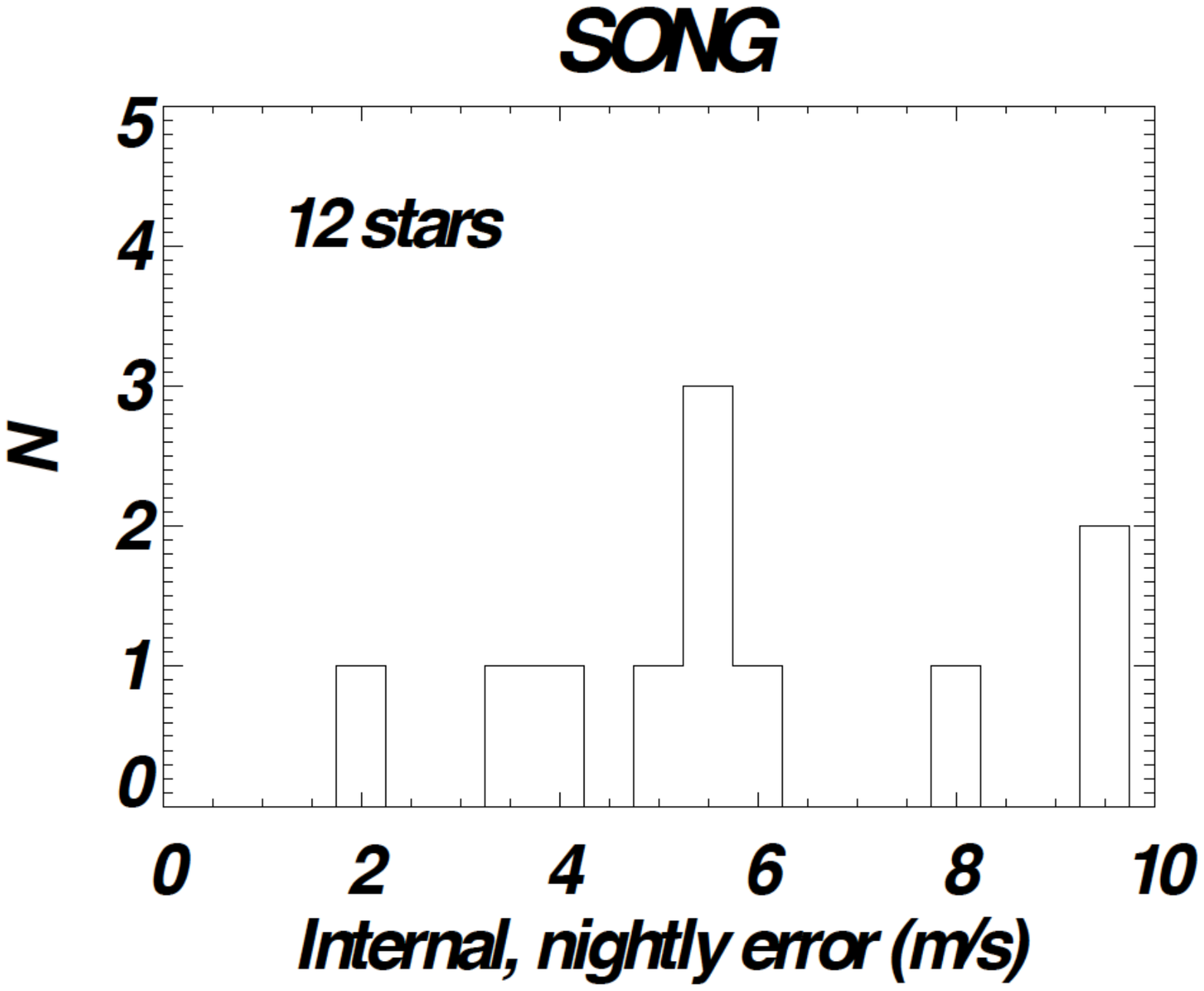}
\caption{Radial velocity precision as a function of SNR for SONG and the 
histogram of nightly errors for 12 stars. One star has internal nightly errors that are greater than 10 \ms. (Courtesy of Frank Grundahl) 
\label{SONG}}
\end{figure*}

The science program focuses on a relatively small sample of twelve stars with the goal of obtaining asteroseismology data, reaching a better understanding of stellar physics and detecting exoplanets. Except for time lost to weather, this is a dedicated facility that obtains very high cadence observations every night.  The telescope will be part of a network that spans the globe for nearly continuous observations of stars. 
The estimated SMP error (Figure \ref{SONG}) is about 2 -- 4 \ms\ with an rms of about 4 \ms. The best precision is obtained for K~dwarfs, but there is still large scatter in the RV errors. 

SONG is optimal for bright targets.  However, the program is new, so the data analysis pipelines are still being developed. The iodine cell is in a collimated beam and demonstrated fringing; this cell has just been replaced with a cell that has wedged optical windows that are counter-rotated. The small 1-m telescope aperture means that the stars must be brighter than V of 6.5 for reasonable exposure times and high SNR.
 
\subsection{Future RV Programs}
In addition to these ongoing Doppler planet searches, there are several funded new spectrometers that will come online in the few years, including the first node of Network of Robotic Echelle Spectrographs 
(NRES) on the LCOGT telescopes in 2015 with a target RV precision of 3 \ms\ \citep{Eastman2014}, CARMENES in 2015 at the 3.5-m Calar Alto Observatory \citep{Quirrenbach2014}, HPF in 2016 on the 9-m HET \citep{Mahadevan2012}, MINERVA 1-m telescope array in 2016 at Mt. Hopkins \citep{Swift2015}, {\it Veloce} in 2016 at the AAT; ESPRESSO in 2017 at the 8-m VLT \citep{Pepe2013}, EXPRES in 2017 on the 4.3-m Discovery Channel Telescope at Lowell Observatory, SPIRou in 2017 at the 3.6-m CFHT \citep{Artigau2014b}, the 
NNEXPLORE spectrograph at the 3.5-m WIYN telescope in 2018, iLOCATOR at the 8.4-m Large Binocular Telescope in 2018 \citep{Bechter2015} and G-CLEF in 2020 at the GMT \citep{Szentgyorgyi2012}.  For a more comprehensive list of future instruments, see Tables 2 and 3 in \citet{Plavchan2015}.

\section{Instrumentation Challenges} 
\label{ic}
Efforts to identify confusing sources of velocity scatter from stellar photospheres will be more effective if the instrumental 
precision is securely below the target RV precision for the stars\footnote{Presentations by Andrew Szentgyorgyi and Suvrath Mahadevan}.  
These requirements set the top level specifications for observing cadence and spectral resolution, which in  
turn dictates the wavelength bandpass, type of wavelength calibration, signal-to-noise 
ratio (SNR), precision for the barycentric correction, and required stability for the spectral line spread 
function \citep{Connes1985, Bouchy2001, Podgorski2014}.  

Asymmetric white pupil spectrograph designs offer several advantages for PRV spectrometers. 
The overall size of the 
optics and the instrument is reduced, and vignetting and aberrations are reduced. The 
white pupil design offers higher efficiency and better image quality than conventional designs with the same 
resolution.  White pupil designs have intrinsic cylindrical field curvature that is traditionally mitigated with a toroidal (cylindrical) field flattener that must be placed very close to the detector. A curved CCD would 
be an ideal solution, 
but these are not commercially available. Instead a Mangin mirror can be used in place of the folding transfer 
mirror to compensate for cylindrical field errors. However, care must be taken since the 
Mangin mirror can produce significant ghosts if it is not wedged. 

High velocity precision requires high SNR and broad wavelength bandpass, so a telescope of at least a modest 
aperture is required for all except the brightest stars.  The size of spectrometer optics increases with 
the size of the telescope unless clever designs are implemented (e.g., the anamorphic design of ESPRESSO or single mode
fibers behind adaptive optics systems). The need for both high resolution and broad bandwidth 
often implies that two (or more) cameras will be required, each with its own challenges. Often, the efficiency 
for the blue arm of the instrument is challenging because of the lower transmission of optical glasses below 
420 nm and the fibers must be kept short to minimize attenuation of light. Spectra obtained from 
the red arm of the spectrometer have telluric contamination that shift across the rest frame of the 
stellar spectra because of the barycentric velocity of the Earth. 

Camera designs for spectrometers fall into one of two categories. Standard designs have 7-9 lenses that are 
independently testable. Fabrication and alignment of these (mostly spherical) optics is straightforward; however, 
there can be a significant hit in terms of efficiency, ghosting and scattered light. New high-dispersion 
glasses are now available (Ohara and Nikon i-Line) that can reduce the number of optical elements needed for standard 
camera designs. The second type of camera is a pupil-folded camera that uses aspheric components.
These are not independently testable and carry some risk in manufacturing and alignment; however, the optics are 
very efficient with few ghosting and scattered light problems. 

The need to precisely centroid spectral lines and to track line profile variations from stellar photospheres 
(granulation, spots, plage, variations in long-term magnetic fields) sets top-level requirements for environmental stability.  
The solution from HARPS was to put the instrument in a vacuum chamber with vibration isolation. 
Temperature and pressure in vacuum enclosures are routinely controlled at the level of 10 mK and $10^{-2}$ 
Torr respectively; at this level, refractive index variations are no longer a problem.  However, temperature 
stability below 1 mK and pressure stability below $10^{-7}$ Torr is desirable because this will stop molecular transfer, 
providing a system 
that is almost completely radiatively coupled.  This is the state of the art for infrared instruments with 
liquid nitrogen tanks and charcoal or zeolite getters for cryo-pumping.  

The ease of environmentally stabilizing a spectrometer is inversely proportional to the size of the 
vacuum enclosure. Most enclosures are high grade 
steel, which is easy to weld, or aluminum, which is more difficult to weld. 
Optical benches have been made out of aluminum (NRES, KiwiSpec, HPF), mild steel (HARPS, ESPRESSO), 
invar and composite material. All have virtues and drawbacks.  Aluminum is light weight, low cost, and 
low technical risk, but may pose a high performance risk for radial velocity precision because of the thermal 
properties that must be carefully controlled.  Steel is heavy, but easy to process, with high thermal inertia and low 
cost. Invar is heavy with high thermal inertia; it is difficult to manufacture, posing a moderate technical and 
RV performance risk. The design trade study for G-CLEF found that carbon fiber epoxy met the systems 
engineering requirement with excellent conductivity and low thermal inertia.  There is some technical risk since this is a new 
material and it is known to outgas water. However, in a vacuum, outgassing will decrease 
over time to acceptable levels (may need to pump for about a year). 

As the spectrometers become more complex, a higher level of systems engineering is required with 
well-considered budgets for throughput, radial velocity precision, sequencing, timing, 
and delivery \citep[e.g.,][]{Podgorski2014}. 
New software programs are being developed to integrate these error budgets and to track 
compliance; this makes it easier to understand the design impact if, for example, 
a vendor delivers a component that does not meet the 
specification (many components are still contracted to vendors with a best effort 
commitment). The error budget should 
distinguish between error sources that can be calibrated out and those that cannot. In their 
error budget for G-CLEF, \citet{Podgorski2014} describe error sources that 
affect the calibration and observations equally and therefore can be calibrated out with an 
extremely stable instrument. These include: 

\begin{itemize}
\item Thermal stability: thermo-elastic distortions of the spectrograph can shift the spectra and affect focus. For 
G-CLEF temperature loads of 0.001C were applied in the FEA model (using the Smithsonian Astrophysical 
Observatory ÒBisensÓ software) and a lateral motion of 110  (55 cm/s) was identified; the solution was to 
adopt lower CTE materials and improve thermal control. 
\item Mechanical stability: the spectrometer is affected by changes in atmospheric pressure outside the enclosure 
that result in deformation of the spectrograph causing lateral shifts and focus errors; vibrations can also cause 
lateral shifts, focus errors, and PSF broadening; instability of materials can occur; for example from moisture 
desorption of the composite bench with G-CLEF. 
\item Pressure changes: because the wavelength calibrator follows the same path and column density as the 
program observations this can be calibrated out if the drift is slow compared to the observation times. 
\item CCD stitching errors: the lithographic process for making CCDs steps the mask in discrete steps, leaving 
positional errors after blocks of 512 pixels. As Doppler shifts or barycentric motion causes the spectral line 
to cross one of these regions, a spurious wavelength shift will be measured \citep{Dumusque2015a}. According to Paul Jorden, 
the stitching errors have been reduced by a factor of three with the current e2v devices.    
\end{itemize}

\noindent
Other error sources that cannot be calibrated out \citep{Podgorski2014}, include: 
\begin{itemize}
\item Barycentric errors: current state of the art algorithms, when combined with precise exposure meters, provide 
corrections good to 2 \cms. 
\item Software fitting errors: affect both the calibrator and extraction of the science spectra.
\item Micro-vibration: it is possible to obtain $\sim 1$ mg vibration control, but smaller vibrations can be coupled to the detector and would broaden the lines. 
\item Detector errors: including controller readout errors, intra-pixel QE variations, non-identical pixel sizes, detector heating on readout, dependence of CTE on amount of carried charge. 
\item Stray light: to minimize this, use baffling and ghost or scattered light analysis in the instrument design 
and keep the echelle orders well-separated. 
\item Tracking: this includes both telescope and on-instrument tracking. For negligible contribution to the RV 
error, displacements should be less than a quarter of the fiber diameter with a scrambling gain of 10,000. This is 
possible with octagonal fibers and a double scrambler to invert the near and far field. 
\item Telescope focus errors and variable seeing conditions will lead to variable instrument illumination, though translating this variation int an RV error is difficult. 
\item Imperfect atmospheric dispersion compensator: will vary the spatial illumination of the optics in a wavelength dependent manner. 
\end{itemize}

Imperfect atmospheric dispersion compensation can potentially introduce systematic effects in the measurements. For example, for 
fiber-fed spectrometers, the photocenter of the image might not fall at the center of the fiber at all the wavelengths. This could 
change the measured spectral energy distribution, introducing time-correlated changes in the slopes of the echelle orders. 
Data reduction and extraction should account for this, and algorithms must be tested against the range of changing conditions 
during commissioning. A second effect is that chromatic dependence in the fiber illumination may couple with scrambling 
efficiency and produce variations in the spatial illumination of the optics in a wavelength dependent manner.

\subsection{Detector Technology 
\label{ccd}} 

Charge-coupled devices (CCDs) are the detector of choice for optical spectrographs due to their 
high quantum efficiency (QE), low noise, large dynamic ranges, excellent linearity, and large pixel format 
substrates\footnote{Presentations by Andy Szentgyorgyi and Suvrath Mahadevan}. CCD detector technology 
continues to provide a steady improvement in performance 
along with new and innovative features, which help to make extreme precision RV measurements 
possible. This section summarizes the current state of the art for CCD technology and highlights 
some of their limitations in optical spectrograph performance.

\begin{figure}[htp]     
\hspace*{-1.cm}\includegraphics[width=7cm, angle=-90, clip]{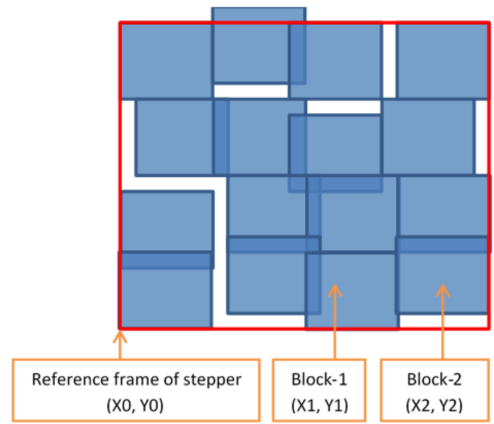}
\caption{Example of the pixel image blocks required to create a 2k x 2k pixel CCD 
with exaggerated random position errors relative to the reference frame (image credit to e2v technologies).                                                                                                                                                               
\label{stitching_blocks} }                                                                                                          
\end{figure}  

CCDs are sensitive over a broad wavelength range from 350 -- 950 nm, but require anti-reflection 
(AR) coatings 
to achieve high QE. Simple coatings must be optimized for specific wavelength ranges and 
thus are not ideal for the broad wavelength ranges of some spectrographs. More sophisticated 
graded coatings are available, which have spatial gradients in the AR coating to match the 
spectrograph bandpass. The spatial pattern of the AR coating can be customized for the spectrograph 
design --- optimized spectral response is ideal for fixed format spectrographs. In addition to 
improving the QE, the AR coatings will minimize reflections and thus minimize ghosts and fringes.

Inherent to CCD fabrication there are slight variations in the pixel spacing and geometry that 
can introduce errors in extreme precision RV measurements. Construction of the CCDs manufactured 
by e2V\footnote{Presentation by Paul Jorden} starts 
with a metal mask with a pattern of amplifiers, registers and image region that defines the 
structure of the device. The image area of the CCD is fabricated with a photolithography process 
by projecting a block of 512 x 512 pixels onto the silicon image region. The number of pixels in 
a block is limited by the field flatness of the camera. To produce CCDs larger than 512 x 512 pixels 
requires that the block mask be stepped and reimaged (stitched) multiple times to form larger 
area devices. A stepper motor moves the camera to position for projecting the adjacent block 
images and each block has a random alignment error relative to the stepper reference frame as illustrated 
in Figure \ref{stitching_blocks}.   
    
Two machines are used by e2V for the stitching process. The Ultratech stepper will lay down 
up to two blocks in x and multiple blocks in y with an alignment precision of 250 nm. The stitching 
errors for these devices can be seen by eye where the adjacent blocks overlap. The newer 
Nikon stepper will lay down multiple blocks in x and y with a stepper alignment precision of 
90 nm. While each stitched block is the same by design, there will be alignment differences 
between the blocks. Even with perfect alignment at the edges of the blocks, there can be a 
measurable boundary effect because of overlapping of the masks.  Both pixel size variations 
and stitching errors can be calibrated with a laser frequency comb or laser metrology \citep{Shao2013}.

Pixels on either side of the boundary can have slightly different dimensions. There are 
magnification and distortion errors in the photolithography process that can influence 
alignment. The pixel sizes have random uncertainties of $\pm150$ nm (or 1\% for a 15-micron pixel) 
from the tolerances of the lithographic process. 

CCDs and complementary metal-oxide-semiconductor (CMOS) chips exhibit pattern noise caused by QE variations, multiple output 
amplifiers, surface defects, and process variations. The pattern noise is small in modern 
detectors and fixed structure can be removed with calibration images.  CCD output 
amplifiers can be manufactured with low read noise ($<$2 e- rms) and consistent 
electrical gain so that multiple outputs do not contribute significantly to pattern noise. 
Process variations in detector fabrication, such as pixel size, thin film thickness, doping, 
metallization, and impurities cause a pixel response non-uniformity (PRNU). The pixel 
response uniformity in back-thinned CCDs is best at 650 nm, with the non-uniformities 
increasing at UV wavelengths from laser annealing process and at red wavelengths due 
to fringing. The PRNU can be improved with a good AR coating and 
thick silicon devices.       

To improve the QE response in the red and near IR, deep depletion CCDs are fabricated with 
high-resistivity silicon material (Òhigh $\rho$Ó) and have less silicon removed during the thinning 
process to produce ÒthickÓ (50-200 micron) devices. This results in the pixel Òsky scraperÓ effect 
where the height of the pixel is very large compared to its width: for example, the LSST chips are 
100 microns thick with 10-micron pixels. The high $\rho$ silicon allows a greater depletion layer 
within the tall pixel that minimizes the field-free region near the surface and allows electrons 
produced there to be more readily captured and collected within the pixel. This minimizes 
charge diffusion effects --- electrons wandering into neighboring pixels --- and improves the 
modulation transfer function (MTF), which is a measure of the resolution capabilities of the 
CCD. Conversely, the high $\rho$ silicon also introduces electrostatic fields in the CCD that 
could adversely affect charge collection: when one pixel fills up with signal charge it can 
influence the collection of charge in neighboring pixels because of these electrostatic 
effects \citep{Weatherill_2014}. In other words, the high $\rho$ silicon improves the MTF, but 
subtle variations in the MTF are possible as pixels fill up with charge and change the 
electric field lines in surrounding pixels. Deep depletion devices are also available and 
can exhibit tree ring patterns from resistivity 
variations in silicon boule \citep{Plazas2014}. 
 
 Charge transfer inefficiency (CTI) is a measure of charge transfer across each pixel and is 
 typically between 0.0010\% and 0.0001\%. CTI losses occur in ÒtrapsÓ that collect and hold 
 some number of electrons before releasing them at a later time. For a 10,000 e- signal, a CTI 
 of 0.001\% corresponds to a loss of 100 electrons per 1000 pixel transfers. 
 This generally results from 100 single electron traps within a 1000-pixel row or column. CTI is an 
 average of charge trapping events over the whole detector array. CTI is a function of temperature 
 and transfer frequency (clocking) as well as the signal level per pixel \citep{Bouchy2009}. Higher clock voltages can 
 improve CTI but also introduce the risk of clock-induced charge (CIC) trapping, although tri-level 
 clocking will reduce CIC. On the larger format devices that are required for high-resolution 
 spectrometers, the process of transferring charge can dissipate power in the silicon chip 
 and cause warping that can affect the measured radial velocity. The ESPRESSO team 
 estimates that 2 nm rms stability is required for 10 \cms\ RV precision. 
 
 Charge traps occur in discrete locations and are stable so that it is possible to 
 identify and characterize them using the pocket-pumping technique \citep{Wood2014}. 
 Because radiation in space  produces ongoing deterioration of the detector by creating traps, 
 EUCLID plans to use pocket-pumping to identify the charge traps and then apply post 
 processing corrections. Detectors used for ground-based facilities will not suffer from 
 ongoing degradation and can probably be characterized in the lab prior to installation in 
 the instrument; the measurement of charge traps would allow a correction to be applied to 
 correct for CTI. 
 
Silicon has a thermal coefficient of expansion of {$2.6 \times 10^{-6}/$C} and thus contracts on 
 cooling; a 4k 15-micron thick sensor will shrink by 12 microns when cooled. For large 
 format CCDs it is very important that the silicon is bonded to a package with a similar 
 expansion rate to prevent stresses from building up as the devices are cooled. 
 Silicon carbide (SiC) is an ideal material for CCD packages because it has a high thermal 
 conductivity and low density with high strength, and a coefficient of expansion {($4.0\times10^{-6}/$C)} 
 that is well matched to the silicon CCD. Both e2V and STA offer SiC packages for their large format CCDs.
    
The detector technology described in this summary was focused on specific detectors currently 
 in use in precision optical RV spectrographs. It is by no means an exhaustive analysis of all detector 
 options and features available today. This summary instead highlights some of the important 
 types of considerations that need to be made when using CCD detectors in extreme precision 
 RV spectrographs. Through a combination of advancements in detector technology and better 
 calibration techniques, the detector performance required to achieve 10 \cms precision is 
 within reach.

\subsection{Wavelength calibration} 
Historically, wavelength calibration for astronomical observations has been carried out with 
emission lamps or absorption cells. These wavelength calibrators have a price performance 
that is hard to beat\footnote{Presentation by Suvrath Mahadevan}. A crisis has been precipitated by a decision by all suppliers, globally,  of hollow cathode lamps to cease the manufacture of Thorium-Argon (ThAr) hollow cathode lamps with metallic cathodes. 
All ThAr lamps are now made with thorium oxide, which introduces impurities into the cathode. 
These impurities produce undesirable spectral features --- sometime referred to as ``grass" --- which 
quantifiably compromise wavelength calibration of cross dispersed echelle spectrographs\footnote{Abhijit Chakraborty has a program to develop ThAr lamps for astronomers in collaboration with IndiaÕs 
atomic energy program}.  
These traditional calibrators have inherent limitations that preclude a Doppler precision of better 
than about 1 \ms. The ThAr spectrum changes as the lamp ages and the emission lines are broad, 
saturated and irregularly spaced. If the ThAr spectrum passes through a fiber that is adjacent 
to the science fiber, the calibration 
is made for pixels that are offset from the science spectrum and an ultra-stable spectrometer is 
required for a reliable interpolation.  Furthermore, there is no conventional lamp that provides 
adequate calibration in the near infrared; charge from argon lines diffuses into neighboring pixels 
and the thorium lines do not have adequate density.  

The iodine reference cell technique is ideal for an unstabilized or general-purpose 
spectrograph, because the forward modeling technique accounts for variations in the spectral 
line spread function and modeling of the wavelength and dispersion with every observation. 
However, the iodine technique has a precision that is limited to $\sim1$ \ms\ for several reasons. First, 
the wavelength region is limited to 510 -- 620 nm, where the molecular iodine lines form, and there are 
light losses of about 25\% from the iodine cell.  The superimposed absorption spectrum also masks 
line profile variations that might be diagnostic of stellar activity. In addition, the intrinsic 
iodine spectrum is required for the forward modeling process; at the level required for sub-meter 
per second precision, the PSF of the high-resolution, high-SNR 
Fourier transform spectrometer (FTS) scans cannot be precisely deconvolved. 
Furthermore, there are degeneracies between parameters used in the forward modeling 
(Spronck et al. 2015) to measure Doppler shifts. Even if the spectrometer SLSF is stabilized, the Levenberg-Marquardt 
algorithm used to model shifts in the stellar lines is driven by chi-squared minimization and 
will fit out weak telluric contamination or some of the photospheric noise; this 
adds scatter to the measurement of the center of mass velocity of the star. 

Both the absolute wavelength of the calibrator and the spectrometer wavelength dispersion must be known 
with incredible precision if instruments are to achieve the precision necessary to detect Earth analogs. The reflex velocity 
induced by the Earth on the Sun is 9 \cms, or equivalently, a 100 kHz shift in frequency over 
one year. The ideal wavelength calibrator for the detection of Earth analogs must have a 
single measurement precision of about 1 \cms\ and must be stable for the expected duration 
of the observations (years). 

Ideally, light from the calibrator would follow the same light path through the spectrometer 
as the starlight. However, because a superimposed wavelength calibrator will obscure the stellar spectrum and 
make it more difficult to extract astrophysical Doppler signals, the calibration source must be spatially 
offset or temporally interleaved with the stellar observations. 
The ideal calibrator would have spectral features that are uniformly spaced (to maximize information content), 
unresolved (making it useful for caliibrating instrumental errors and the instrumental profile), 
well separated (ensuring features can be cleanly resolved by a high-resolution spectrometer),
and robust to ensure long-term accuracy is maintained. 

\begin{figure*}[htp]     
\includegraphics[width=0.9\linewidth,angle=0, clip]{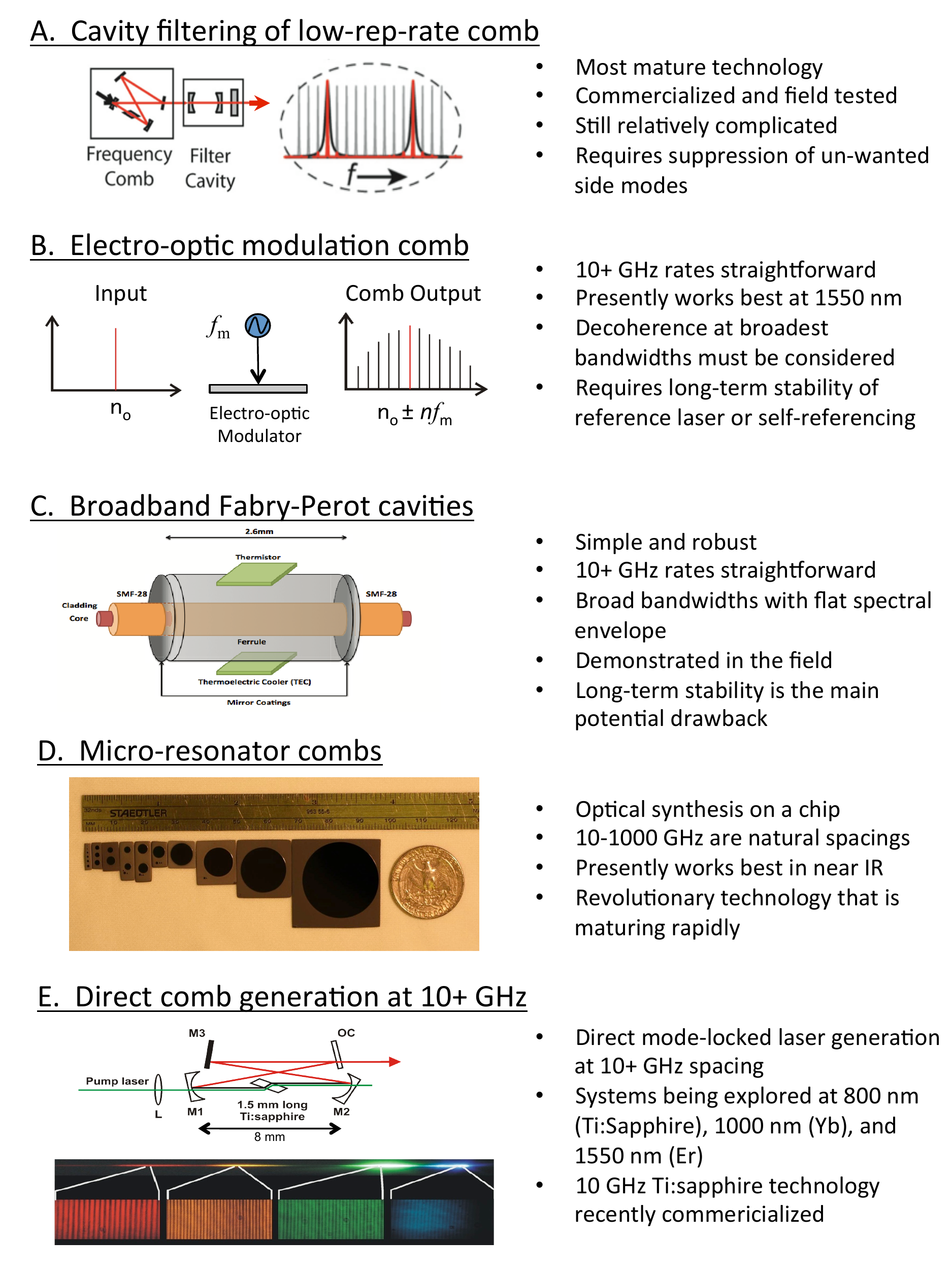}
\caption{Overview of various photonic sources that have been employed, or are 
being studied, as calibration sources for precision radial velocity spectroscopy.  Frames (A), (C), (D), 
and (E) are adapted with permission from \citet{Osterman2007,Halverson2014,Bartels2009,Li2012}, respectively.                                                                                                                                                              
\label{lfc_tech} }                                                                                                          
\end{figure*}   

\subsubsection{High Precision Wavelength Calibrators}
To improve the wavelength calibration precision from lamps or reference cells, 
several photonic sources, including laser frequency combs, stabilized etalons,
line-referenced electro-optical combs, tunable Fabry-P{\'e}rots, and chip-scale microcombs are 
being studied for precision radial velocity spectrometers.\footnote{Presentation by Scott Diddams} 
These methods are summarized in Figure \ref{lfc_tech}. 

LFCs can meet the requirements of an ideal wavelength calibrator. LFCs use 
femtosecond mode-locked lasers to produce extremely narrow emission lines 
\citep{Murphy2007, Li2008, Osterman2007, Steinmetz2008}. The LFC has perfectly uniform spacing 
of its individual emission line modes; this has been tested to 20 digits of precision \citep{Ma2004}.  
Moreover, 
LFCs with broad octave-spanning spectra can employ self-referencing to measure 
the carrier envelope offset, which provides a straightforward means to stabilize the 
absolute optical frequency of every comb element, in addition to the mode spacing. 
This creates an optical synthesizer where the frequency of every comb tooth is 
known with the same precision as the microwave standard (such as cesium clocks 
or GPS signals) that provide the reference signal for the LFC.  And because such 
microwave standards can be traced to the internationally defined SI second, the 
absolute frequencies of the comb are stable for decades. For example, when 
referenced to the free and wide-spread signals from the GPS (global positioning 
system) clocks, the long-term absolute frequency of the comb could be stable to 
better than $10^{-11}$ (i.e., $< 0.3$ cm/s) \citep{Quinlan2010, Lombardi2008}, meeting 
or exceeding the precision and stability required for RV planet searches. The 
advantage of the self-referenced  LFC is significant, in that it directly ties the 
comb frequencies to international standards (not artifacts or secondary references) 
and leverages the tremendous technological infrastructure of GPS and other 
global navigation satellite systems (GNSS).

Among the challenges for LFCs is the innately close mode spacing of typically 
0.1 -- 1 GHz, which requires spectral filtering of $\sim 98$\% of the modes 
with one or more Fabry-P{\'e}rot (FP) 
cavities to produce a calibration grid with 10 -- 40 GHz spacing (commensurate 
with the resolving power of a high-resolution spectrometer). Although filtered, the 
unwanted modes do not completely go away \citep{Sizer1989, Osterman2007, 
Braje2008, Quinlan2010, Ycas2012}. 
More important than the fact that residual side modes remain, is that the suppressed 
remnants can be asymmetric Ð and thus potentially shift the line center as seen by 
the spectrograph. Careful attention to this matter is required in order to provide a 
comb with sufficient side mode suppression and symmetry such that it does not 
affect the calibration at the level of desired RV performance \citep{Probst2014}.

Another challenge with LFCs is that spectral broadening with nonlinear optics 
is required to generate a comb that spans a broad bandwidth or operates at 
optical wavelengths.  This is straightforward at the ÒnativeÓ mode spacing of 
0.1-1 GHz, but due to a corresponding reduction in pulse peak power, significant 
broadening is more challenging at 10+ GHz mode spacing. A complication of this 
is that the side modes that one would like to suppress can be parametrically 
amplified during nonlinear spectral broadening and can even overtake the 
principal comb line \citep{Chang2010}. In some systems, this problem has 
been overcome by using up to 3 concatenated filter cavities in order to provide 
sufficient suppression of unwanted side modes \citep{Probst2014}.  Another 
approach to dealing with parametric sidemode reamplification is to first spectrally 
broaden at the native repetition rate, and then spectrally filter the broadened 
comb with a purely linear Fabry-P{\'e}rot filter cavity. This approach has been 
undertaken by the teams providing LFC calibrators for HARPS-N 
\citep{Li2015} and the Penn State Habitable zone Planet Finder (HPF).  
For example, with the HPF instrument the original 250 MHz laser at 1550 nm is 
broadened into the Y and J band with a highly nonlinear fiber (HNLF) at the 
native repetition rate to avoid parametric amplification of the side modes; the 
signal is then filtered using cavities with low dispersion mirrors.  While 
sidemodes may still exist, their impact on the spectral calibration can be 
calculated from simple linear optics and measured in-situ \citep{Li2008}.   

As an example of how far the LFC technology has matured in the past few years, 
Menlo Systems\footnote{Presentation by Tilo Steinmetz} now offers a commercial 
GPS-locked LFC that is designed to reach an accuracy of 4 mm/s at optical wavelengths \citep{Probst2014}. 
The system uses an state of the art IR fiber laser as a robust seed source. The low repetition 
rate (corresponds to a small mode spacing) of the fundamental femptosecond-laser is increased by mode 
filtering through three FP cavities with finesse of about 3000. Any mode spacing between 1GHz and 25GHz - in multiple 
integers of 250MHz - can thus be choosen to best fit the spectrographs resolving power. The 
filter narrowband IR spectrum is then amplified in an fiber amplifier and subsequently broadened 
in a specially designed photonic crystal fiber (PCF).

The system uses an IR fiber laser with mode filtering through three FP combs, 
signal amplification and spectral broadening. The use of FP cavities with a 
high finesse of 3000 significantly reduces the sidemodes and thereby 
minimizes their reamplification during spectral broadening. The spectrum 
is flattened to better than a few dB by using a liquid crystal display with 
active feedback loop, and the line intensities have low fluctuations over 
a broad optical wavelength band of 400 -- 700 nm with 15 to 25 GHz mode spacing.

Two Menlo system LFCs (with mode spacings of 18 GHz and 25 GHz) were 
tested during a campaign at HARPS from April 8 -- 18, 2015. The comb 
illuminated 48 echelle orders with 12,000 lines per channel. The HARPS pipeline 
produces 1-d extracted spectra for every order and a Gaussian was used to fit the 
LFC line centers. Modal noise in the fibers changes the intensity profile of the comb line recorded 
on the detector.  Fiber agitation is required to ensure each comb mode has a smooth, stable SLSF that 
is not dominated by the speckle interference pattern associated with multimode fibers illuminated 
by coherent sources. The wavelength solution was determined with an accuracy of 5 \cms\ after 
binning 5 exposures, and improved to better than 1 \cms\ by binning 20 exposures. 

The Menlo LFC systems have been developed for long term fail-safe turnkey operation. 
The system takes 10 minutes to change from the off state (laser amplifiers off) to 
standby mode with the laser on, locks enabled, and high power amplifier off. 
From the standby mode, it only takes 10 seconds to switch to full operation. There 
are some expected maintenance costs: the broadening fiber needs to be changed 
once per year and the pump diodes can fail every year or two. 
These components are built in as modules that can be easily replaced at a cost 
of a few thousand dollars. 

\subsubsection{Line-referenced electro-optical frequency combs (LR-EOFC)} 
In contrast to the mode-locked laser at the heart of the LFC, the LR-EOFC 
provides a spectrum of lines generated by electro-optic modulation. A pump 
laser is locked to an atomic or molecular transition and is then phase modulated 
and amplified to produce tunable sidebands with a spacing set by the modulation 
frequency \citep{Ishizawa2011}. \citet{Yi2015} have 
demonstrated a stability of $<200$ kHz ($\sim 0.3$ \ms) with a LR-EOFC at H-band 
wavelengths. The advantage of this technique is that there are no side modes; the 
system only generates the carrier with frequency sidebands that can be referenced 
to a microwave standard. Comb generation makes use of commercial off the 
shelf telecommunication components, but has currently been operated in 
wavelength bands where optical amplifiers and phase modulators are 
available (1-2 $\mu$m) from the telecom industry.  

There is still technology development needed to create EOFC broadband 
combs with constant power per comb mode, but the same techniques 
employed by \citep{Probst2014} should be adaptable to such combs. Another 
challenge is that although the line spacing of the LR-EOFC is fixed to a 
microwave reference, the seed laser for the comb needs to be locked to prevent 
wavelength drifts. In present systems, this has been provided by a laser that 
has its frequency locked to transitions in hydrogen cyanide or acetylene 
\citep{Yi2015}.  While this is straightforward, robust and commercially available, 
the long-term stability (e.g. months to years) of such molecularly-stabilized lasers 
is not known. Alternative atomic references for such LR-EOFCs could include 
narrower saturated absorption transitions in acetylene \citep{Edwards2004} or 
the well-studied rubidium D1 and D2 lines.  The spectral broadening is also 
more difficult at the high mode spacing of the EOFC due to the commensurate 
reduction in pulse energy.  However, very recently a 10 GHz EOFC centered 
at 1550 nm has been broadened to octave span and self-referenced such that 
the frequency of its seed laser can also be referenced to a microwave standard, 
and thereby inherit the stability advantages similar to the LFC described above \citep{Beha2015}. 

\subsubsection{Micro-combs}
One of the most exciting developments in laser frequency comb technology is a 
new approach to generate broad bandwidth combs in chip-scale devices. 
Such ÒmicrocombsÓ use the nonlinear Kerr effect in a dielectric micro-ring to 
produce a comb with line spacing that can be tuned between 10 and 800 GHz 
without the need for filtering \citep{DelHaye2007, DelHaye2008, 
Kippenberg2011, Jung2014}. 
While still at a relatively early stage of technical development, such microcombs 
hold the promise of frequency comb stability and accuracy in a very robust, compact, 
and potentially inexpensive silicon chip package. At present, the main challenges 
associated with microcombs are related to deterministic low-noise operation and 
increased spectral span, as required for self-referencing.  Additionally, the broadest 
bandwidth microcombs presently operate with mode spacing of ~200 GHz, which 
may too high for many high-resolution spectrographs \citep{Brasch2014}, although 
microcomb technologies exist down to mode spacings of a few GHz \citep{Li2012}.  
Finally, microcombs generally have native wavelengths in the near IR (1550 nm) 
but doubled or tripled frequencies can shift the comb to red and green wavelengths 
and self-referencing to microwave standards can be used for 2/3 or full 
octave-spanning combs \citep{Jost2015, DelHaye2015} . The 
development of microcombs is progressing rapidly, and it is likely that 
in the near future this new integrated technology could offer an attractive 
alternative for precision RV applications.

\subsubsection{Fabry-P{\'e}rot Interferometers} 
Fabry-P{\'e}rot interferometers provide a dense grid of lines with a technological approach 
that is simpler and a price tag that is 
lower than laser frequency combs\footnote{Presentation by Ansgar Reiners}. 
However, the systems need to be referenced 
to LFCs, lamps or atomic features \citep{Bauer2015, McCracken2014, 
Reiners2014, Schwab2015, Halverson2014}. Precise vacuum and temperature 
controls are generally required to stabilize the devices and it is important to track 
and correct for drifts in the dispersion.

Etalons are fixed length Fabry P{\'e}rot interferometers that produce broadband optical 
combs when illuminated with a broad bandwidth white light source. The HARPS 
etalon has Zerodur spacers to control the index of refraction, 
scrambled uniform illumination, and stringent temperature and pressure control 
(Wildi et al. 2012). These broadband sources have adjustable mode spacing, simplicity 
and good short-term stability. The night-to-night drift of the etalon lines is about 10 \cms 
and the stability over 60 days is about 1 \ms\ \citep{Wildi2012}. Near-IR fiber etalons have achieved 
better than 1 \ms\ stability over 12 hour timescales. In most cases, 
the frequency drift is not directly measured or corrected. \citet{Reiners2014} 
have suggested including a gas cell (e.g., Rb) stabilized laser to track frequency drifts 
and to derive wavelength offset corrections. 

Tunable FPIs that are white-light illuminated also produce a broadband comb of 
lines like the etalon, however they employ feedback to stabilize the wavelength 
zero point of the comb with locking to optical atomic standards like the rubidium 
D2 line at 780 nm. In order to provide sufficient precision, the FP cavity finesse 
needs to be high (at least 50 -- 100) at the wavelength of the laser lock, while the finesse 
can be decreased via engineering of the coatings to provide broad bandwidth 
coverage at other wavelengths.

Reaching 1 \cms\ wavelength calibration precision
is a grand challenge. By way of comparison, the Ramsey fringe of the cesium 
clock has a line width of about 1 Hz 
and a microwave oscillator can be locked to the central fringe with a precision 
of $10^{-13}$ in one second; and with averaging one can achieve a fractional measurement 
uncertainty of about $10^{-16}$ \citep{Heavner2014}.  In terms of Òsplitting the lineÓ, this is 
the most precise spectroscopy that has ever been done and allows the line centroid to 
be determined to one part in 1,000,000. However, this measurement is carried out in 
the lab with state-of-the-art microwave sources and laser-cooled cesium atoms.  
By comparison, astronomers who aim to reach 1 \cms\ precision 
(equivalent to ~15 kHz resolution in the visible) are using stellar spectral features 
that are many GHz wide.  If successful, this grand challenge would rival the 
line-splitting precision of work done by the AMO physicists; however, with the 
additional challenges of the accurate collection, dispersion and recording 
(at the photon counting level) of the relatively broad and faint spectral absorption lines.

\subsection{Fibers and scrambling}
The spectrograph projects the image of the entrance slit onto the 
detector as a function of wavelength.  Images of the telescope pupil (far field) also illuminate 
the echelle and other optics; alt-az telescopes that employ image derotators can end up with 
rotating spider mounts on the echelle. Any variability in the 
illumination of the spectrograph optics (say, because of guiding errors or changes in seeing) 
manifests as variability in the spectrum. This is a severe limitation for precise radial velocity 
measurements. Optical fibers provide an important benefit; not only do they offer a flexible way to couple 
light from the telescope into a spectrograph\footnote{Presentations by Paul Fournier and 
Suvrath Mahadevan
and breakout session led by Gabor F\H{u}r\'{e}sz and Paul Fournier}, they also offer the 
additional advantage of providing a more stable input illumination \citep{Heacox1986}. 

However, light will exhibit modal noise when traveling through a multi-mode fiber; this is particularly important for 
coherent light sources like lasers or laser frequency combs.  Modal ``noise'' is a manifestation of mode coupling that 
produces an irregular intensity distribution emerging from a fiber \citep{Baudrand2001}. When a laser is coupled 
into a multi-mode fiber, the light travels in many modes along the fiber; these modes are 
out of phase with each other, resulting in the complicated speckle interference pattern that 
illuminates the spectrograph optics. If nothing moves, the spatial speckle pattern is time invariant. 
However, stresses or motion of the fiber cause temporal shifts in the speckle pattern. When multi-mode 
fibers are used, some effort is required to scramble, or mix the modes. This can be done by shaking or 
agitating the fiber \citep{Lemke2011, McCoy2012, Plavchan2013b}.

When multi-mode fibers are used, some effort is required to scramble, or mix the modes.  
Many experiments have been carried out showing that when light is injected into a multi-mode 
fiber with a circular cross-section, the output intensity distribution is not completely independent 
of the input illumination. Circular fibers have reasonably good azimuthal scrambling, but poor 
radial scrambling. This results in variable illumination of the slit and spectrograph optics, and 
these undesired perturbations are imprinted in the spectrum, adding noise to the data.  
 
The ideal optical fiber has high throughput and a well-scrambled output that 
is independent of the way that light enters the fiber.  To achieve high transmission, 
the fiber length should be minimized; AR coatings can be applied to the fiber faces, and 
the focal ratio degradation (FRD) can be minimized by controlling the light injection 
cone to match the numerical aperture or acceptance angle of the fiber.  
Some FRD will still occur during the fiber fabrication and mounting process, from 
micro-roughness and micro-bending, subsurface damage and other induced 
stress in the fiber \citep{Ramsey1988, Carrasco1994, Avila1998}. The use of connectors can also be a significant source of FRD 
and soft, low shrinkage glue must be used to avoid strains. Fiber manufacturers do not 
typically measure FRD.

The use of fibers with non-circular cross-sections (octagonal, square or rectangular) 
break the symmetry of the wave guide and provide an easy way to scramble modes \citep{Chazelas2010, Avila2012, Spronck2012}.
Scrambling performance is improved by exchanging the in-focus near field (the image) and 
the out of focus far field (pupil illumination) between two fibers \citep{Hunter1992}. Variations in the fiber near field 
directly lead to fluctuations in recorded spectral features while variations in the far field change the illumination of 
spectrometer optics, leading to distortions in the PSF. Both effects lead to spurious velocity shifts and are 
detrimental to extreme precisoin measurements.

\begin{figure}[ht]
\begin{center}
\includegraphics[width=8cm, clip]{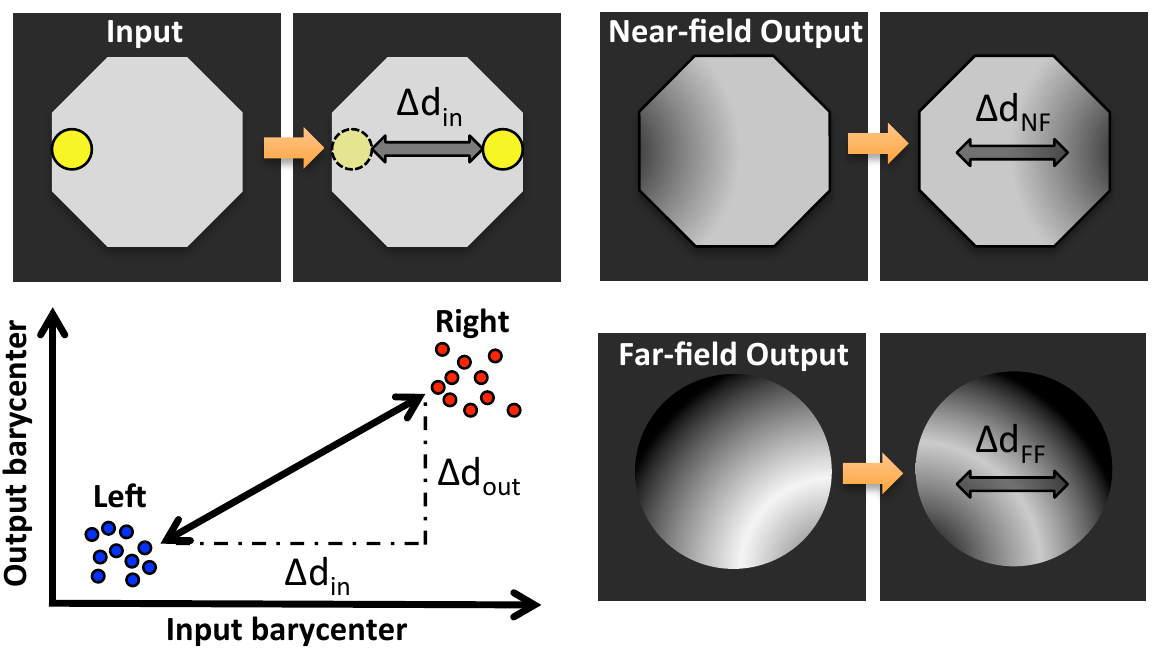}
\caption{Cartoon illustrating measurement of scrambling gain by translating the test fiber 
across a fixed illumination spot and recording the output at each position. 
Measurements are made for the fiber near field and far field. (Courtesy of Sam Halverson) 
\label{scramble} }
\end{center}
\end{figure}

As shown in Figure \ref{scramble}, the scrambling gain of a system --- for the near field --- can be measured by 
taking the ratio of the displacement of the light source at the entrance of the fiber to the 
residual shift in the center of mass intensity distribution of the fiber
output \citep{Avila2008}. Scrambling gains of up to a few hundred are achieved for pure circular fibers; 
octagonal core geometry yields a scrambling gain of up to about 1000, while octagonal fibers 
with double scramblers exhibit $\sim 10-20$k scrambling gain \citep{Feger2012}. Effects of 
inadequate scrambling in the far field can only be evaluated, by modeling how the observed pupil 
illumination irregularity propagates through an optical system \citep{Sturmer2014}. This can be done by ray-tracing 
through the instrument optical model, which leads to an estimate of the RV measurement sensitivity to far field variations. 
It has to be noted, though, that predicting the exact 
behavior and effects of the far field for an as-built instrument is almost impossible. It is because the 
as-built aberrations can significantly effect how the far field propagates through an optical system. 
Eliminating sensitivity of the as-built spectrographs, or variations in the far field of fiber systems, is 
key to further improve PRV performance.

Single mode fibers (SMFs) offer an interesting alternative to multi-mode fibers \citep{Crepp2014}.
True to their name, SMFs propagate a single mode (actually, two polarization modes are possible) 
of light; as a result, the fibers deliver a uniform 
intensity distribution to the spectrometer --- the SLSF has unparalleled stability --- and there is no modal 
noise from the fiber.  The challenge is coupling light into the small 5$\micron$ diameters of the SMF, 
however, extreme adaptive optics systems with high strehl ratios like the 
Subaru Coronagraphic Extreme Adaptive Optics (SCExAO) have reached fiber coupling 
efficiencies of 68\% with an $8\micron$ fiber core \citep{Jovanovic2014}.  

\section{Extracting Keplerian Signals} 
\label{jit}
Once the Doppler shift has been measured in a spectrum, the time series velocities 
are analyzed for Keplerian signals.  At this point, there are systematic error sources that 
add to the formal errors to obscure the detection of low amplitude systems.  
Telluric lines (from the Earth's atmosphere), 
stellar oscillations, sunspots, plage, granulation and meridional flows, all modify the shape of spectral lines as 
a function of time, and produce obscuring velocity signals. 

In many cases, astronomers have not been able to physically distinguish between stellar jitter and 
center of mass velocity with activity diagnostics; instead, some magnitude of the jitter is added in 
quadrature when using chi-squared minimization for fitting for Keplerian signals in radial velocity 
time series data. Regardless of whether the magnitude is assumed (e.g., based on spectral type) or 
inferred from the data, this approach is clearly insufficient; it implicitly assumes that jitter can be 
represented as a white noise source and does not leverage information about the time-coherence in 
the stellar activity signal.  

Another approach for mitigating stellar activity is to first measure apparent radial velocity shifts 
(neglecting activity) and then apply a bandpass filter in the Fourier (frequency) domain with the aim
of removing signals due to stellar activity.  This approach has been used with some success 
\citep{Howard2011b, Howard2013, Fulton2015}. However, the Fourier components are fixed;  
since the underlying noise sources change in time, the 
extracted signal will be distorted. In the best case, this introduces small systematic errors in the orbital 
parameters. In the worst case, true planetary signals are attributed to stellar activity or the analysis 
process introduces spurious coherent velocity variations that are mistaken for planets. 

The following cases offer a representative sample of putative planetary systems that have been debated, and 
the different approaches applied by the community: 

\begin{itemize}
\item GJ 581, with emphasis in the case of GJ 581d planet candidate initially described \citet{Mayor2009}. 
Subsequent arguments relating the signal to stellar rotation, activity and the statistical treatment of the problem 
can be found in \citet{Robertson2014a}, and \citet{Hatzes2015}.
\item $\alpha$ Cen Bb, which corresponds of a signal of $\sim 0.5$ \ms\ on a time-series strongly dominated 
by noise with structure. The initial claim was made in \citet{Dumusque2012}, and some doubts on the series 
whitening approach and caveats about statistical significance can be found in \citet{Hatzes2013} and \citet{Raipaul2016}.
\item HD 41248b \& c. These two Doppler signals were initially attributed to a pair of resonant planets in \citet{Jenkins2013}. 
Arguments in favor and against of activity induced variability were given in \citet{Santos2014} and \citet{Jenkins2014}.

\end{itemize}

\subsection{Statistical Analysis Techniques}
The interpretation of radial velocity data increasingly relies upon statistical 
analysis techniques\footnote{Presentation by Tom Loredo}. In medical fields, there has been a near crisis in the ability 
to replicate the statistical analysis of scientific results \citep{Ioannidis2005}. 
This is in part because scientists often know what results they expect and this can influence their 
interpretation, but it is also because of the misuse of null hypothesis tests, such as 
false alarm probability (FAP) and Kolmogorov-Smirnoff analysis. The repercussions can include 
financial cost, wasted time re-assessing false discoveries, and public mistrust of science. 
Our community seems to appreciate the importance of reaching out to statisticians and 
developing more sophisticated approaches to data analysis. 

The FAP test is a common tool for testing the reality of a Keplerian signal in radial velocity data \citep{Marcy2005}.   
This statistical test is usually carried out with a bootstrap Monte Carlo simulation. A Keplerian fit 
to the initial prospective signal is made and the $\chi_\nu^2$ value is saved. The velocity data 
are then scrambled, while retaining the times of the observations and the errors; for each realization, 
a blind search is made for the best Keplerian 
value and the resulting $\chi_\nu^2$ is recorded. If a signal was really contained in the original 
velocities, then the subsequent scrambled data are likely to have poorer $\chi_\nu^2$ values. 
This approach has the advantage of characterizing the impact of non-Gaussian 
systematic errors in a data set.  The reported FAP is the fraction of Monte Carlo realizations that yielded 
$\chi_\nu^2$ values lower than the original set.  As noted by \citet{Marcy2005}, the FAP test specifies 
the probability that incoherent noise could have yielded a superior Keplerian fit by chance. It does not 
necessarily show that the planet interpretation is correct --- the signal could arise from window functions 
or stellar activity or systematic instrumental errors. 

Statisticians prefer to call this a p-value because the language ``false alarm probability" seems 
to carry an implication about the reality of the signal or the planet interpretation. The p-value tells us 
about the probability of obtaining a particular signal given the null hypothesis (i.e., given the assumption 
that the data are merely incoherent noise).  

\subsubsection{Bayesian Analysis}
The field of radial velocity exoplanet detection has undergone a dramatic shift over the past two 
decades\footnote{Presentation by Eric Ford}.  
For many early years, ``discovering a planetÓ meant that astronomers had: 1) identified a possible
model to explain a series of Doppler observations as due to a planet and measurement noise, 
and 2) rejected the null hypothesis that a simpler model without the planet could reasonably explain the 
data.  This approach worked well for planet surveys that focused on well-characterized and quiet 
FGK dwarf stars with radial velocity amplitudes that were much larger than 
individual Doppler errors and plausible stellar activity signals.  
Astronomers did not want to risk incorrect detections near the threshold of their errors. 
When in doubt, the solution was to collect more observations.  

More recently, astronomers have begun adopting a Bayesian framework for characterizing the masses 
and orbits of exoplanets, as well as performing rigorous model comparison to quantify the evidence for 
planet detections.  In this framework, the criteria for detecting a planet is that the model including a planet 
has a higher posterior probability than the model or models without a planet.

The paradigm shift to adopt Bayesian inference has been fueled by the appreciation for greater 
statistical rigor, more accurate estimates of parameter uncertainties, and the greater sensitivity 
possible with a rigorous statistical analysis.  Perhaps most importantly, a Bayesian approach 
allows us to quantitatively answer questions that we actually want to ask.  For example, we can 
quantify the evidence for there actually being $N$ planets, rather than just rejecting the null hypothesis.   
Many statisticians like to think of Bayesian model comparison as a quantitative "Occam's razor" for 
comparing competing models.  Others prefer to make principle decisions using a utility function after 
after marginalizing over uncertainty in models and parameters.  

Of course, simply switching to a Bayesian framework does not solve all our problems.  One barrier 
to widespread adoption of Bayesian method is the necessity of computing multidimensional integrals.  
This can be computationally expensive, particularly for models with many parameters and when comparing 
the evidence for multiple competing models. This barrier has been largely overcome for parameter 
estimation applied to Doppler surveys thanks to increased computational power and improved 
sampling algorithms.  A second barrier is the intellectual investment required to understand the 
underlying algorithms.  For example, the choice of the proposal distribution in Metropolis-Hastings Markov chain 
Monte Carlo (MCMC) algorithms often makes the difference between fast and extremely slow 
convergence.  Therefore, one should always check multiple convergence diagnostics to identify 
any signs of non-convergence. Bayesian model comparison is particularly computationally expensive, 
so it is tempting to use short-cuts, such as the Bayesian Information Criterion (BIC).  
The BIC was designed to identify the most probable model in abundant data settings, but its 
approximations are not sound for quantifying model {\it uncertainty} in small or modest data settings, 
as arise in Doppler planet hunting.   Fortunately, the exoplanet community is gradually gaining experience with successful 
application of Bayesian methods to interpret observations. 

A third barrier is the misperception that the need to explicitly choose priors is an inherent weakness 
of the Bayesian approach.  If anything, it is a feature that the Bayesian framework clarifies the 
assumptions before entering into an analysis and provides a straightforward path for testing the 
sensitivity of conclusions to those assumptions.  Of course, it is useful to choose priors wisely.  
Often physical intuition, such as symmetries, or knowledge of previously identified planetary systems, 
can inform the choice of priors.  A general principle is to err on the side of overly broad priors.  When 
in doubt, one should perform a sensitivity analysis to quantify whether key results would change 
for different reasonable choices of priors \citep{Tuomi2013}.  

For complex models it is also wise to verify that one's 
model works well on simulated data to ensure that the chosen priors are appropriate for a given 
application. For model checking refer to \citet{Gelman2013}. Too often people using Bayesian methods 
ignore model checking, because it doesn't have a neat and tidy formal expression in the Bayesian 
approach. But it is no less necessary to do goodness-of-fit type checks for a Bayesian analysis than it 
is for a frequentist analysis.

A more fundamental challenge is that one must identify an appropriate model for interpreting
observations, regardless of whether using a frequentist or Bayesian approach.
Analysis is straightforward when the physical model is known, but
inference is much more challenging when the model is physically incomplete.
Early searches for giant planets appropriately considered only KeplerÕs laws,
but modern searches for low mass planets are significantly affected by stellar activity,
especially when surveys include active stars.
Unfortunately, we do not currently have a practical model of stellar activity,
nor even a clear path to such a model.
Activity indicators sometimes correlate with velocities, but this is not always the case.  

The problem of model incompleteness has been one cause for apparent contradictory results concerning 
signals that seem to be highly significant but are close to the measurement uncertainty level. For example, 
a discontinuity of a few \ms\ (eg. from a change of the gas cell or Th Ar lamp) will typically be much better 
fit by a sinusoid than a model with white noise only. If the model could account for real discontinuities, 
it would be clear that the addition of a sinusoid yielded a poorer match to the data. In a Bayesian context, this 
leads to an overestimation of the evidence because the sum over the models is only done for a subset of them. 
Thus, evidence ratios should not be interpreted as definitive proof for detections. This issue affects the model 
probabilities even if the multidimensional integrals are executed exactly. It is also related to the discussion 
about prior choices, in the sense that zero probability is implicitly assigned to a large number of alternative 
models. One way to mitigate this confusion is to secure minimal evidence ratio thresholds based on 
simulated data that is as close as possible to the real data. A more informative, but more resource intensive approach, 
is to confirm or refute signals for the same star using different instruments or techniques. For example, 
highly significant candidate signals were spotted in UVES data, but could not be confirmed with HARPS 
measurements \citep{Tuomi2014}. When applied to a large enough sample of objects, it is possible to 
construct more complete models or design a properly calibrated probability threshold system to 
identify reliable exoplanet signals.

It is useful to keep in mind that ``All models are wrong; some models are useful" (attributed to George Box 1979).  
Even after one verifies that a model and algorithm perform well when analyzing simulated data, one 
must still verify that they are useful for actual astronomical observations.  This will require extensive 
astronomical observations, extracting additional information from each spectrum to probe stellar activity, 
new models for joint radial velocity and activity data, and improved sampling algorithms. 
Therefore, future Doppler detections of low-mass planets will require very large amounts telescope time.  
In an effort to make efficient use of precious astronomical resources, it is natural to seek out the most 
powerful statistical methods to increase the sensitivity to low-mass planets and scientific impact of 
Doppler surveys.  Our community must identify realistic science goals, design observing strategies 
optimized to match those goals, and persuade time allocation committees to allocate sufficient time 
to characterize long-term and short-term stellar variability and systems of multiple planets.  

We will likely be forced to learn to live with uncertainty.  Given our upcoming requests for large investments 
of astronomical observatories, we will need to publish observations of both strong and marginal detections 
and learn to responsibly convey the limitations of our studies to other scientists and the public.  This greater
cultural shift towards emphasizing reproducible research is essential to the credibility of planet detections, 
astronomy and science in general.  

\subsubsection{Gaussian processes}
A Gaussian process regression is a powerful statistical analysis technique (beyond simply minimizing 
chi-squared in our models) that can be applied to the detection of weak signals in the 
presence of red noise\footnote{presentation by David Hogg}. Both the systematic and 
correlated noise sources can be included in the model and should be fit simultaneously with the 
signal (not sequentially) and marginalized out to correctly remove or minimize the parameters that are 
not of interest and to thereby avoid distorting the underlying signals.  

Chi-squared is defined as the sum of the squared differences between the data points, $y_i$, and the 
model ($m_i$ are theoretical points for a Keplerian signal), divided by the independent uncertainties ${\sigma_i}^2$, as shown 
in Equation \ref{chisq}.  It is also possible to switch from summation to vector notation 
and to relate chi-squared to the natural log of a Gaussian 
likelihood function, $L$,  by writing the difference of the data\footnote{Also see the presentation by 
Guillam Anglada-Escude} and model as a vector difference 
with the noise represented as a diagonal co-variance ($V$) matrix (Equation \ref{vector}). 

\begin{equation}\label{chisq}
\chi^2  \equiv \sum_{i} \frac{[y_i - m_i]^2}{{\sigma_i}^2} 
\end{equation}

\begin{equation}\label{vector}
\begin{split}
\chi^2 & = [y - m]^T \cdot V^{-1} \cdot [y - m] \\
V & = C \\
C_{ij} & = {\sigma_{i}}^2 \delta_{ij} \\
-2 \ln L & = [y - m]^T \cdot V^{-1} \cdot [y - m] + \ln \det V  + N\\
\end{split}
\end{equation}

If there are multiple components to the (Gaussian) noise, their covariances add in the noise 
covariance tensor, $V$.  For example, if a second noise source is added through the 
covariance tensor, $Q$, the variance tensor becomes 
$V = C + Q$ as shown in Equation \ref{monoise}.  This approach is straight forward and allows for 
the analysis of very high dimensional spaces and high fidelity signal recovery. 

\begin{equation}\label{monoise}
\begin{split}
-2 \ln L & = [y - m]^T \cdot V^{-1} \cdot [y - m] + \ln \det V + N \\
V & = C + Q \\
C_{ij} & = {\sigma_{i}}^2 \delta_{ij}  \\
\end{split}
\end{equation}

Systematic sources of error can be modeled as the sum of $M$ basis vectors with off-diagonal terms. If 
a Gaussian prior can be placed on the linear amplitudes, the systematic noise can 
be represented by a rank-M covariance matrix and marginalized to obtain a best fit. 
For example, in the case of \kepler photometric time-series data, the systematic noise 
sources (e.g. from the spacecraft jitter) can 
be modeled as the sum of M basis vectors; in this case, the basis vectors might contain 
information about the behavior of other stars. 

\begin{equation}\label{margnoise}
\begin{split}
-2 \ln L & = [y - m]^T \cdot V^{-1} \cdot [y - m] + \ln \det V + N \\
V & = C + B \cdot \Lambda \cdot B^T \\
B & = {\rm block \, of \,} M \,{\rm basis \,vectors}   \\
\Lambda & = M \times M \, {\rm prior \, variance} \\
\end{split}
\end{equation}

The noise tensor, $V$, encodes beliefs about the contaminating Gaussian noise and it 
can be designed to include a component for complex systematic noise. 
The models may be enormous, but it is important to fit for all of the nuisance parameters. 
For expressions like $V$ in Equation \ref{margnoise}, it is numerically safer to factorize the matrix
than to compute inverse matrices or determinants directly.\footnote{In the python numpy package, 
never use \code{inv}, always use \code{solve}; never use \code{det}, always use \code{slogdet}.} 

\citet{Baluev2013} treated time-correlated red noise sources as a Gaussian 
random process with an exponentially decaying correlation function and found that 
some of the announced planets around GJ 581 were illusions of red noise\footnote{Presentation 
by Roman Baluev}. Good statistical methods for handling red noise sources from stellar 
photospheres and instrumental errors are needed before we can determine the ultimate limitation 
to Doppler precision from stellar noise \citep{Baluev2015}. In recent years, Bayesian inference has become 
a central tool for the study of planetary or stellar signals in any RV dataset \citep[e.g.,][]{Faria2016}.

\subsection{Photospheric velocities: stellar jitter}
The contributions to radial velocity measurements that arise from photospheric 
motions are generally termed ``stellar jitter" and are currently an obfuscating source of 
time-correlated noise for RV measurements that aim to detect weak Keplerian 
signals. The typical magnitude of stellar jitter is 1 -- 3 \ms\ for chromospherically 
quiet stars \citep{Dumusque2011}. Stellar jitter is caused by a variety of physical 
processes\footnote{Presentation by Damien S{\'e}gransan}.
Unfortunately, there is no general analytical model for photospheric 
signals. Activity signals manifest as both incoherent and quasi-periodic variability. 

\begin{figure*}[!h]
\includegraphics[height=8cm, angle=90, clip]{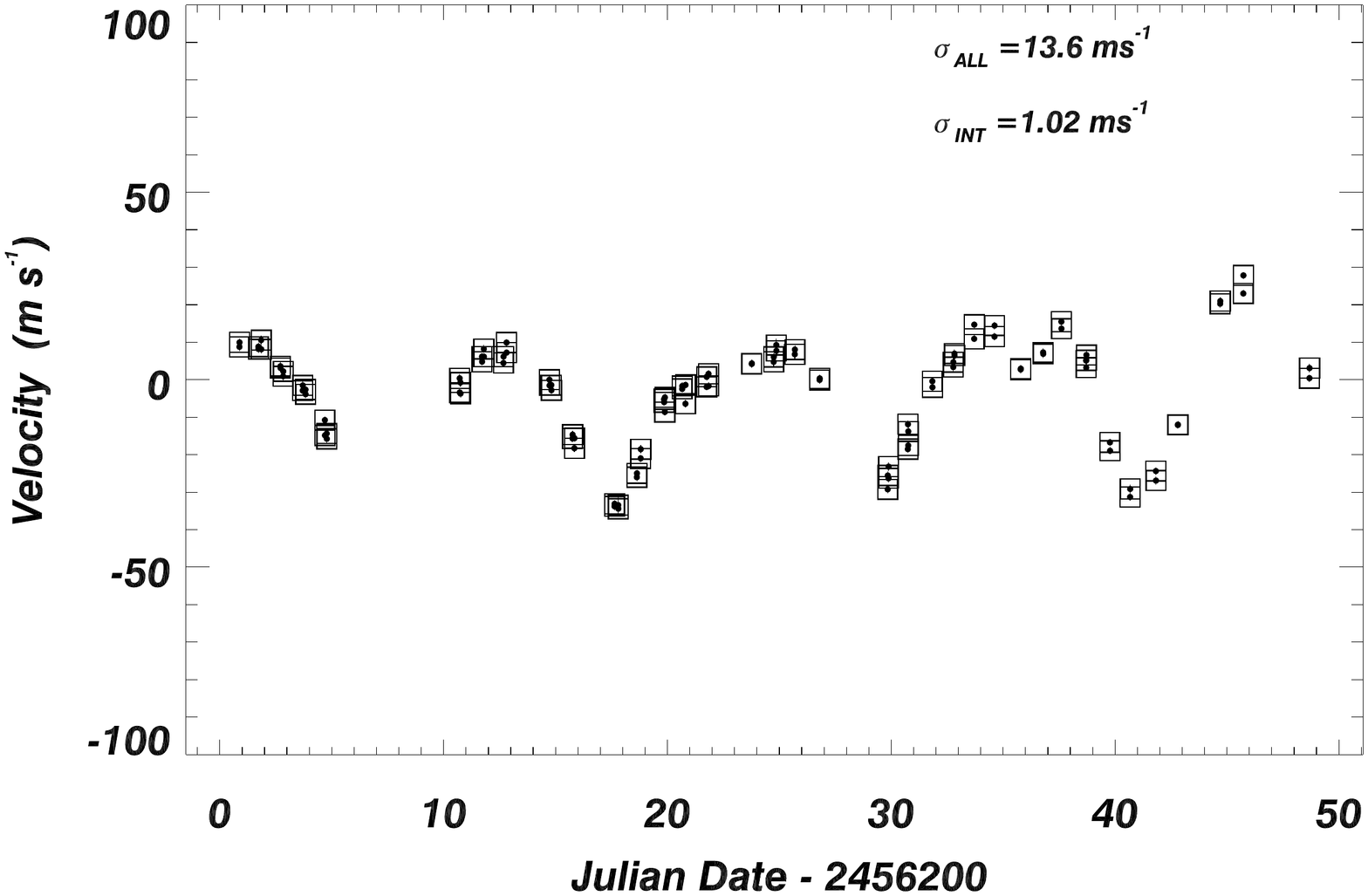}
\includegraphics[height=8cm, angle=90, clip]{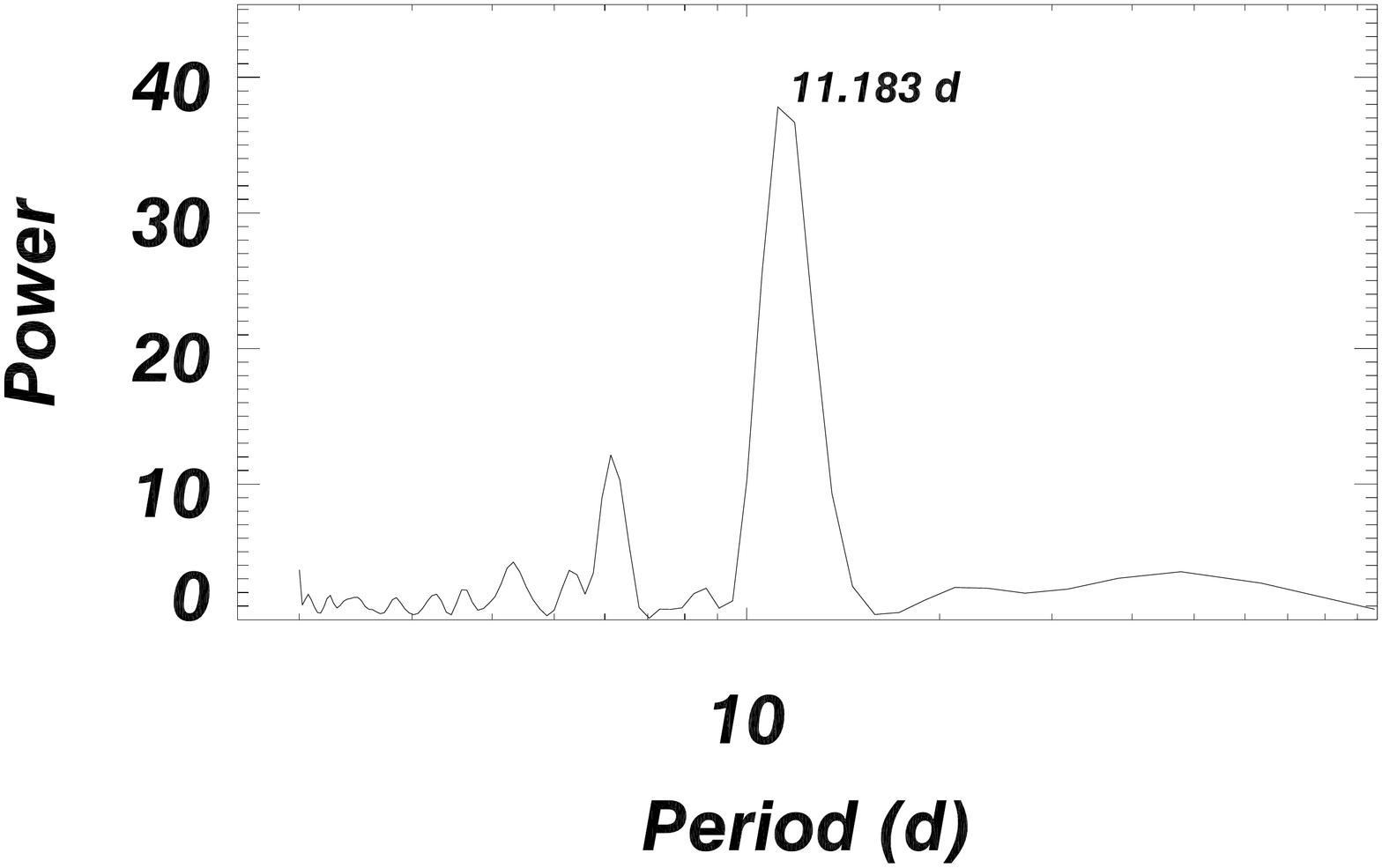}
\includegraphics[height=8cm, angle=90]{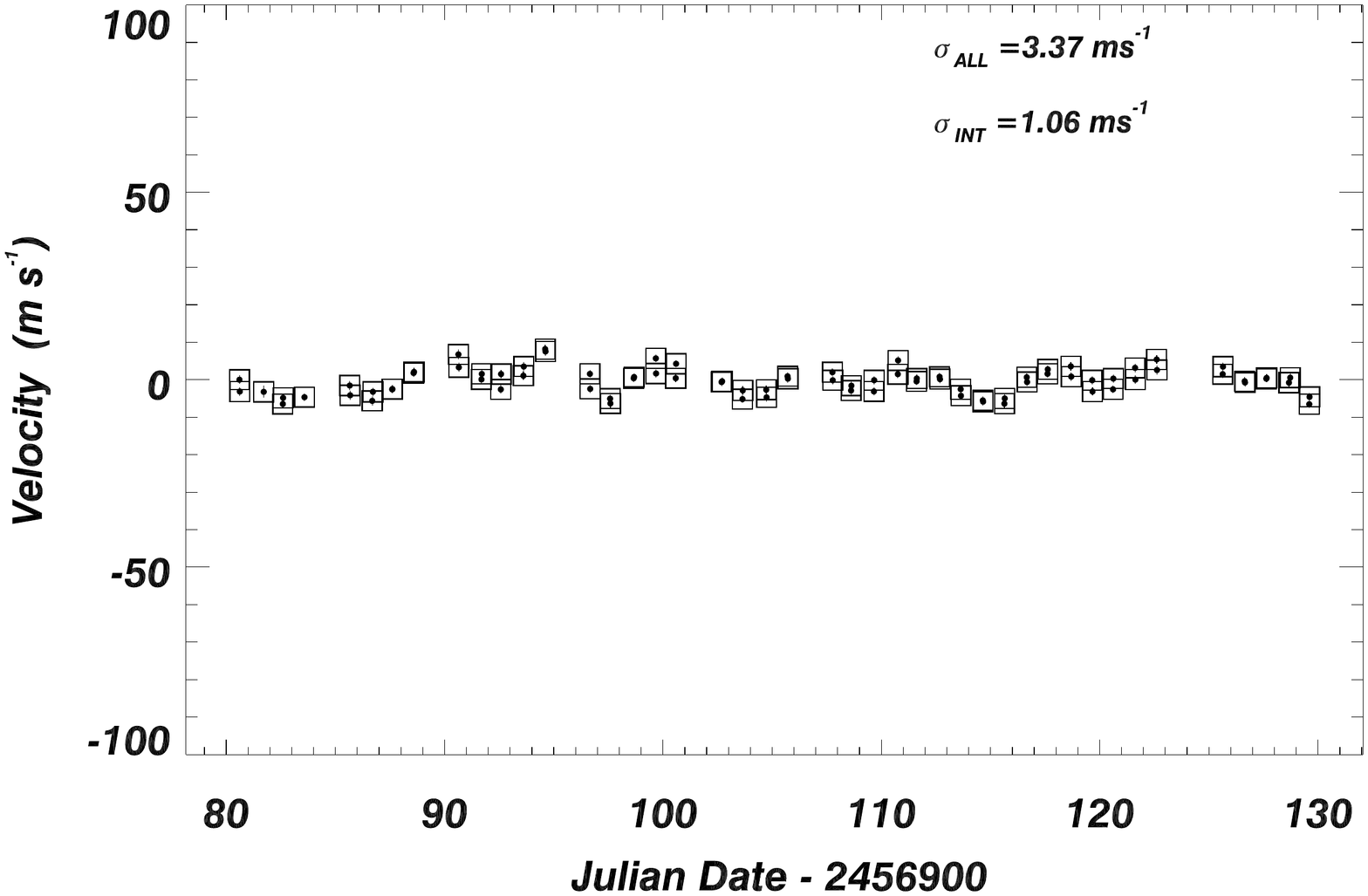}
\includegraphics[height=8cm, angle=90]{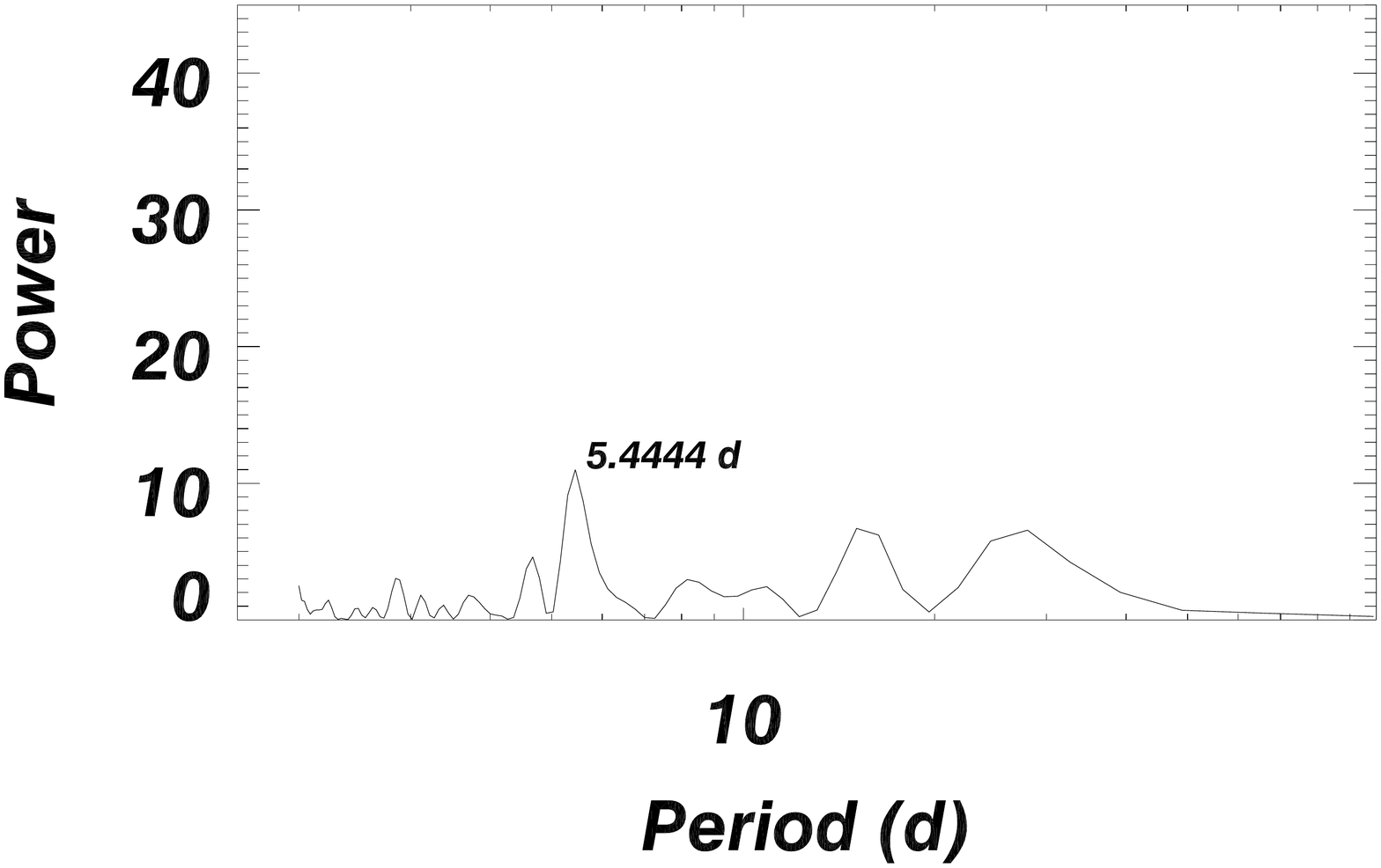}
\caption{Radial velocity measurements of Epsilon Eridani using the CHIRON 
spectrometer at CTIO have internal errors of 1 \ms. In 2012 (top panel) a clear 11.4-day periodicity is seen, 
which matches the rotation of the star. The rms variability is 13.6 \ms\. This signal attenuates over time and 
in 2014 (bottom panel) was not detectable. (Courtesy of Debra Fischer)
\label{epseri} }
\end{figure*}

Stars that are quiet can rather suddenly become more active \citep{Santos2014}. 
For example Figure \ref{epseri} shows radial velocity measurements of the moderately 
active star, Epsilon Eridani. Velocities were obtained with the CHIRON spectrograph 
and are shown for two different epochs. In 2012, a 50 day time series
of radial velocities (from JD 2456200 to JD 2456250) has an rms of 13.6 \ms\ and shows periodogram 
power that peaks at 11.18 days, consistent with the 11.35 and 11.55-day spot rotational period 
observed in 2005 using the Microvariability and Oscillations of STars (MOST) satellite \citep{Croll2006}. 
The 11 day radial velocity signal in Epsilon Eridani has been shown to correlate with photometric variability using MOST satellite and concurrent radial velocity measurements with the CHIRON spectrograph in 2014 \citep{Giguere2016}. This signal weakens over time with the rms dropping to a 3.37 \ms\ and is not seen in the periodogram of velocities obtained in 2014 (from JD 2456980 to JD 2457030).  

Cool stars have convective envelopes that support acoustic modes (p-modes) with velocity variations 
at the level of a meter per second on timescales of several minutes \citep{Kjeldsen1995}. Granulation 
in the photosphere is a manifestation of thousands of rising warm gas cells surrounded by a 
network of descending cool gas \citep{DelMoro2004}; kilometer-per-second velocities average out to some 
extent, leading to a net blueshift of a few meters per second in Sun-like stars \citep{Gray2009} and operate on 
timescales from hours to a few days. The granulation blueshift depends on stellar properties 
and for a given star varies as photospheric magnetic fields evolve \citep{Dumusque2011}. 

Active regions on stars arise from magnetic fields that thread the photosphere.  Magnetic fields 
coalesce into flux tubes, producing faculae (plage is the chromospheric counterpart) that are 
bright when small and dark when large (spots). The typical active structure that is 
observed on the Sun is composed of a dark spot in the center with strong magnetic 
fields and large flux tubes, surrounded by faculae. On average the faculae region is ten 
times larger than the dark spot region \citep{Chapman2001}. Magnetic 
fields act locally to inhibit convection, which suppresses the net convective blueshift induced 
by convection \citep{Dravins1982}. A dark spot and a faculae are therefore redshifted regions 
relative to the quiet photosphere \citep{Cavallini1985}.

Magnetic flux tubes form and decay on timescales comparable to the stellar 
rotation period, typically several days to a few months. 
Dark cool spots manifest as missing flux at the position of the spots. This breaks the symmetry of a rotationally 
broadened line, producing an east-west line asymmetry as a function of the rotational phase. This 
produces a net velocity shift and also changes the FWHM of all the spectral lines, and therefore the 
FWHM of the spectral CCF profile. As dark spots and bright plage evolve and rotate across the visible 
hemisphere, they alter the weighting of projected 
velocities \citep{Dumusque2014, SaarDonahue1997}. 

Long term magnetic cycles change the convection patterns (e.g. inhibiting convection), thus changing 
line-bisectors and line-shifts \citep{Dravins1982}. These cycles 
have timescales from a few hundred days to several years \citep{Meunier2013, Santos2010}, comparable to the 
orbital periods of real Jupiter like planets. 

It is easier to obtain high precision on very short timescales where the changes in the 
photospheric velocities that imprint variability in the spectral line profile are minimized.
\citet{Bourrier2014} devised an empirical correction for an observed wavelength dependent velocity 
trend and obtained residual velocity rms of 30 \cms\ when fitting the Rossiter-McLaughin effect 
for 55 Cancri e\footnote{Presentation by Guillaume H{\'e}brard}.  Long term precision is more 
difficult to maintain because the many sources of stellar jitter add spurious temporally correlated scatter to the 
center of mass Doppler velocities. These noise signals are a function of spectral type (lowest jitter for K dwarfs 
and larger for warmer stars) and evolutionary state of the stars (higher for 
subgiants or more evolved stars). Fortunately, stellar jitter has two important properties 
that can be exploited: 

\begin{itemize}
\item{it is not a persistent Keplerian signal; it waxes and wanes, it is not perfectly coherent, and it varies 
on timescales that are different from center of mass (COM) radial velocities}   
\item{the underlying physical phenomena that spawn jitter have detailed spectroscopic, photometric, 
wavelength dependent, and polarization signatures that are in principle distinguishable from simple 
wavelength shifts due to Keplerian Doppler shifts.}
\end{itemize} 

\begin{figure}[h]
\begin{center}
\includegraphics[width=8cm, clip]{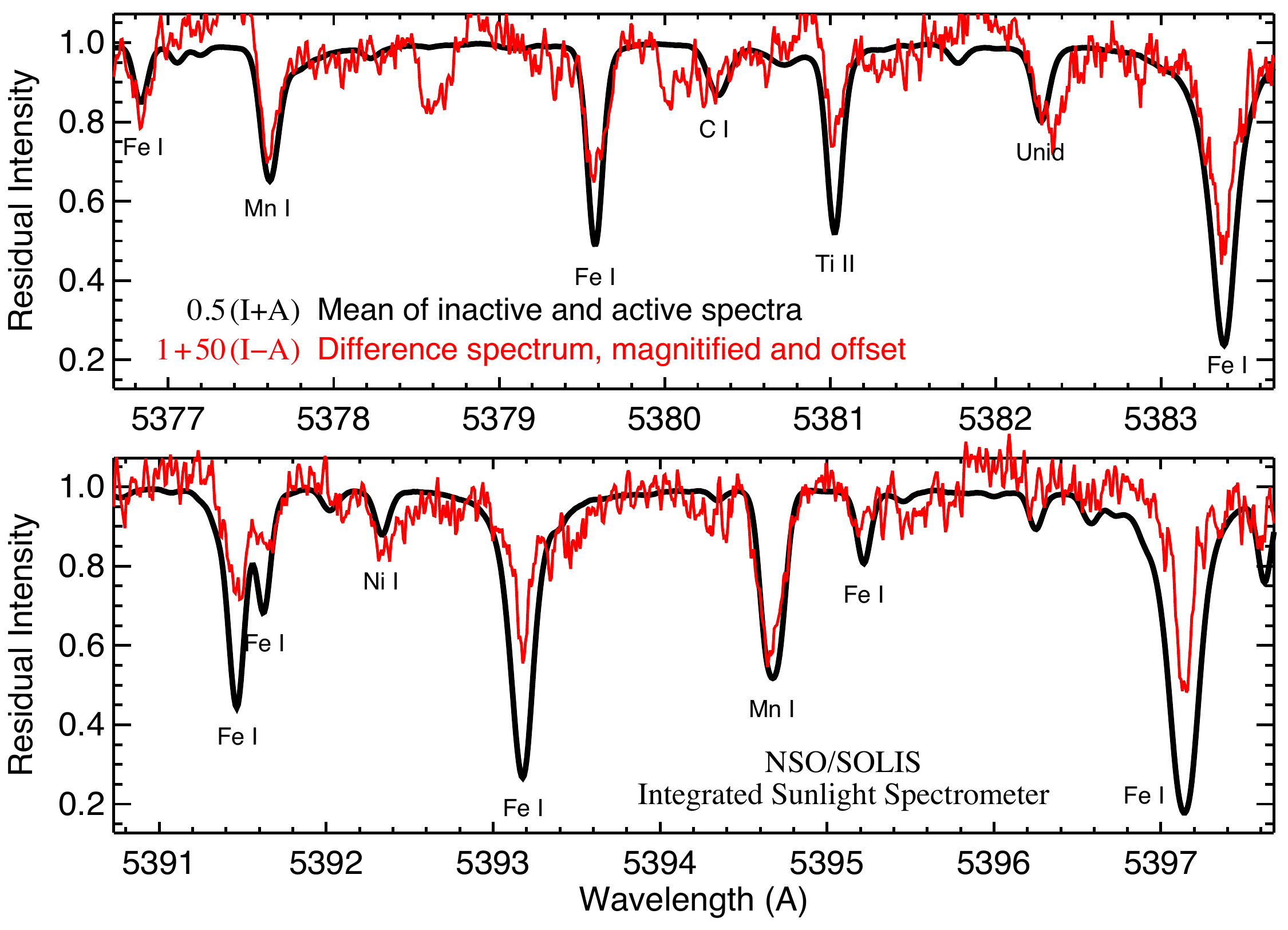}
\caption{Full disk solar spectrum and the magnified difference between active and inactive
states. Individual lines respond differently to activity changes, providing a way to help
distinguish surface phenomena from Keplerian velocity shifts (Courtesy of Jeff Valenti).
\label{activesun}}
\end{center}
\end{figure}

Figure \ref{activesun} compares the full disk solar spectrum at times of low and high activity, 
overplotting the magnified difference spectrum (red) on the average solar spectrum (black). 
Individual lines repond differently to changes in photospheric activity and provide a way to 
distinguish surface phenomena from Keplerian velocities. 

\subsubsection{Spectral Diagnostics of Jitter}
Astronomers have used diagnostics of stellar activity to decorrelate photospheric signals 
and radial velocities\footnote{Presentation by Nuno Santos}. 
These diagnostics include measurements of emission in the cores of Ca II H\&K spectral lines, 
changes in the spectral line bisector, or variability in the full width half maximum (FWHM) of the 
spectral cross correlation function \citep[CCF;][]{Dumusque2014}.  Emission in the cores of H $\alpha$, the Na-D lines, 
or the sodium doublet at about 819 nm \citep{Schlieder2012} provides more information about 
stellar activity for M dwarfs, which do not have much flux at the Ca H\&K 
wavelengths\footnote{Presentation by Paul Robertson}. 

The variations in the line bisector and FWHM are induced by active regions and are 
increasingly pronounced with stellar rotation. However, when looking at slow rotators like 
the Sun, it is more difficult to resolve the tiny variations in the bisector and FWHM, making 
these ineffective diagnostics of stellar activity \citep{Desort2007, Dumusque2014, Santos2014}. 
Observing with higher spectral resolution and with high sampling should help and additional 
spectroscopic diagnostics must be found. Currently, if significant signals can be detected in 
activity diagnostics, it is at least a warning that prospective Doppler signals may be 
spurious \citep{Queloz2001}. 

Two important upgrades were made to the HARPS pipeline that improved the precision 
of spectroscopic activity diagnostics. First, it was found that subtraction of scattered light 
(especially from the ThAr calibration fiber) near the Ca II H\&K lines improved 
the $\log R'_{HK}$ measurements \citep{Lovis2011}.  Second, the blue-to-red flux ratio 
varies because of weather conditions; imposing a fixed spectral energy distribution to 
the star before calculating the CCF vastly improved the CCF bisector and FWHM 
diagnostics\footnote{Presentation by Christophe Lovis} \citep{Cosentino2014}. 
The very stable spectral line spread function has been a decisive advantage
for high-fidelity spectrographs like HARPS that allows 
for recovery of spectroscopic activity indicators, even for chromospherically quiet stars. 

A caveat is that the activity indicators can be affected by instrumental effects and should not be 
used blindly. Over the past decade, there has been a small focus drift in HARPS that might have been 
mis-interpreted as a drift in FWHM from instrinsic stellar variability. Because this was seen 
for many stars, it is clearly an instrumental 
effect and was fitted out. In another example, low frequency modulation in the radial velocities was observed as a 
function of air mass in the star 18 Sco (and has also been seen in other stars). The 
source of this variation is ambiguous; is this granulation? seeing? SNR? 

The stellar spectrum contains a wealth of information that remains to be explored. The physics of 
stellar atmospheres should guide us in devising clever spectroscopic diagnostics and 
different stars may require different approaches. Regions of faculae 
and spots in the Sun have been modeled and demonstrate unequivocally that these features 
produce velocity variations of a few meters per second over one activity cycle \citep{Lagrange2010, 
Meunier2010, Borgniet2015}. Similarly, Haywood et al. (2015; poster at the EPRV meeting) shows 
correlations for simultaneous Solar Dynamic Observatory data and HARPS velocity measurements. 
Other useful approaches might include Zeeman modeling of the magnetic field strength \citep{Reiners2013} 
and magneto-hydrodynamic simulations to better understand the effect of granulation 
on line formation \citep{Cegla2015}. The HARPS-N solar telescope will provide excellent data to 
understand further the effect of stellar activity on radial velocity measurements \citep{Dumusque2015b}.
Studies like these associate physical phenomena with 
radial velocity variations and offer an important means for disentangling photospheric signals 
from center of mass velocities. 

\subsubsection{Identifying Jitter with Doppler Imaging}
Stellar astronomers carry out Doppler imaging on rapidly rotating stars by inverting time series spectral 
line profile information\footnote{Presentation by Thorsten Carroll}. Conventional wisdom suggests that 
it is not possible to carry out Doppler imaging on slowly rotating stars, but progress has recently 
been made by employing principal component analysis of the CCF for time 
series radial velocity data from HARPS. 

To motivate this analysis, solar synoptic magnetograms from June 2012 were used
as a model for the magnetic field on the surface of the Sun. Magneto-convection simulations were 
then used to simulate the temperature and velocity structure with one degree spatial resolution 
using the MURaM code \citep{Voegler2005} and a radiative transfer code was used to produce 
composite spectra for 156 absorption lines at each of 40 rotation phases. A singlular value decomposition 
(SVD; analogous to the CCF) of the time series spectra revealed an apparent radial velocity 
variation (induced by the photospheric surface phenomena) with the same period as the stellar 
rotation and an amplitude of $\sim1.5$ \ms. Contributions to the simulated photospheric 
velocity variations came from the magnetic field (from Zeeman splitting; $< 1$\% effect), 
spots ($< $10\% effect) and plage ($\sim 90$\% effect). The 40 time series CCF profiles were 
then inverted to produce a reasonably realistic reconstruction of the surface features that were 
initially used to generate the spectra.  

In his workshop presentation, Thorsten Carroll reported promising progress using 
principal component analysis (PCA) on the CCF of HARPS spectra for the slowly rotating star HD~41248 (rotation period 
of 25 days). This star is particularly interesting because of confusion about whether radial velocities 
are Keplerian signals or a red noise source from photospheric 
signals \citep{Jenkins2013, Santos2014, Jenkins2014}.
Carroll found that the first PCA eigenvector 
contains almost the entire radial velocity signal. The spectral lines from HD~41248 were 
then divided according to low ($< 1.25$eV) and high ($> 2.5$eV) excitation potential (EP). 
The CCF of both sets of lines showed essentially the same phase-folded periodicity in 
the radial velocity; however, an offset in the velocity zero point was observed. Velocities derived
from the CCF for the high EP lines were blue-shifted by about 150 \ms\ relative to the low EP lines 
because they form deeper in the stellar atmosphere and where the convective velocities are higher.
The eigenfunctions of the CCF contained additional interesting information; the first eigenfunction 
for the high EP lines showed an asymmetry (a broadening effect), while the first eigenfunction 
for the CCF of low EP lines was symmetrical. Thus, the spectral line shape varied with depth in the stellar 
atmosphere in a way that is consistent with the radial velocity signature --- this is an impressive 
demonstration that the CCF velocity signature here comes from convection, not from a center of mass 
velocity of the star. Thus, radiative transfer with magneto-convection simulations may ultimately 
offer an important foundation for understanding photospheric velocities as we move toward 10 \cms. 

\subsection{Near Infrared Radial Velocities}
For sunlike stars with spots, there may be an advantage to observing at near infrared 
wavelengths\footnote{Presentation by Lisa Prato}. The temperature of a spot can be
500 to 1500K cooler than the surrounding photosphere. If we adopt a toy model of 
two blackbodies with temperatures representing the spot and the stellar photosphere, the 
contrast between the emitted light is much higher at blue wavelengths than at red 
wavelengths.  As a result, the radial velocity modulation for spotted stars has larger 
amplitudes in the optical, compared to near infrared wavelengths, an effect that has been 
well-studied in young active stars \citep{Mahmud2011, Prato2008}.  Because of the importance 
and potential of RV measurements at NIR wavelengths, many future instruments, including 
CARMENES \citep{Quirrenbach2014}, the Infrared Doppler (IRD) spectrograph for high-precision 
radial velocity measurements \citep{Tamura2012}, SPIRou \citep{Artigau2014b}, and 
HPF \citep{Mahadevan2012}, are being designed to operate in the NIR. 

Most of the activity indicators that are used today are simply
not as precise or as efficient as we need them to be\footnote{Presentation by Pedro Figueira}. 
There is not a single activity indicator that is 100\% efficient and it is unrealistic to assume 
that all signals without activity correlations are correct. One way to confirm the planet interpretation
is to measure Doppler velocities in the near infrared; the precision for the NIR radial velocities 
is now at the level of about 2 \ms\ \citep{Figueira2010, Crockett2012}. 

\citet{Huelamo2008} showed that 
the radial velocity variations using optical spectroscopy of a star in the TW Hydra 
association disappeared when observed with CRIRES using near infrared
wavelengths. Infrared CRIRES data were also used to confirm that variations in several K giants
were caused by planetary companions \citep{Trifonov2015}.
\citet{Bean2010} used an ammonia gas cell for wavelength calibration 
on CRIRES to achieve a Doppler measurement precision of 5 \ms\ on 
M dwarf stars\footnote{Presentation by Jacob Bean}. 
They found that the noise floor for NIR precision was set by telluric contamination; there 
was no NIR wavelength region in the CRIRES bandpass that was free of telluric lines at the level needed to detect 
habitable planets, even around M dwarfs.  

CRIRES is fed with light from an AO system, 
which helps to stabilize the slit illumination. While the slit fed spectrograph is ideal for 
background sky subtraction, intensity scrambling is not possible, so the 
spectrograph has variable illumination of the optics. The spectrograph is 
cryo-cooled, however, it is not a stabilized instrument.  \citet{Bean2010} found that 
the use of an ammonia gas cell calibration for the K-band (2.3 $\micron$) required 
many free parameters to forward model the instrumental point spread function and the 
system was less efficient because the K-band was not at the peak of the spectral energy distribution 
of their early M dwarf stars. However, the NIR spectral range does contain significant Doppler information for mid to
late M dwarfs \citep{Figueira2016, Beatty2015} and advantages of working at NIR wavelengths 
to mitigate activity signals make it worthwhile to continued efforts to improve precision in the NIR. 

With faculae or plage, the associated magnetic fields suppress
convection and this perturbs the spectral line profile and produces photospheric 
Doppler shifts. Although there is only a small temperature difference between faculae and the photosphere; 
\citet{Marchwinski2015} used the FF' technique \citep{Aigrain2012} on
photometry from the SORCE spacecraft to show that the plage-dominated activity 
in the Sun exhibits smaller radial velocity scatter at near infrared wavelengths 
than optical wavelengths.  So, the NIR may be useful as a diagnostic even in stars 
that have small spots but significant faculae and plage. 

Interestingly, \citet{Reiners2013} found that the Zeeman effect, which increases with wavelength, 
can spuriously increase the radial velocity signal at redder wavelengths. 
The amplitude of the radial velocity signal caused by the Zeeman effect 
alone can be comparable to that caused by temperature contrast; a spot magnetic field 
of about 1000G can produce a similar RV amplitude as a spot temperature
contrast of about 1000K. For the active M dwarf AD Leo, \citet{Reiners2013} found that 
the radial velocity signal increases at longer wavelengths, consistent with a strong influence
of the Zeeman effect. Therefore, the RV signal depends on the combination 
of spot temperature and magnetic field.

There are several instrumental challenges for EPRV with infrared spectrometers; 
these include detector technology, thermal and environmental 
control, modal noise, wavelength calibration and sky background \citep{Plavchan2013a, 
Plavchan2013b, AngladaEscude2012}. For most of these 
issues, progress has lagged behind optical instruments. The use of adaptive optics 
may be the one area where NIR spectrographs have an advantage over optical instruments
and the use of single mode fiber feeds to IR spectrographs \citep[e.g., iLocator,][]{Crepp2014} offers 
an interesting path forward. 

\subsubsection{Simulating Stellar Noise}
The community has developed simulation tools for understanding and modeling stellar activity, 
including the SOAP \citet{Boisse2012}, 
SOAP 2.0 \citet{ Dumusque2014} code for simulating realistic spectra that captures the effect of 
spots and faculae. 
A new code, StarSim \citep{Herrero2015} produces realistic synthetic time series spectra 
that accounts for limb darkening, spots, faculae and convection using a surface 
integration and includes photometric information\footnote{Presentation by Enrique Herrero}.  The inputs 
to StarSim include stellar parameters (effective temperature, $\log g$, metallicity, rotation, differential 
rotation, and inclination) information about the active regions (spot positions and sizes, temperature contrast,
faculae temperature contrast, faculae to spot area factor, and evolution of the active region) and information
about planets (size, ephemeris, spin-orbit angle, impact parameter).  The spectroscopic contribution from 
the photosphere, spots and faculae are synthesized from the Phoenix BT-Settl models) and stored in a 
spherical grid representing the star. The synthetic spectrum is calculated by integrating the surface 
elements that contain temperature, limb darkening, radial velocity and geometry (size and projection). 
Time series observations are produced as a function of the rotational phase angle. By using the HARPS 
mask template, the CCF models can be produced and analyzed to try to recover the underlying simulated
noise. 

\subsubsection{RV Fitting Challenge \label{RVchallenge}}
In advance of the workshop, Xavier Dumusque organized 
a challenge\footnote{\url{https://rv-challenge.wikispaces.com}} for fitting 
Keplerian signals in radial velocity data \footnote{Presentation by Xavier Dumusque}.  He provided simulated 
RV data sets, using the time stamps of HARPS observations for some well-sampled stars. 
The data contained stellar signals from from oscillations, granulation, spots and plages, magnetic cycles 
and Keplerian signals. Eight teams participated in this simulation and used different 
techniques to recover planetary signals, as detailed in Table \ref{tab:RV_players}. The results of the RV fitting challenge 
are fully described in \citet{Dumusque2016}. Here, we summarize the take away messages of this exercise.  

\begin{figure*}
\begin{center}
\includegraphics[width=14cm]{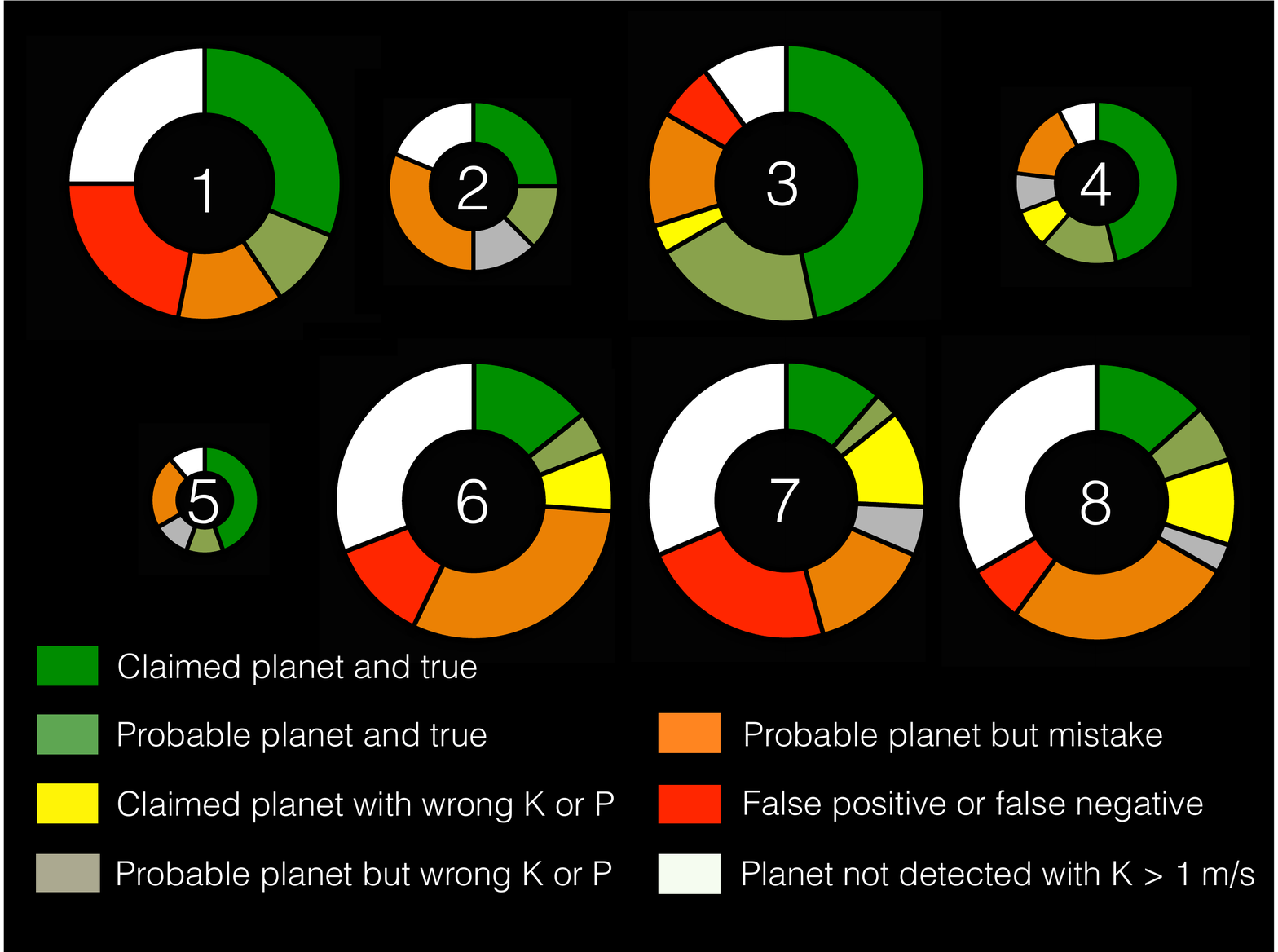}
\caption{Summary of the result of the RV fitting challenge for teams 1 to 8. The size of the circles represents 
the number of systems analyzed: 14 systems for the large size circle, 5 for the medium size and 2 for the smallest size. The meaning 
of the colors are defined in the legend and in the text (Courtesy of Xavier Dumusque) }
\label{fig:RV_chal}
\end{center}
\end{figure*}

For the RV fitting challenge, 14 data sets were provided, including 4 data sets from real 
HARPS observations and 10 simulated data sets. A total of 48 planetary signals were present in the data, 
with semi-amplitudes ranging from 0.16 to 5.85 \ms. In Fig. \ref{fig:RV_chal}, we summarize the 
results of the different teams. For each signal detected, we assign a different 
color flag depending on the true signals present in the data. The different possibilities are:

\begin{itemize}
\item dark green: the group recovered a planetary signal that existed in the simulated data, and the group was confident in the detection, 
meaning that they would have published this result.
\item light green: the group recovered a planetary signal that existed in the data, however the group assigned a low confidence to the detection.
\item yellow: the group recovered a planetary signal that existed in the data and was confident enough for publication; however, the amplitude 
or period is slightly wrong compared to the true signal, or an alias of the true signal was detected.
\item grey: the group recovered a planetary signal that existed in the data, but was not confident in the detection. The amplitude or 
period was slightly wrong compared to the truth, or an alias of the true signal was detected.
\item orange: the group recovered a prospective planetary signal that did not exist in the data, but assigned a low confidence level to the detection 
\item red: false positive or false negative, i.e., the group recovered a planetary signal that does not exist in the data, but they were confident enough 
in the detection for publication (false positive), or the group rejected the detection of a true signal (false negative).
\item white: a simulated planetary signal was present in the data that has an amplitude larger than 1 \ms, but this signal was not detected.
\end{itemize}

To study the statistics of exoplanetary populations, it is most important that published planets with correct 
parameters (dark green) are found. It is also important to understand when false positive or false negative signals (red) are reported 
and when relatively large signals with amplitudes that exceed error bars are missed (white). 
Signals that are published with modest errors in their parameters (yellow) will not strongly 
corrupt the planet population statistics. The RV challenge also helped to illucidate trends regarding the signals (light green, grey, orange) 
that were either true or false, but that would not be published. 

In Figure \ref{fig:RV_chal}, the most successful groups should have a larger dark green region, and smaller red and white 
regions. Given this metric for success, we separate the teams in two different groups: teams 1 to 5 (groups that used a Bayesian framework 
to compare between models) and teams 6 to 8 (groups that did not use Bayesian models).  
Groups 1 to 4 also used red noise models in addition to Keplerians to account for stellar signals, and group 5 used a 
white noise model that was free to vary depending on the activity level of the star in addition to Keplerian models.   
The other groups compared RV signals with signals present in the activity observables (${\rm \log R^\prime_{HK}} $, FWHM, BIS SPAN) and rejected 
significant signals it they were also  in the RVs or in any of the stellar activity diagnostics. 

\begin{deluxetable*}{llll}
\tablecolumns{4}
\tablewidth{0pt}
\tabletypesize{\scriptsize}
\tablecaption{Fitting Teams\label{tab:RV_players} }
\tablehead{
\colhead{ } &  \colhead{Teams}  & \colhead{People}  & \colhead{Techniques}   \\  
}
\startdata
1 & Torino       & M. Damasso, A. Sozzetti, R. Haywood             & Bayesian framework with \\
   &                    & A.S. Bonomo, M. Pinamonti, and P. Giacobbe & Gaussian Process to account for red noise \\
2 & Oxford         & V. Rajpaul and S. Aigrain                                  & Bayesian framework with \\
    &                     &                                                                                 & Gaussian Process to account for red noise \\ 
3 & M. Tuomi      &  M. Tuomi and G. Anglada-Escud\'e              & Bayesian framework with \\
     &                     &                                                                                & Moving Average to account for red noise \\
4 &  P. Gregory  & P. Gregory                                                          & Bayesian framework with \\
    &                       &                                                                               &  Apodised Keplerian to account for red noise \\
5 & Geneva        & R. Diaz, D. S\'egransan and S. Udry              & Bayesian framework with white noise                \\
6 & H. Hatzes    & H. Hatzes                                                             & Pre-whitening                                                          \\
7 & Brera           & F. Borsa, G. Frustagli, E. Poretti and M. Rainer  & Filtering in frequency space               \\
8 & IMCCE        & N. Hara, F. Dauvergne and G. Bou\'e               & Filtering in frequency space                                \\
\enddata
\end{deluxetable*}

According to \citet{Dumusque2016}, among the teams that used a Bayesian framework to compare between models and red noise models to account for stellar 
signals, team 3 was the best at finding true planetary signals; they were confident in the majority of their detections and they made a 
small number of mistakes. This team detected 85\% (17 out of 20) of the true planetary signal present in the data that had a 
semi-amplitude larger than 1 \ms. The remaining 3 signals that were not detected all have periods longer than 500 days and 
their non-detection could be explained either by magnetic cycle RV effects, or simply by the fact that team 3 only fitted up to a 
second order polonomial to remove any RV trends in the data. This team would have only published 60\% of these 
large amplitude signals, leaving the remaining signals as tentative. When analyzing planetary signals with semi-amplitudes 
smaller than 1 \ms, team 3 did also the best job at recovering those signals, with 14\% detected (4 out of 20). However the 
team would have only published 2 of those planets, with amplitudes of 0.48 and 0.69 \ms. Finally team 3 refrained from announcing 
non-existing planets with semi-amplitudes larger than 1 \ms, and would have only announced 2 non-existing systems with smaller semi-amplitudes. 

The next two best performing teams were Team 1 and Team 4.  While Team 4 ranked third, they actually did a better job than 
Team 1 on the fitting first five systems. However it was difficult to make a fair comparison because they did not analyze the other systems.

In conclusion, the RV fitting challenge \citep{Dumusque2016} showed that teams using a Bayesian framework that allows for model comparison, and that 
includes a red noise model to account for stellar signals were more efficient at finding true planetary signals, while limiting the 
number of false claims. Among the different red noise models, it the moving average model used by M. Tuomi and G. Anglada-Escud\'e 
(Team 3) was the most efficient. The use of Gaussian Processes was also an efficient approach, however this technique seems to be 
sensitive to the {\it a priori} determination of the rotation period; it seems likely that some improvement can be made here \citep{Dumusque2016}. 
The apodized Keplerian technique \citep[Team 4][]{Gregory2016} also yields good results. \citet{Gregory2016} find that it is possible to 
achieve a factor of about six reduction in the stellar activity noise in the simulated radial velocity sets with the apodized Keplerian technique. 
Unfortunately, Team 4 only analyzed only the first 5 systems as they were developing their technique, and this ultimately penalized their ranking. 
Finally, planetary signals with amplitudes above 1 \ms were detected almost all of the time, 
while only 14\% (4 out of 28) of true smaller signals were discovered. Out of these 4 true small signals, 2 would have been published; however,  
2 false planetary signals would have been published as well. Therefore, even with the best models of stellar signals, planetary signals with 
amplitudes less than 1 \ms\ are rarely extracted correctly with current precision and current techniques. 

\subsection{Aliasing} 
Radial velocity data sets will generally include aliased signals, which are interactions 
between gaps in data sampling and signals\footnote{Presentation by Rebekah Dawson}. 
A sinusoidal signal in time series 
data is a delta function in the frequency domain. Because observations are taken at specific times, 
a window function is applied to the data; in the time domain this results in sampling gaps. 
The convolution of the sinusoidal frequency and the window function produces multiple peaks 
in the frequency domain. 

How do we know what the time sampling is for complex sampling patterns? Solar and sidereal 
day and annual sampling gaps naturally appear in most radial velocity data sets and 
these aliases and their harmonics produce peaks at positions that can be calculated: 

\begin{equation}\label{alias}
f_{alias} = | f_{true} \pm f_{sample} |
\end{equation}

\citet{Dawson2010} show that aliases imprint a set of predictable peaks; by modeling a noise 
free data set, it can be possible to confirm Keplerian signals in the presence of aliases.  However, 
low SNR data is more ambiguous when trying to identify aliasing, and chromospheric 
activity can produce a forest of peaks with associated aliasing signals.  Stellar activity does 
not produce an exact sinusoidal signal; instead, the imprinted perturbations
are stochastic and quasi-periodic. The potent combination of variable intensity in the stellar 
noise signal and poor sampling leads to ambiguity in identifying aliases that can sabotage the 
identification of weak Keplerian signals. 

\subsection{Barycentric corrections}
Corrections for the velocity of the Earth must be applied in order to recover precise 
radial velocity measurements of stars for exoplanet detection.\footnote{A breakout session 
on barycentric corrections was led by Jason Eastman and Lars Buchhave.} 
\citet{Wright2014} outline the magnitude of error contributions to the 
barycentric velocity correction, including rotational and orbital motion of the Earth, 
second order corrections to the non-relativistic Doppler formula, secular acceleration, 
precession, nutation, gravitational redshifts, blueshifts, and the Shapiro delay (for light 
bending around the Sun).  In their publicly available code
for barycentric correction,\footnote{http://astroutils.astronomy.ohio-state.edu/exofast/} 
the authors show that their calculations match those from the 
TEMPO2 \citep{Hobbs2006, Edwards2006} code at a rms level of $0.3 \,{\rm mm\, s^{-1} }$. 
The accuracy of TEMPO2 is a gold standard with a precision better than 1 \cms\ 
that was demonstrated when measuring the Doppler effect of pulse arrival times for 
the detection of Earth-mass exoplanets orbiting a neutron star \citep{Wolszczan1992}. 

The barycentric correction for radial velocity measurements at the 10 \cms\ level requires a precision  
of about one second in the flux-weighted exposure time. This is generally accomplished by 
monitoring the flux of starlight into the spectrograph at regular intervals over the exposures. 
The barycentric velocity correction should be determined for the time the signals arrive at the 
solar system barycenter --- i.e., at the Barycentric Julian date in Barycentric Dynamical 
Time (${\rm BJD_{TDB} }$). An error of 1 second in time introduces a semi-amplitude error of 
about 3 \cms, so to reach an error of less than 1 \cms\ the midpoint time must be accurate to 
about 250 milli-seconds \citep{Eastman2010, Wright2014}.  

\begin{figure*}[htp]     
\begin{center}
\includegraphics[width=8cm, clip]{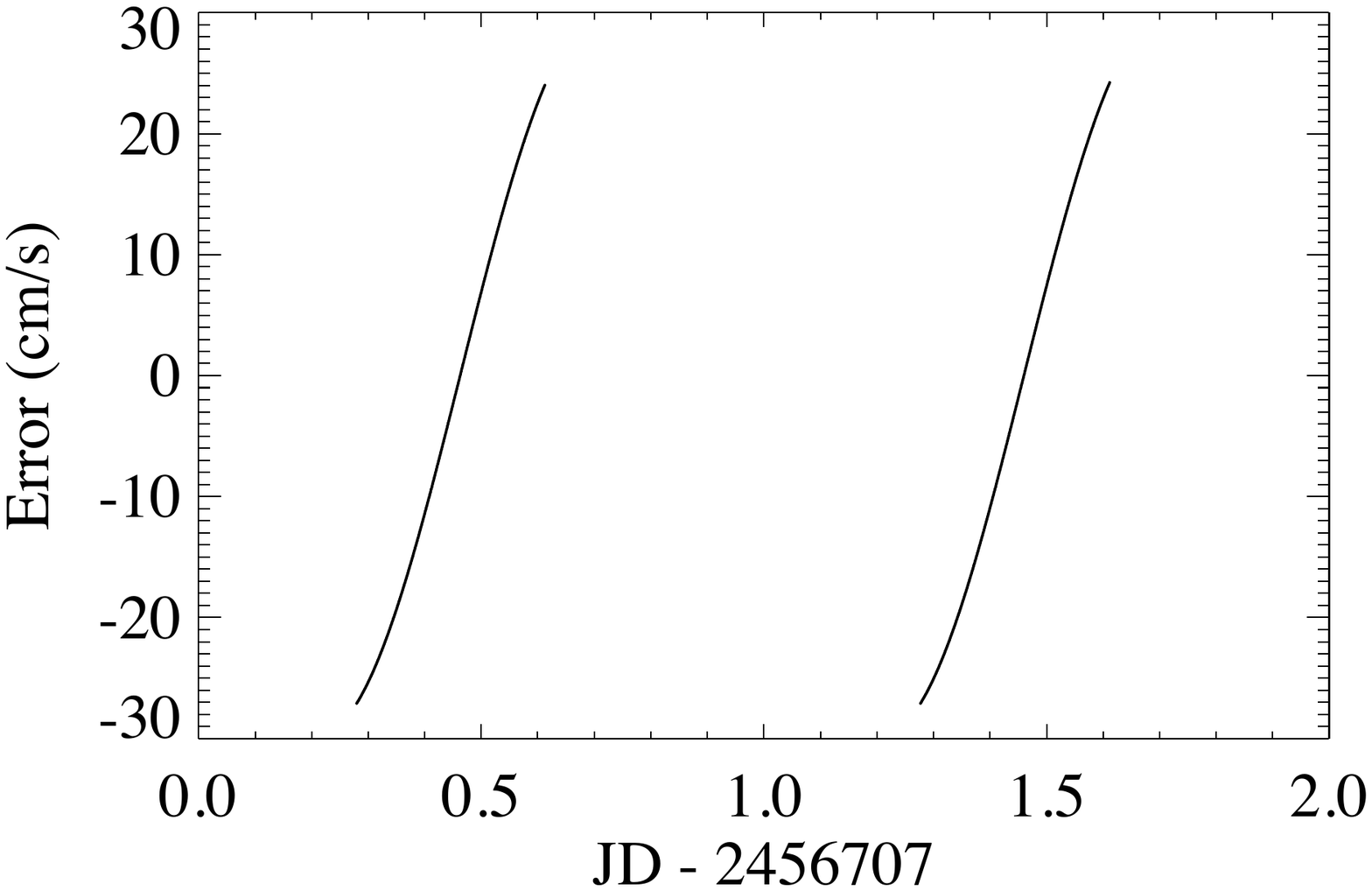}
\includegraphics[width=8cm, clip]{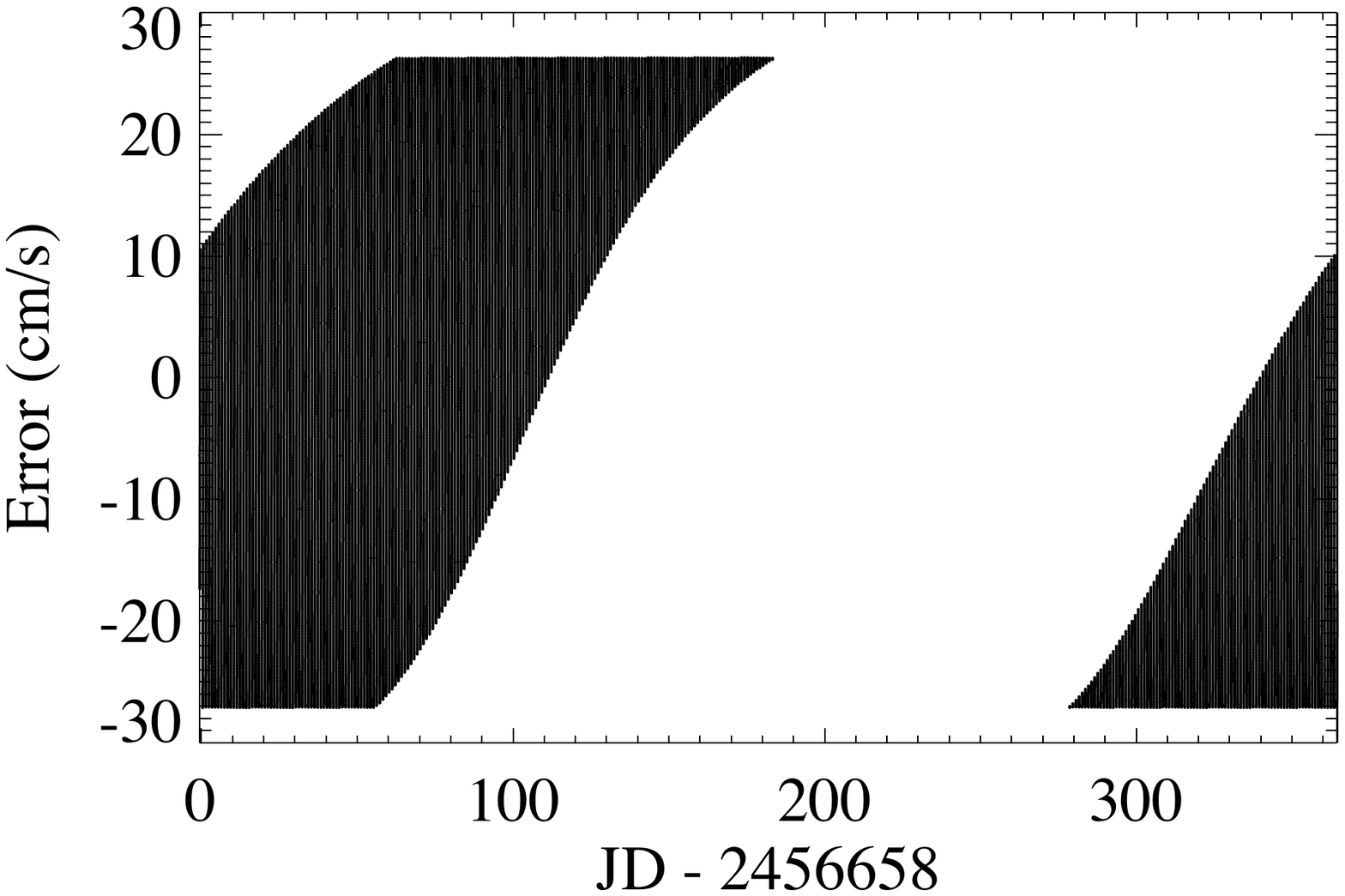}
\caption{Radial velocity error when acceleration of the barycentric velocity is not properly handled 
(i.e., when a single barycentric correction is derived for the flux-weighted midpoint time, which is 
current standard practice in the community). This simulation assumes 30-minute observations of Tau Ceti 
taken from CTIO with uniform flux during the observation. Velocity errors are only calculated for 
times when the star would be observable (i.e., airmass less than 2 and the Sun below $-18$ degree twilight).
(left) RV errors that could be incurred over a 2-day diurnal cycle. (right) RV errors over 
a full year when the same observing constraints are applied. (Figure courtesy of Lars Buchhave, Jason Eastman, Jason Wright)                                                                                                                                                          
\label{fluxweight} }   
\end{center}                                                                                                       
\end{figure*}

An effect that was not addressed by \citep{Wright2014} is the need to integrate 
the barycentric correction over the exposure time (Figure \ref{fluxweight} is provided from an article 
in prep, Buchhave, Eastman \& Wright 2016). A barycentric correction should be calculated for 
each sampled flux measurement and then weighted by the photon counts. 
This updates the older procedure of calculating a single barycentric correction to the 
photon-weighted midpoint time. The older procedure was deemed as acceptable for most state of the art 
Doppler surveys because the relatively short exposures incurred an error from the acceleration 
term in the barycentric velocity that was far below other terms in the error budget. 
However, this is an error source that must be handled more carefully for higher precision surveys. 
Even assuming uniform flux during an exposure, the older procedure can introduce a velocity 
error of $\pm 25$ \cms\ for a 30 minute exposure; this scales as the square of the exposure 
time.  

The optimal flux sampling rate depends, unfortunately, on the apriori unknown rate of variability in the atmosphere. 
If the atmospheric opacity is constant, then the midpoint time precision is independent of sampling
and lower sampling rate is balanced by the benefit of extra photons. Even in an extreme scenario 
with a 6 second "blackout" at the end of an otherwise uniform 5-minute exposure, integrating flux 
in 1 minute bins would produce an error in the flux-weighted midpoint time of only 0.6s.  Although 
more detailed analysis is warranted on the subject, binning flux every $\sim$10 seconds is 
reasonable for almost every imaginable weather condition and 
still provides good sampling for short exposure times of about one minute. 
For extreme RV precision, it is recommended that the exposure times are limited to about 20 minutes to 
minimize the error in the barycentric velocity calculation from nonlinear terms discussed above. 

For an accuracy in the barycentric velocity of $\sim 1$ \cms\, it is also important to 
sample photon counts as a function of wavelength. For example, if observations are 
taken at large zenith angle without an atmospheric dispersion compensator (ADC) the 
image of the star will be chromatically dispersed; if guiding favors 
collection of the red light for the first half of the exposure and then collection of 
blue light for the second half of the exposure, the flux-weighted midpoint would be
different for the red and blue photons. However, even with a good ADC, variable 
wavelength-dependent extinction can occur as the telescope tracks the star through 
a range of hour angle, even for short exposures. A basic exposure meter design might use a 
dichroic to divide the light between two channels; however, a low resolution spectrometer 
with 8 - 10 channels is probably preferable for providing a better correction as long as 
a fast readout can maintain $\sim$10 second sampling with high SNR.  Once the wavelength-dependent 
barycentric correction has been derived, it should be interpolated (linearly or weighted with an 
extinction curve) for each wavelength chunk or order and applied when weighting and combining the 
velocities from different parts of the spectrum.  For example, if each order is cross-correlated, 
then a barycentric correction can be calculated for the central wavelength and applied on an 
order-by-order basis before calculating the average velocity. 

Recommendations for good practices in deriving precise barycentric corrections include 
sampling the flux during the exposure (at least every minute, but ideally every few seconds), 
recording all times explicitly as a function of wavelength, ensuring the accuracy of the time 
stamp in the file headers, and including the time standard (usually GMT or UT).

\subsection{Telluric contamination}
Telluric absorption lines from many species (e.g., ${\rm H_2O, O_2, CH_4, CO_2}$) are 
imprinted in a stellar spectrum when starlight passes through the Earth's 
atmosphere. Some molecular species are well-behaved with only small seasonal 
variations in column density, while the lines associated with water can vary in depth with changing 
humidity throughout the night (temporally) or with different positions on the sky (spatially).  
There is some variability in the wavelengths and line profiles of telluric lines because of atmospheric 
winds, generally limited to Doppler shifts of about 10 \ms\ \citep{Figueira2010b}, corresponding 
to a fraction of a pixel.  More important, the telluric lines raster across the stellar spectrum with 
time because of the barycentric velocity of the Earth.

\begin{figure*}[htb]
\begin{center}
\includegraphics[height=15cm, width=0.9\linewidth]{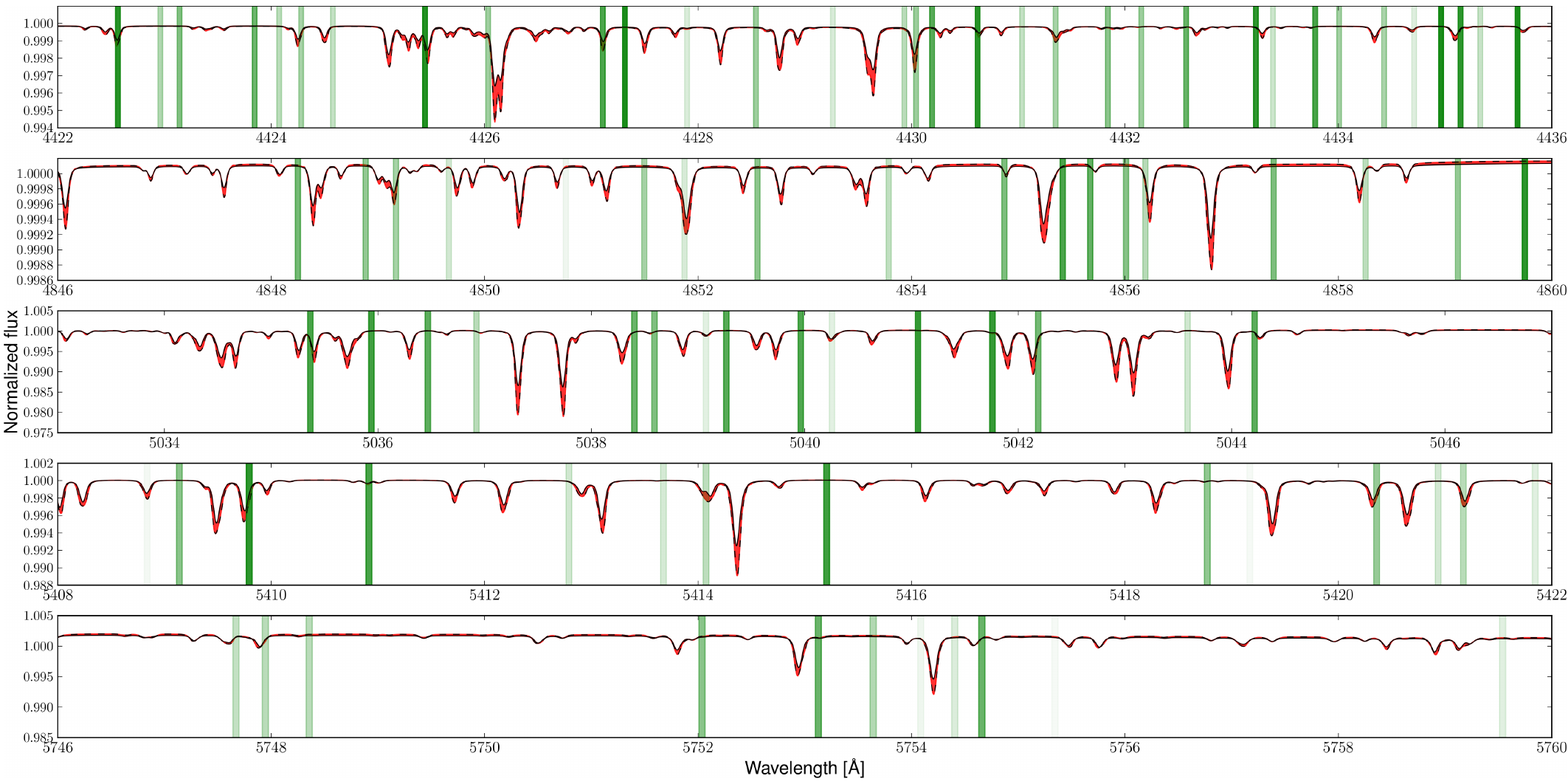}
\caption{Reproduced Fig 1 from \citet{Cunha2014}. Spectrum of micro-telluric lines in the optical spectrum 
between 442 and 570 nm. (Figure courtesy of Diana Cunha and Nuno Santos).
\label{micro-tell}}
\end{center}
\end{figure*}

Like any attribute that perturbs the shape of the SLSF, telluric contamination
must be properly treated for high precision radial velocity measurements.\footnote{The telluric 
contamination breakout session was organized by Sharon Wang with panel members Jason Wright, 
Cullen Blake, Pedro Figueira, Nuno Santos, Peter Plavchan, Andreas Seifahrt, Claire Moutou, 
Francois Bouchy.}  Left untreated, telluric contamination is most important in the near infra-red (NIR) 
where it has an impact of at least a few meters per second \citep{Bean2010} and where telluric lines 
can be quite opaque (with $\tau > 0.2$). The impact at optical wavelengths is of order $\sim 0.2 -- 1$ \ms\ where 
(Figure \ref{micro-tell}) micro-telluric lines have depths of 0.2 -- 2\% \citep{Cunha2014, Artigau2014a}.

\begin{figure*}[htb]
\begin{center}
\includegraphics[width=0.9\linewidth, angle=0]{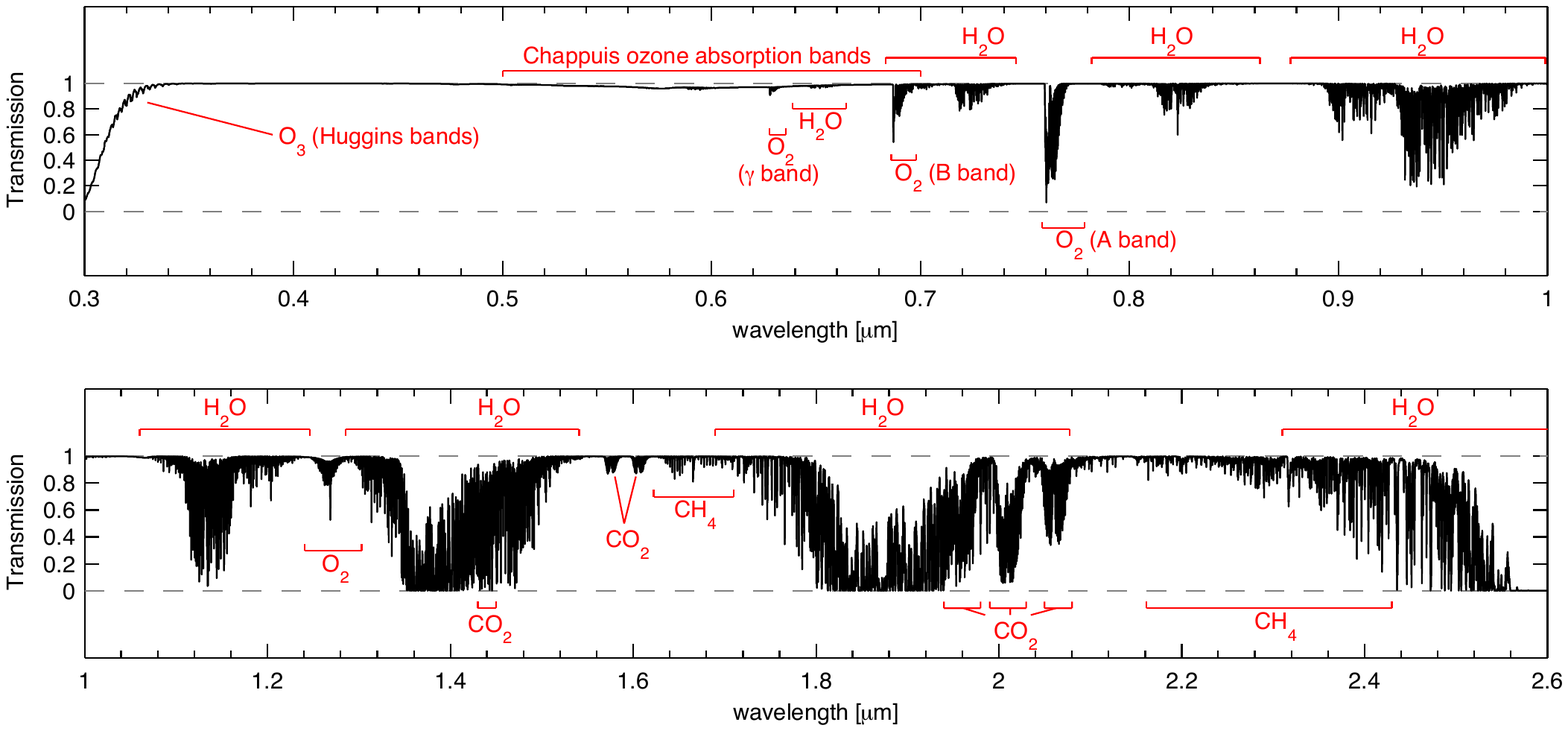}
\caption{Reproduced Fig 1 from \citet{Smette2015}. Spectrum of telluric lines synthesized with a line-by-line radiative transfer code using the 
annual mean profile for Cerro Paranal at a resolution of about 10,000. (Figure courtesy of Wolfgang Kausch, Alain Smette, Stefan Kimeswenger, Stefan Noll).
\label{full-tell}}
\end{center}
\end{figure*}

Most groups correct or model telluric lines with physically motivated synthetic models that 
have a comprehensive line list and use radiative transfer with accurate atmospheric 
profiles, which is demonstrated to be more accurate than empirical correction using telluric 
calibration frames \citep{Gullikson2014,Smette2015}. Most codes are base on the line-by-line 
radiative transfer model \citep[LBLRTM;][]{Clough2005} and the High Resolution Transmission 
(HITRAN) line database \citep{Rothman2013}. Examples of published codes are: 
TAPAS \citep{Bertaux2014}, TelFit \citep{Gullikson2014}, Molecfit \citep{Smette2015}, 
and TERRASPEC \citep{Bender2012}. A model of the telluric spectrum from 0.3 to 30 microns 
is shown in Figure \ref{full-tell}. These codes model telluric lines to a precision of 2 -- 5\%; 
however, for deep lines or lines with large uncertainties in the HITRAN database, the precision is 
not as good.  Poorly fitted lines are masked out or rejected \citep{Bean2010, Seifahrt2010, Blake2010}. 

The current telluric modeling precision is limited by: missing lines in the HITRAN database, uncertainties or 
errors in attributes such as the line position, strength, or shape, limitations in current modeling codes 
for deriving correct line profiles (i.e., velocity dependence, line mixing effects), insufficient knowledge of 
real time atmospheric conditions (e.g. water column density variations), and wind-induced line shifts.  There are 
annual meetings on HITRAN where help could be requested for specific lines or species under a 
range of physical conditions (excitation levels, temperatures). It is also possible that telluric modeling 
can be improved with empirical corrections using on-sky data or a Fourier Transform Spectrograph 
(FTS) observation although micro-tellurics will be difficult to observe with high SNR. Perhaps a 
version of HITRAN can be designed for astronomy applications through compilation and modeling 
of a large volume of telluric calibration data. This ``Astro-HITRAN" would be simpler but more robust 
for astronomical uses for certain molecular species, and it would allow for scaling of line strengths, 
for example, for different amount of water vapor content. The variable water species still 
present a modeling challenge, and \citet{Blake2011} have suggested a clever use of 
global positioning satellite measurements of precipitable water vapor as a prior when 
scaling these features.    

\section{Synergy with other fields}
Precision radial velocities are critical to the success, efficiency and 
scientific yield of future space missions\footnote{Presentation by Alessandro Sozzetti}. 
There is obvious synergy with transit missions (TESS, CHEOPS, PLATO) in deriving 
bulk densities for thousands of exoplanets. New connections will be made with Gaia, where 
of order 10,000 new gas giant planets are expected to be detected. The combination of Gaia 
astrometry and radial velocities will illuminate several questions about planet formation: the 
true mass distribution of jovian planets beyond the ice line, the frequency of solar system 
analogs, whether super Earths are regularly accompanied by giant planets and how gas giant 
planets affect the presence of terrestrial planets. 

Radial velocity measurements can also enhance the productivity of 
direct imaging missions by providing orbital phase information and information about inner 
companions. The ideal targets for direct imaging are young stars where the gas giant planets 
will be brighter; these will be amenable to RV follow-up if successful techniques 
are developed for mitigating photospheric noise in these active stars.  

\section{Summary}
A comparison of several ongoing Doppler surveys shows that most teams have reached about 
1 \ms\ in single measurement radial velocity precision. The best performing spectrometer is the 
purpose-built vacuum-enclosed HARPS instrument, with R=115,000, a broad wavelength range 
from 400-700 nm, and a fiber feed system that utlizes octagonal fibers and an optical double scrambler. 
HARPS reaches a precision 
of about 0.8 \ms\ for observations with SNR of 200 in 3 \kms\ bins at 500 nm. The next generation 
instruments will need to improve upon HARPS with better illumination stability,  
higher fidelity spectra (more stable spectral line spread function, higher resolution and higher SNR), 
broader wavelength bandpasses and more stable and precise wavelength calibration, and improved 
detector performance and characterization efforts. 

Tremendous progress has been made over the past five years on technology development for 
high-resolution optical spectrometers. In particular, the development of frequency 
combs for the wavelength calibration of spectrometers for Doppler planet searches is now mature 
technology. Laboratory experiments and initial tests on HARPS indicate that wavelength calibration 
is likely not the primary limitation in achieving 10 \cms\ Doppler precision. 
Frequency combs have become powerful tools for characterizing spectrometer drift, estimating the 
spectral line spread function across the instrument bandpass, and characterizing and calibrating 
detector imperfections.  The remaining work to be done with frequency comb technology is primarily technical implementation, including making systems robust and reliable, and mitigating modal noise when coupling comb light into the spectrograph. A turn-key laser frequency comb is now available from Menlo Systems that shows great promise as a long-term calibration device. The Menlo combs have been in operation at the VTT solar telescope in Tenerife (since 2012), at HARPS in La Silla (permanent since 2015), at FOCES in Munich at USM (since 2015) and at Xinglong China (since February 2016). At USM the LFC has been running continuously for several months. Beyond the more classical astro-comb calibration technologies, there is significant innovation in the fields of microcombs, electro-optical modulation combs, tunable Fabry-Perots, and stabilized etalons as well. 

Fiber optic cables are also a well-understood technology. Not only does optical fiber
technology allow for spectrometers to be placed in convenient locations relative to
telescope focus, but modern fibers can also improve instrument illumination stability
significantly while maintaining high efficiency. Care must be taken when connectorizing
instrument fiber cables to ensure stress-free mounting (thereby minimizing FRD.)
Standard fibers used in astronomical instruments do incur modal noise penalties, as only
a finite number of modes can be populated within the fiber, but much of this can be
mitigated through mechanical agitation. However, modal noise is a significant issue for
coherent sources, such as laser frequency combs, and our community must develop new
mitigation techniques if the exquisite precision of next generation wavelength calibration
sources is to be fully realized by future spectrometers

Telluric contamination is another area that the community will need to address for high precision
Doppler measurements. The problem is most challenging for infrared observations, but micro-tellurics 
at optical wavelengths are also a concern at sub meter per second precision. 
Current approaches involve forward modeling to fit for 
tellurics or masking out contaminated pixels in the spectrum.  Routine monitoring of atmospheric 
conditions (water vapor distribution and column density, wind etc) is recommended and this information 
should be stored as meta-data with the spectra. 

Barycentric corrections have been demonstrated to be accurate to about 2 \cms\ \citep{Wright2014}. 
The flux-weighted exposure midpoint time should not be used for the correction; instead, barycentric velocity
corrections should be calculated throughout an exposure and combined by weighting each
point by the exposure meter data.  
Color-dependent barycentric corrections should also be calculated by collecting the chromatic flux time series as a 
function of wavelength (in a 
few or several wavelength bins).  Any stellar exposure with sufficient SNR to achieve $< 1$ \ms\ precision 
should provide enough photons for time series at several wavelength bins.
The final Doppler velocity is the combination of measured velocities from a set of lines or wavelength 
bins; each bin or line would then be corrected according to the BC from the appropriate wavelength region.
A high performance atmospheric dispersion corrector
(ADC) will reduce the sensitivity to flux-weighted barycentric corrections.

The community is now exploring correct statistical techniques with the goal 
of improving reproducibility and extracting weaker signals in the presence of time-correlated noise. 
Frequentist techniques are faster, but Bayesian techniques are more reliable in estimating the 
true errors and uncovering underlying exoplanet populations from our data.  There is common 
misuse of null hypothesis tests like the p-value or Kolmogorov-Smirnoff test. Astronomers
are encouraged to team up with statisticians to avoid the pitfalls of applying statistical tests 
as black boxes. Many of the techniques that seem new to astronomers are well understood techniques 
for statisticians. 

There was extensive discussion about how to handle stellar noise. In the past, it was adequate to 
identify and decorrelate trends in radial velocities based on calcium H \& K emission, 
or the FWHM or line bisector of the cross correlation function.  However, we now appreciate 
that photospheric velocities, or jitter, is fundamentally imprinted in the spectrum differently 
from a Keplerian Doppler shift.  There are many techniques 
that show promise in distinguishing jitter from center of mass velocities, including principal 
component analysis of spectral lines or the cross correlation function. Stellar jitter is currently the main
limitation in the detection of small rocky planets and it merits significant effort by the community. 
It will not work to avoid stars with jitter (because stellar activity varies over time). We are 
advancing our understanding of the physics of stellar noise with solar data and photometry from Kepler. 
Plage is a more important contributor (90\% level) to stellar jitter than spots (a 10\% problem) 
for slowly rotating stars.  

It is not an adequate strategy to average down hundreds of observations; this wastes precious 
telescope time and systematic errors don't necessarily average down as the square root of the number of observations. 
The RV challenge led by Xavier Dumusque showed that the community has not been successful in 
detecting velocity amplitudes that are smaller than the single measurement precision, even when 
hundreds of observations are obtained. We must improve single measurement precision as a first 
step toward detecting low amplitude exoplanetary systems. We must learn how to distinguish 
stellar noise from Keplerian velocities. 

There has been negative advocacy by some in the exoplanet field who claim 
that a fundamental floor of precision is imposed by photospheric velocities, or stellar jitter 
at the level of 1 \ms. That statement may be true for many of the current spectrometers and 
analysis techniques.  At a resolution of 50,000 on spectrometers that are not stabilized, 
stellar noise from chromospherically quiet stars cannot be distinguished from Keplerian velocities. 
Fortunately, the community at large has continued to press forward on this issue and it seems likely 
that an instrumental precision of 10 \cms\ will be achieved with the next 
generation of high fidelity spectrometers. We do not yet know how this will translate into  
detectability for low amplitude signals in the presences of unavoidable stellar noise; however, it is 
premature to speculate that we cannot do better. This is an area of active research in the 
community and there has been promising progress. Research to distinguish stellar activity from 
center of mass Doppler shifts must continue to be a high priority, simply because the exoplanet 
endeavor cannot expand\footnote{Survey input from Rosemary Mardling} if we do not solve 
this problem and extract weaker signals in the presence of time-correlated noise.  

\section{Findings and Conclusions}
There have been several reports highlighting the importance of precise radial velocities for
supporting NASA missions, including the 2006 Exoplanet Task Force \citep{Lunine2008}, the 
2010 Decadal Survey \citep{Blanford2010}, and the 2012 NSF Portfolio Review report \citep{Eisenstein2012}. 
The cancellation of the \$2B Space Interferometry Mission 
was accompanied by a statement that ground-based radial velocity measurement precision could be 
pushed to 10 \cms, enabling at least partial characterization of the architectures of nearby 
planetary systems.  
The NASA ExoPAG commissioned a report "Radial Velocity Prospects 
Current and Future" \citep{Plavchan2015} with findings from the precision radial velocity community\footnote{Presentation by Scott Gaudi} that complements and foreshadows many of 
the findings from the EPRV workshop. This report highlights a key challenge: EPRVs have 
transitioned from small PI-based programs to ``big" science that requires dedicated resources;
however, the NASA and NSF budget models have not kept up with this. 
While radial velocity precision has been stuck at 1 \ms\ for 
the past several years, there has been progress in technology and analysis techniques that offers 
promise for moving toward 10 \cms\ precision and this progress needs to be communicated to 
the astronomy community at large and to the funding agencies.  

The ``Big Three" science goals for extreme precision radial velocities are: 
\begin{itemize}
\item Confirmation and characterization of the many transiting planets discovered by transit surveys 
(ground-based, Kepler, K2, TESS, CHEOPS, PLATO). EPRV measurements will uniquely provide 
mass measurements needed for bulk density measurements. 
\item Identification and orbital characterization of planets down to ice giant masses around 
FGK stars to be imaged by WFIRST-AFTA. 
\item The discovery of potentially habitable planets around the nearest and brightest stars. 
\end{itemize}

Resources will be required to meet these ambitious goals.  Habitable planets 
around FGK type stars have velocity semi-amplitudes of order 10 \cms\ and periods of 
about one year. The discovery of these planets requires high statistical precision 
and high cadence. Additional work needs to focus on reducing any systematic errors or astrophysical 
noise to considerably less than 10 \cms.  In additional to controlling systemic errors resulting from instrumental 
and calibration effects, considerable additional work needs to focus on the critical and difficult task of removing, 
suppression, or separation of intrinsic stellar noise to considerably better than 10 \cms. 
This is {\it in principle} a tractable problem because the Doppler variation due 
to orbiting bodies has a unique signature: all of the lines move by a known amount without 
changing their shape. Current instruments and detection algorithms are not going 
to solve this problem; the RV fitting challenge by Xavier Dumusque (Section \ref{RVchallenge}) 
showed that no one can reliably detect planets with amplitudes below 1 \ms\ in current data sets. 
We need higher fidelity spectra to make progress and this is a problem that will require the next generation 
instruments.  

\subsection{Recommendations}
In support of the key role that precision spectroscopy has for space missions, 
NASA has commissioned the Extreme Precision Doppler Spectrograph (EPDS) 
for the WIYN 3.5-m telescope\footnote{Presentation by Mario Perez}. This is a start, but 
one new instrument is not enough. There is a role for small telescopes with high cadence 
observations and larger aperture telescope that obtain high resolution, high SNR observations.  
Participants at this workshop discussed the big ideas that could catapult progress in the field 
and enable success for the highest priority science goals of precision Doppler surveys.

\begin{enumerate}
\item {\bf Dedicated moderate to large aperture telescopes.} High SNR and very high cadence data are 
required over long time baselines; this could be achieved for hundreds of stars using dedicated 4 to 10 meter 
class telescopes.  Most of the struggles we have with aliasing and stellar activity is complicated 
by under-sampled data sets.  In the same way that Kepler transit detections benefited from high 
cadence sampling, Doppler measurements would benefit from a paradigm shift with many more 
observations, perhaps a world-wide effort with coordinated observations at several longitudes for 
a few stars. 
\item {\bf Stable Spectrometers.} Instruments with exquisite long-term stability are required: spectrometers that are fiber
fed with high illumination stability, excellent wavelength calibration, and precise
temperature and pressure control represent the immediate future of precision RV
measurements. Optical spectrometers contain significant amounts of Doppler information
for solar-type stars, but radial velocities from NIR spectrometers have less contamination
from stellar activity (spots as well as plages). Based on current technology, solutions to
the technical challenges presented at the workshop seem within reach for the
spectrometers that are being built today.
\item {\bf Proper treatment of stellar noise.} It is likely that distinguishing between stellar 
activity and Keplerian velocities will 
require very high resolution (perhaps $\sim$150,000), and expanded wavelength coverage 
(optical plus near infrared). The high resolution is not required for measuring line 
centroids; the line centroid information is resolved at $R\sim 60,000$. However, the need 
to measure higher-order line shape variations, which are a clear signal of stellar activity or changes in the 
instrumental line spread function rather than a true Doppler shift, likely require much higher resolution and 
sampling, as well as higher SNR.  Simultaneous high-resolution optical and infrared monitoring can mitigate 
any pathological cases where stellar activity or instrumental effects may cause line centroid shifts without strong 
line shape variations, since these shifts are neverthless likely to wavelength dependent.  In particular, the 
ionization and line formation depth in the photosphere depends on the wavelength, and thus so do the 
absorption line shapes through the photospheric velocities.   
\end{enumerate}

All of this leads to a vision for precision radial velocity ``dream machines" requiring 
large aperture telescopes with dual optical and infrared channels (0.4 to 1.7 $\micron$), 
very high optical resolution ($R > 150,000$), high IR resolution ($R > 50,000$), fiber-fed with 
a very high scrambling gain, environmentally stabilized, and with advanced wavelength 
calibration (e.g., laser frequency combs).  
The cost would be of order \$20 M per instrument and ideally there would be few of these at 
various longitudes. In order to carry out the science, these machines would require an assurance 
of a significant allocation of observing time over several years. 
 
This is a cost that at first glance seems out of reach. However, if precisions approaching 10 \cms\ can 
be achieved, Doppler surveys may be able to locate the host stars of planets that would be observed by 
the next generation of flagship direct imaging missions, such as the Habitable Exoplanet Imaging Mission 
and the Large UV-Optical-IR (LUVOIR) mission, both of which will be studied by NASA over the next few years 
\citep[see PAG Reports:][]{Sembach2015, Boch2015, Gaudi2016}.
The cost for such missions is to be determined, but is certainly greater than one billion (by definition for a 
flagship), and may be up to ten(s) of billions of dollars for the more ambitious architectures that will be 
considered. The choice of the flagship mission, the technical specifications and 
the scientific productivity of a flagship mission may well be dependent on whether exoplanets 
are already known or whether they need to be discovered and it still be helpful to measure the mass 
of the planet after they are discovered by these direct imaging missions. EPRV is likely to be the only techniques that 
offers promise for detecting low mass planetary systems around nearby stars before the launch of the next 
flagship mission.  An investment at the \$100M level would be a wise investment that 
could ultimately save billions of dollars toward the goal of characterizing small rocky planets. 

\acknowledgements
We thank the anonymous referee for providing an exceptionally careful review of this paper and for 
many suggested changes which improved the quality of the paper. 
We are grateful to Yale University, to the National Science Foundation grant AST-1458915 and to the NASA Exoplanet Science 
Institute (NExScI) for providing financial support that enabled this workshop and provided support for participants.  
D. A. Fischer thanks NASA NNX12ACG01C for inspiring the study of extreme precision radial velocity measurements. 
R. I. Dawson acknowledges the Miller Institute at the Univ of California, Berkeley for Basic Research in Science.
S. A. Diddams acknowledges support from NIST and the NSF grant AST-1310875.
X. Dumusque would like to thank the Society in Science for its support through a Branco Weiss Fellowship.
E. B. Ford was supported in part by NASA Exoplanet Research Program award NNX15AE21G. 
Work by B. S. Gaudi was partially supported by NSF CAREER Grant AST-1056524. 
G. Laughlin acknowledges support from the NASA Astrobiology Institute through a cooperative agreement between NASA Ames Research Center and the University of California at Santa Cruz, and from the NASA TESS Mission through a cooperative agreement between M.I.T. and UCSC.
A. Reiners acknowledges support from the European Research Council under the FP7 Starting Grant agreement number 279347 and from DFG grant RE 1664/9-1. 
N. C. Santos and P. Figueira acknowledge support by Funda\c{c}\~ao para a Ci\^encia e a Tecnologia (FCT) through the 
research grants UID/FIS/04434/2013 and PTDC/FIS-AST/1526/2014 as well as through Investigador FCT contracts 
of reference IF/01037/2013 and IF/00169/2012, and POPH/FSE (EC) by FEDER funding through the 
program ``Programa Operacional de Factores de Competitividade --- COMPETE." 
P. Figueira further acknowledges support from FCT in the form of an exploratory project reference IF/01037/2013CP1191/CT0001. 
A. Szentgyorgyi thanks the Giant Magellan Telescope Organization for their support for much the work described 
in his contribution to this paper under Contract  No. GMT-INS-CON-00584. 
S. X. Wang is supported by a NASA Earth and Space Science Fellowship (NNX14AN81H). S. X. Wang and J. T. Wright 
acknowledge support from NSF grant AST-1211441 for the work on telluric contamination. J.T. Wright acknowledges 
NSF grant AST1109727 and NASA grant NNX12AC01G for work on barycentric corrections. 
The Center for Exoplanets and Habitable Worlds is supported by the Pennsylvania State University, the Eberly College of 
Science, and the Pennsylvania Space Grant Consortium. 
The PARAS program is fully supported and funded by the Department of Space, Govt. of India and 
Physical Research Laboratory (PRL), Ahmedabed, India. The PARAS 
team would like to acknowledge the support of the Director, PRL, and the Mt. Abu Observatory staff for running the 
program and Francesco Pepe (Geneva Observatory) and Larry Ramsey (Pen State University) for many scientific 
and technical inputs. Two of the PARAS team members, Suvrath Mahadevan and Arpita Roy would like to thank 
the Centre of Exoplanets and Habitable Worlds at Penn State University for their partial support. 
The McDonald Observatory planet search is currently supported by the National Science Foundation under Astrophysics grant AST-1313075, and has been supported in the past by various NASA grants.

\end{document}